\documentclass[%
 reprint,
 amsmath,amssymb,
 aps,nofootinbib,
 prl,superscriptaddress
 ]{revtex4-2}

\usepackage{hyperref}

\usepackage[utf8]{inputenc}

\usepackage{graphicx}
\usepackage{dcolumn}
\usepackage{bm}
\usepackage{multibib}

\newcommand{\ket}[1]{\ensuremath{| {#1} \rangle }}
\newcommand{\bra}[1]{\ensuremath{\langle {#1} |}}

\renewcommand{\vec}[1]{\bm{#1}}

\newcites{SM}{Supplementary Material Bibliography}

\begin{document}

\title{Probing the energy-smeared $R$-ratio on the lattice}

\author{Constantia Alexandrou}
\affiliation{Department of Physics, University of Cyprus, 20537 Nicosia, Cyprus}
\affiliation{Computation-based Science and Technology Research Center, The Cyprus Institute, 20 Konstantinou Kavafi Street, 2121 Nicosia, Cyprus}

\author{Simone Bacchio}
\affiliation{Computation-based Science and Technology Research Center, The Cyprus Institute, 20 Konstantinou Kavafi Street, 2121 Nicosia, Cyprus}

\author{Alessandro De Santis}%
\affiliation{Dipartimento di Fisica and INFN, Universit\`a di Roma Tor Vergata, Via della Ricerca Scientifica 1, I-00133 Roma, Italy}

\author{Petros Dimopoulos}
\affiliation{Dipartimento di Scienze Matematiche, Fisiche e Informatiche, Universit\`a di Parma and INFN, Gruppo Collegato di Parma, Parco Area delle Scienze 7/a (Campus), 43124 Parma, Italy}

\author{Jacob Finkenrath}
\affiliation{Computation-based Science and Technology Research Center, The Cyprus Institute, 20 Konstantinou Kavafi Street, 2121 Nicosia, Cyprus}

\author{Roberto Frezzotti}
\affiliation{Dipartimento di Fisica and INFN, Universit\`a di Roma Tor Vergata, Via della Ricerca Scientifica 1, I-00133 Roma, Italy}

\author{Giuseppe Gagliardi}
\affiliation{Istituto Nazionale di Fisica Nucleare, Sezione di Roma Tre, Via della Vasca Navale 84, I-00146 Rome, Italy}

\author{Marco Garofalo}
\affiliation{HISKP (Theory), Rheinische Friedrich-Wilhelms-Universit\"at Bonn,
	Nussallee 14-16, 53115 Bonn, Germany}

\author{Kyriakos Hadjiyiannakou}
\affiliation{Department of Physics, University of Cyprus, 20537 Nicosia, Cyprus}
\affiliation{Computation-based Science and Technology Research Center, The Cyprus Institute, 20 Konstantinou Kavafi Street, 2121 Nicosia, Cyprus}

\author{Bartosz Kostrzewa}
\affiliation{High Performance Computing and Analytics Lab, Rheinische Friedrich-Wilhelms-Universit\"at Bonn, Friedrich-Hirzebruch-Allee 8, 53115 Bonn, Germany}

\author{Karl Jansen}
\affiliation{NIC, DESY, Platanenallee 6, D-15738 Zeuthen, Germany}

\author{Vittorio Lubicz}
\affiliation{Dipartimento di Matematica e Fisica, Universit\`a Roma Tre and INFN, Sezione di Roma Tre, Via della Vasca Navale 84, I-00146 Rome, Italy}

\author{Marcus Petschlies}
\affiliation{HISKP (Theory), Rheinische Friedrich-Wilhelms-Universit\"at Bonn, Nussallee 14-16, 53115 Bonn, Germany}

\author{Francesco Sanfilippo}
\affiliation{Istituto Nazionale di Fisica Nucleare, Sezione di Roma Tre, Via della Vasca Navale 84, I-00146 Rome, Italy}

\author{Silvano Simula}
\affiliation{Istituto Nazionale di Fisica Nucleare, Sezione di Roma Tre, Via della Vasca Navale 84, I-00146 Rome, Italy}

\author{Nazario Tantalo}
\affiliation{Dipartimento di Fisica and INFN, Universit\`a di Roma Tor Vergata, Via della Ricerca Scientifica 1, I-00133 Roma, Italy}

\author{Carsten Urbach}
\affiliation{HISKP (Theory), Rheinische Friedrich-Wilhelms-Universit\"at Bonn,
	Nussallee 14-16, 53115 Bonn, Germany}

\author{Urs Wenger}
\affiliation{Institute for Theoretical Physics, Albert Einstein Center for Fundamental Physics,
	University of Bern, Sidlerstrasse 5, CH-3012 Bern, Switzerland}

\collaboration{Extended Twisted Mass Collaboration (ETMC)}

\begin{abstract}
We present a first-principles lattice QCD investigation of the $R$-ratio between the $e^+e^-$ cross-section into hadrons and that into muons. By using the method of Ref.~\cite{Hansen:2019idp}, that allows to extract smeared spectral densities from Euclidean correlators, we compute the $R$-ratio convoluted with Gaussian smearing kernels of widths of about $600$~MeV and central energies from $220$~MeV up to $2.5$~GeV. Our theoretical results are compared with the corresponding quantities obtained by smearing the KNT19 compilation~\cite{keshavarzi2020g} of $R$-ratio experimental measurements 
with the same kernels and, by centring the Gaussians in the region around the $\rho$-resonance peak, a tension of about three standard deviations is observed. From the phenomenological perspective, we have not included yet in our calculation QED and strong isospin-breaking corrections and this might affect the observed tension. 
From the methodological perspective, our calculation demonstrates that it is possible to study the $R$-ratio in Gaussian energy bins on the lattice at the level of accuracy required in order to perform precision tests of the Standard Model.
\end{abstract}

\maketitle

\section{
\label{sec:introduction}
Introduction
}

The $R$-ratio between the $e^+e^-$ cross-section into hadrons with that into muons plays a fundamental r\^ole in particle physics since its introduction in Ref.~\cite{Cabibbo:1970mh}. In recent years, the importance of the $R$-ratio has been mainly associated with the fact that its knowledge, as a function of the center-of-mass energy of the electrons, allows to predict the leading hadronic contribution (HVP) to the muon anomalous magnetic moment ($a_\mu$) via a dispersive approach. The dispersive determinations of $a_\mu^\mathrm{HVP}$, reviewed in detail in Ref.~\cite{Aoyama:2020ynm}, are in strong tension (about four standard deviations) with the experimental determination of $a_\mu$. On the other hand, lattice determinations of (partial) contributions to $a_\mu^\mathrm{HVP}$, obtained without any reference to the experimental measurements of $R$, are in much better agreement with the $a_\mu$ experiment~\cite{Borsanyi:2020mff}. 

\paragraph{The focus of this paper is $R$, smeared with Gaussian kernels, and not $a_\mu$.}
The experiments that measure $R$ are radically different from those that measure $a_\mu$ and, moreover, $R$ is an energy-dependent probe of the theory while $a_\mu$ is natively a low--energy observable. For these reasons a detailed phenomenological investigation of $R$ represents an independent precision test of the Standard Model with respect to that provided by $a_\mu$.   
We address here the theoretical side of this problem by computing the energy-smeared $R$-ratio on the lattice with the required non-perturbative accuracy. 

To this end, we rely on our effort within the ETMC that produced a collection of state-of-the-art lattice QCD ensembles with four dynamical Twisted Mass quark flavours~\cite{Frezzotti:2000nk} at physical pion masses together with the Euclidean correlators with two insertions of the hadronic electromagnetic current (see TABLE~\ref{tab:ensembles} and Ref.~\cite{Alexandrou:2022amy}). From these correlators, by using the method proposed in Ref.~\cite{Hansen:2019idp} and recently validated in Ref.~\cite{Bulava:2021fre} (see also Ref.~\cite{Bulava:2023mjc}), we extract the $R$-ratio smeared with normalized Gaussian kernels, $G_\sigma(\omega)=\exp(-\omega^2/2\sigma^2)/\sqrt{2\pi \sigma^2}$, according to
\begin{flalign}
R_\sigma(E)=\int_{0}^\infty d\omega\, G_\sigma(E-\omega)\, R(\omega)\;. \label{eq:Rsigma}
\end{flalign}
We then compare our theoretical determinations of $R_\sigma(E)$ with experiments by smearing the $R$ measurements with the same Gaussian. In this way, by varying $E$ and $\sigma$, we probe $R$ in Gaussian energy bins of different widths (see also Ref.~\cite{Bertlmann:1984ih}). 
With $E$ around the $\rho$-resonance peak and at $\sigma\simeq 600$~MeV we manage to compute $R_\sigma(E)$ with an accuracy at the $2\%$ level. In these Gaussian bins our results are in tension (about three standard deviations) with experiments. 

From the phenomenological perspective, the observed tension might be ascribed to QED and strong isospin-breaking effects, that we have not included yet in our iso-symmetric QCD calculation, or to underestimated experimental uncertainties (see e.g.\ Ref.~\cite{CMD-3:2023alj}). From the methodological viewpoint, our results clearly demonstrate that it is possible to study the $R$-ratio in Gaussian energy bins on the lattice at the precision level required to perform precision tests of the Standard Model. 

A resolution in energy of $O(600)$~MeV can also be obtained by considering the so-called intermediate window contribution ($a_\mu^{\mathrm{HVP},W}$) to $a_\mu$. Presently, the comparison of the lattice determinations~\cite{Borsanyi:2020mff,Alexandrou:2022amy,FermilabLattice:2022smb,Ce:2022kxy} of $a_\mu^{\mathrm{HVP},W}$ with the corresponding dispersive determinations~\cite{keshavarzi2020g} represents a more stringent test of the Standard Model w.r.t.\ the one performed in this paper. Having demonstrated here that a precise lattice calculation of $R_\sigma(E)$ is possible, we plan in the near future to substantially reduce the widths of the Gaussian bins by increasing the statistical precision of our lattice correlators.

\section{
\label{sec:materialsandmethods}
Methods and Materials 
}
\paragraph{Methods.}In order to compute $R_\sigma(E)$ we start from the two-point Euclidean correlator of the quark electromagnetic current
\begin{flalign}
V(t)=-\frac{1}{3}\sum_{i=1}^3\int d^3x \,\mathrm{T}\bra{0}J_i(x)J_i(0)\ket{0}
\end{flalign}
where $J_\mu=\sum_{f}q_f \bar \psi_f \gamma_\mu \psi_f$ with $f=\{u,d,s,c,b,t\}$, $q_{u,c,t}=2/3$ and $q_{d,s,b}=-1/3$. These correlators are the primary data of our lattice simulations and are connected to the $R$-ratio by the well known formula
\begin{flalign}
V(t)=\frac{1}{12\pi^2}\int_{0}^\infty d\omega \, \omega^2 R(\omega)\, e^{-t\omega}\;.
\label{eq:vtinfinite}
\end{flalign}
%
Theoretically $R(\omega)$ is a distribution, the spectral density of the correlator $V(t)$, and \emph{has} to be probed by using suitable smearing kernels,
\begin{flalign}
R[K]=\int_{0}^\infty d\omega\, K(\omega)\, R(\omega)\;.
\end{flalign}
In this perspective the correlator $V(t)$ itself represents a class of observables, corresponding to $K(\omega)=\omega^2 \exp(-t\omega)/12\pi^2$, whose sensitivity to the energy dependence of $R$ can be varied by changing $t$. The window contributions~\cite{RBC:2018dos} to $a_\mu^\mathrm{HVP}$ are elements of another class of observables whose smearing kernels are natively well localized in the Euclidean-time domain (see e.g. FIG.~1 and FIG.~2 of Ref.~\cite{Alexandrou:2022amy}) but that can also be used to probe the energy dependence of $R(E)$ by changing the parameters that define the time-window (see Ref.~\cite{Colangelo:2022vok} and FIG.~\ref{fig:kernelW} below). 
By choosing $K(\omega)=G_\sigma(E-\omega)$ we provide here results for $R_\sigma(E)$, a class of observables that are natively well localized in the energy domain. 

The determination of $R_\sigma(E)$ on the lattice is possible, with controlled statistical and systematic errors, by using the method\footnote{An alternative, but closely related strategy, has recently been proposed in Ref.~\cite{Boito:2022njs}. We also point out to the readers familiar with the Bayesian literature on the subject that, by using the results of Ref.\cite{Valentine2020}, the method of Ref.~\cite{Hansen:2019idp} can be understood within the language of Gaussian Processes, see e.g.\ Refs.~\cite{Horak:2021syv,DelDebbio:2021whr,Candido:2023nnb} and the explanation provided in the supplementary material. See also the recent review~\cite{Rothkopf:2022fyo} for a critical discussion of the different methods, with emphasis on those based on Bayesian inference.} of Ref.~\cite{Hansen:2019idp}. The starting point of this approach is the following exact representation of the smearing kernel for $\omega>0$,
\begin{flalign}
\frac{12\pi^2G_\sigma(E-\omega)}{\omega^2} =\sum_{\tau=1}^{\infty} g_\tau\, e^{-a\omega \tau}\;,
\label{eq:DeltaRepExact}
\end{flalign}
where $\tau$ is an integer variable and $a$ is an arbitrary scale that, on the lattice, we identify with the lattice spacing. Once the coefficients $g_\tau\equiv g_\tau(E,\sigma)$ are known, $R_\sigma(E)$ can be computed according to
\begin{flalign}
R_\sigma(E)=\sum_{\tau=1}^\infty g_\tau\, V(a\tau)\;.
\label{eq:RRepExact}
\end{flalign}
Although the mathematics is quite simple the game is rather delicate from the numerical point of view. Indeed, since the sums in Eqs.~(\ref{eq:DeltaRepExact}) and~(\ref{eq:RRepExact}) have necessarily to be truncated, the goal is to find a finite set of coefficients such that both the systematic and statistical errors on the resulting approximation to $R_\sigma(E)$ can be kept under control. 
The smoother the kernel is the simpler is the game. The numerical problem rapidly becomes ill-posed for $\sigma\ll E$ (see Refs.~\cite{Hansen:2019idp,Bulava:2021fre} for illustrative numerical evidences of this fact). In this regime, any procedure aiming at minimizing the systematic error due to the imperfect reconstruction of the kernel produces coefficients $g_\tau$ that are huge in magnitude and oscillating in sign. As a consequence, any tiny error on $V(a\tau)$ is amplified when the truncated sum of Eq.~(\ref{eq:RRepExact}) is evaluated. The algorithm of Ref.~\cite{Hansen:2019idp} provides a regularization mechanism to this problem. We refer to Refs.~\cite{Hansen:2019idp,Bulava:2021fre} for extended discussions of this point and to the supplementary material for the details of the numerical implementation performed in this work.

\paragraph{Materials.}The lattice gauge ensembles used in this work, generated by the ETMC, are listed in TABLE~\ref{tab:ensembles} and described in full details in Ref.~\cite{Alexandrou:2022amy} together with the lattice correlators $V(t)$, used there to compute the short and intermediate window contributions to $a_\mu^\mathrm{HVP}$ and here to compute $R_\sigma(E)$.
\begin{table}[t]
\begin{ruledtabular}
\begin{tabular}{lcccc}
\textrm{ID}&
$L^3\times T$&
$a$ \textrm{fm}&
$aL$ \textrm{fm}&
$m_\pi$ \textrm{GeV}\\
\colrule
\textrm{B64} & $64^3\cdot 128$ & 0.07957(13) & 5.09 & 0.1352(2) \\
\textrm{B96} & $96^3\cdot 192$ & 0.07957(13) & 7.64 & 0.1352(2) \\
\textrm{C80} & $80^3\cdot 160$ & 0.06821(13) & 5.46 & 0.1349(3) \\
\textrm{D96} & $96^3\cdot 192$ & 0.05692(12) & 5.46 & 0.1351(3) \\
\end{tabular}
\end{ruledtabular}
\caption{\label{tab:ensembles}%
ETMC gauge ensembles used in this work. The quoted pion masses have been obtained by a direct computation of the small light-quark mass correction that is necessary to match $m_\pi=135.0$ MeV starting from simulations with slightly heavier pions ($m_\pi=0.1402(2)$~GeV on the B64 ensemble, $m_\pi=0.1401(1)$~GeV on the B96 ensemble, $m_\pi=0.1367(2)$~GeV on the C80 ensemble and $m_\pi=0.1408(2)$~GeV on the D96 ensemble, see Ref.~\cite{Alexandrou:2022amy} for more details).
}
\end{table}
In particular, in order to better estimate the systematics associated with continuum extrapolations, we use the same mixed-action setup described in Ref.~\cite{Alexandrou:2022amy,Frezzotti:2004wz}  and analyze both the so-called Twisted Mass (TM) and Osterwalder-Seiler (OS) lattice regularized correlators $V(t)$. The results for $R_\sigma(E)$ obtained in the two regularizations differ by $O(a^2)$   cutoff effects~\cite{Frezzotti:2003ni,Frezzotti:2005gi} and must coincide within errors in the continuum limit.

\begin{figure}[t!]
\includegraphics[width=\columnwidth]{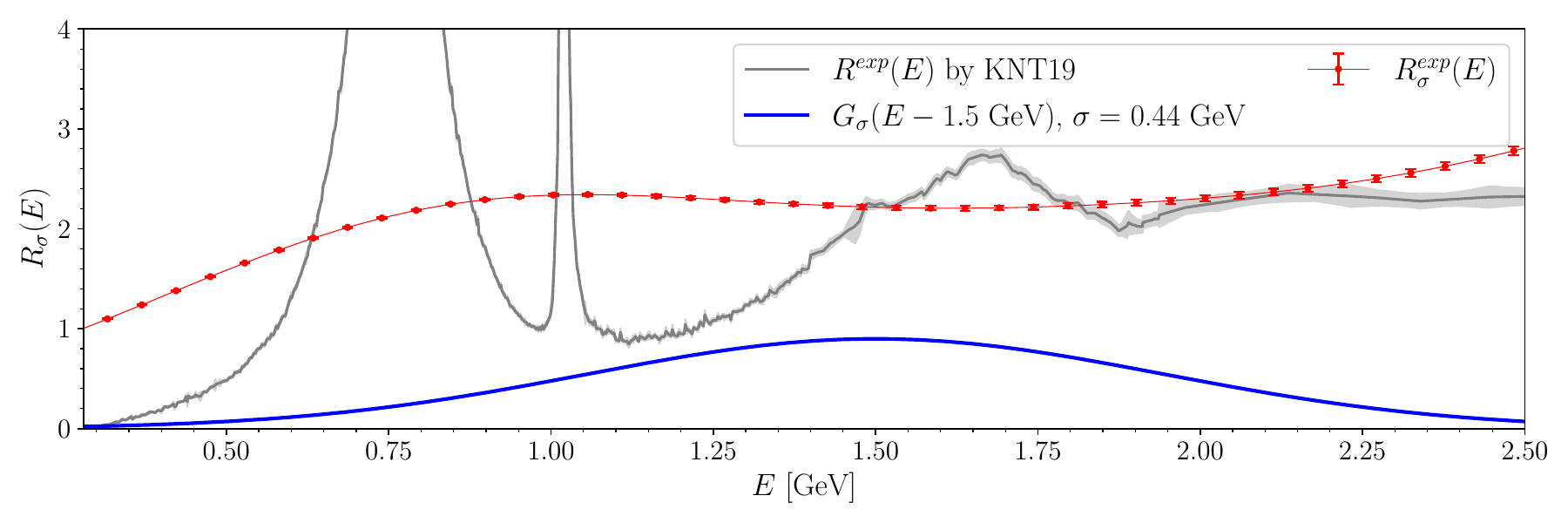}
\caption{\label{fig:smearedrexp} The grey band shows $R^\mathrm{exp}(E)$ from the KNT19 compilation~\cite{keshavarzi2020g}. The red points are the results of the smearing of $R^\mathrm{exp}(E)$ with a Gaussian of $\sigma=0.44$~GeV according to Eq.~(\ref{eq:Rsigma}). The smearing Gaussian corresponding to center energy $E=1.5$~GeV is shown in blue.}
\end{figure}

In order to compare our theoretical results with experiments, we rely on the KNT19 compilation~\cite{keshavarzi2020g} of $R^\mathrm{exp}(E)$, providing data in the range $E\in[0.216,11.1985]$~GeV together with the full covariance matrix that takes into account the correlation between the different experiments, see FIG.~\ref{fig:smearedrexp}. The central values and errors of $R^\mathrm{exp}_\sigma(E)$ quoted below have been obtained by generating bootstrap samples of $R(E)$, each of which simulating an independent measurement, from a multivariate Gaussian distribution using the $R^\mathrm{exp}(\omega)$ central values and covariance matrix. Each sample is then integrated with $G_\sigma(E-\omega)$, see the supplementary material for more details.
\begin{figure}[t!]
\includegraphics[width=\columnwidth]{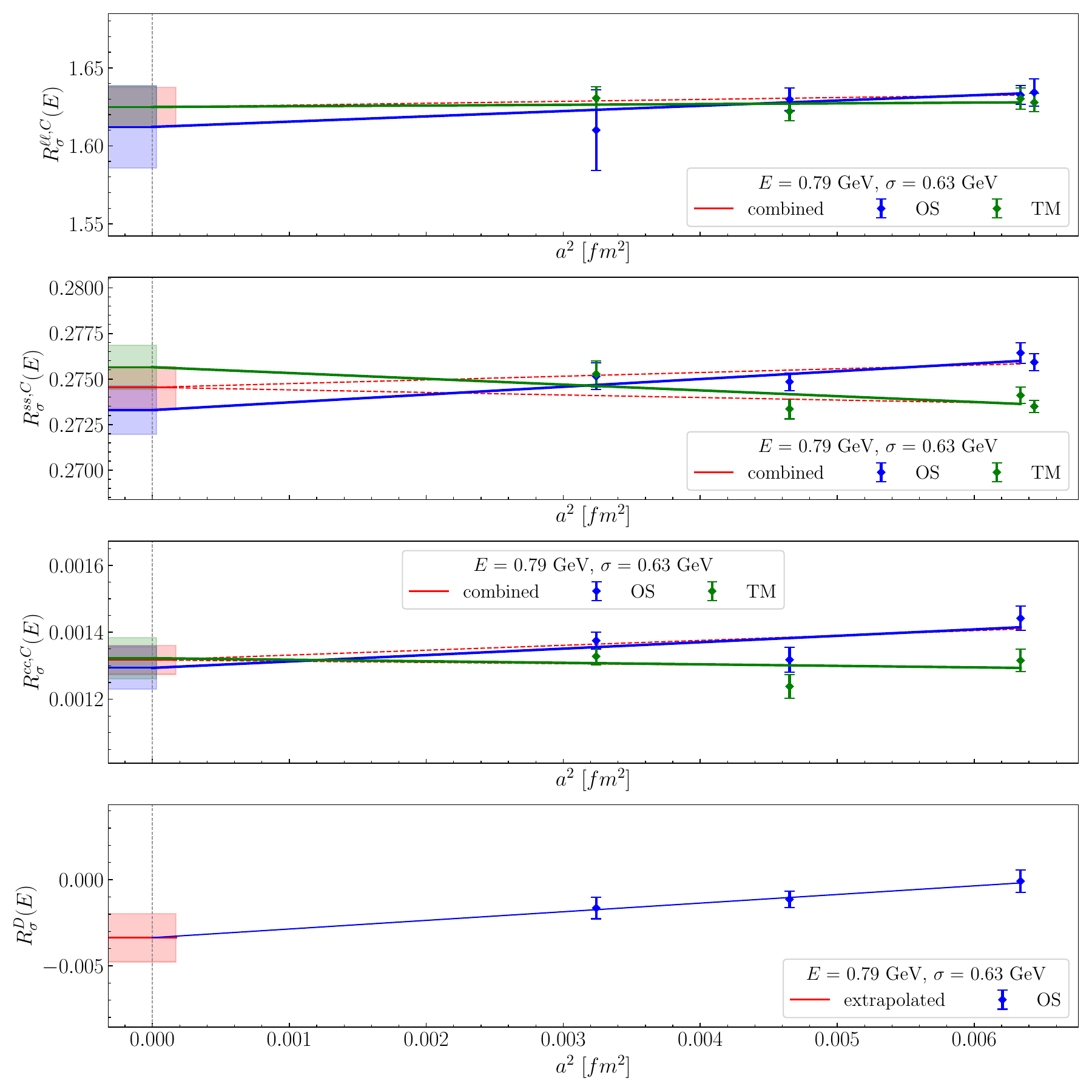}
\caption{\label{fig:contmain} Continuum extrapolations of the different contributions to $R_\sigma(E)$ at $E=0.79$~GeV and $\sigma=0.63$~GeV. From top to bottom, the plots correspond to the connected light-light ($R^{\ell\ell,C}_\sigma(E)$), the connected strange-strange ($R^{ss,C}_\sigma(E)$), the connected charm-charm ($R^{cc,C}_\sigma(E)$) and the disconnected ($R^{D}_\sigma(E)$) contributions. The blue and green points correspond respectively to the OS and TM lattice regularizations. In the case of the connected contributions we performed both correlated-constrained (red) and uncorrelated-unconstrained linear extrapolations in $a^2$ and found them to be compatible within errors in all cases. The disconnected contribution has been computed in the OS regularization only and extrapolated linearly in $a^2$. 
In the case of $R^{\ell\ell,C}_\sigma(E)$ and $R^{ss,C}_\sigma(E)$ there are two points for each regularization at the coarsest lattice spacing (slightly displaced on the $x$-axis to help the eye) corresponding to the ensembles B64 and B96 and, therefore, to different volumes. No significant finite-volume effects have been observed for all considered values of $E$ and $\sigma$.
}
\end{figure}
%

\section{
\label{sec:results}
Results
}
%
\begin{figure}[t!]
\includegraphics[width=\columnwidth]{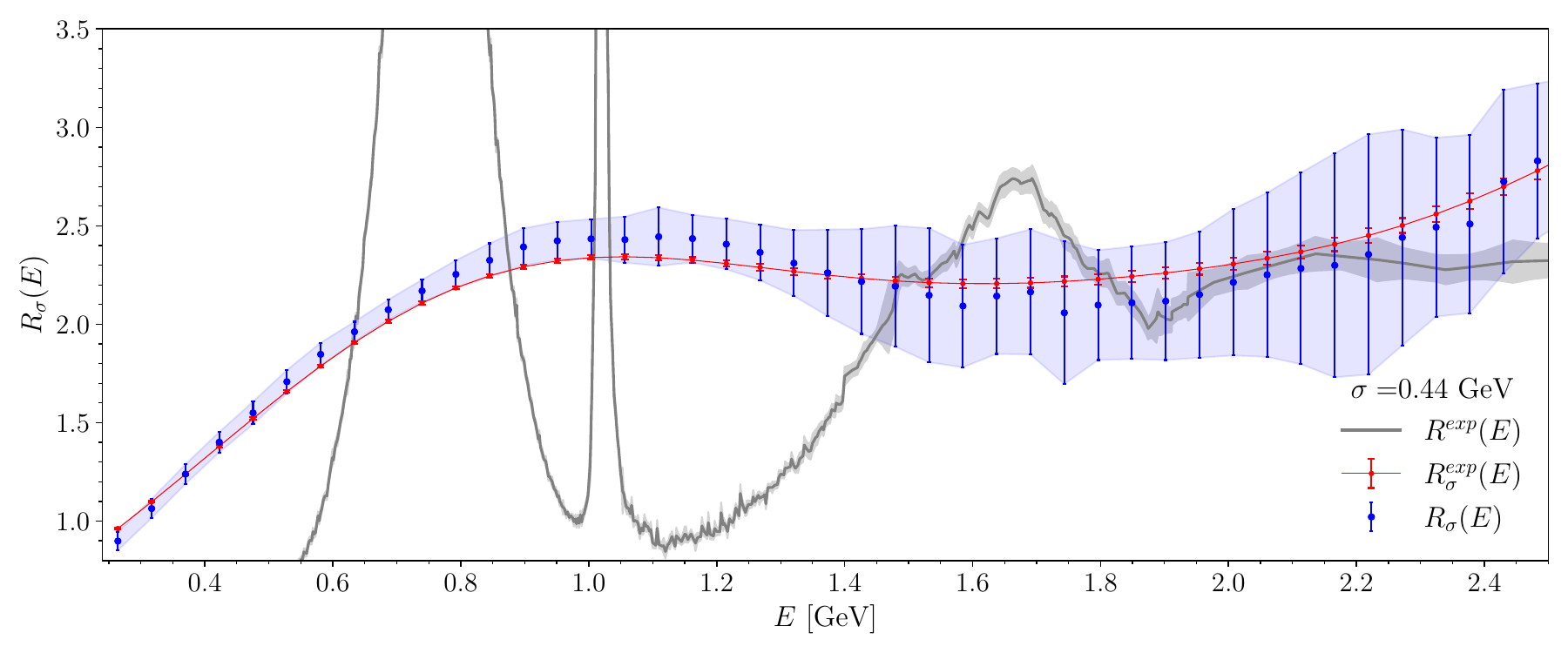}\\
\includegraphics[width=\columnwidth]{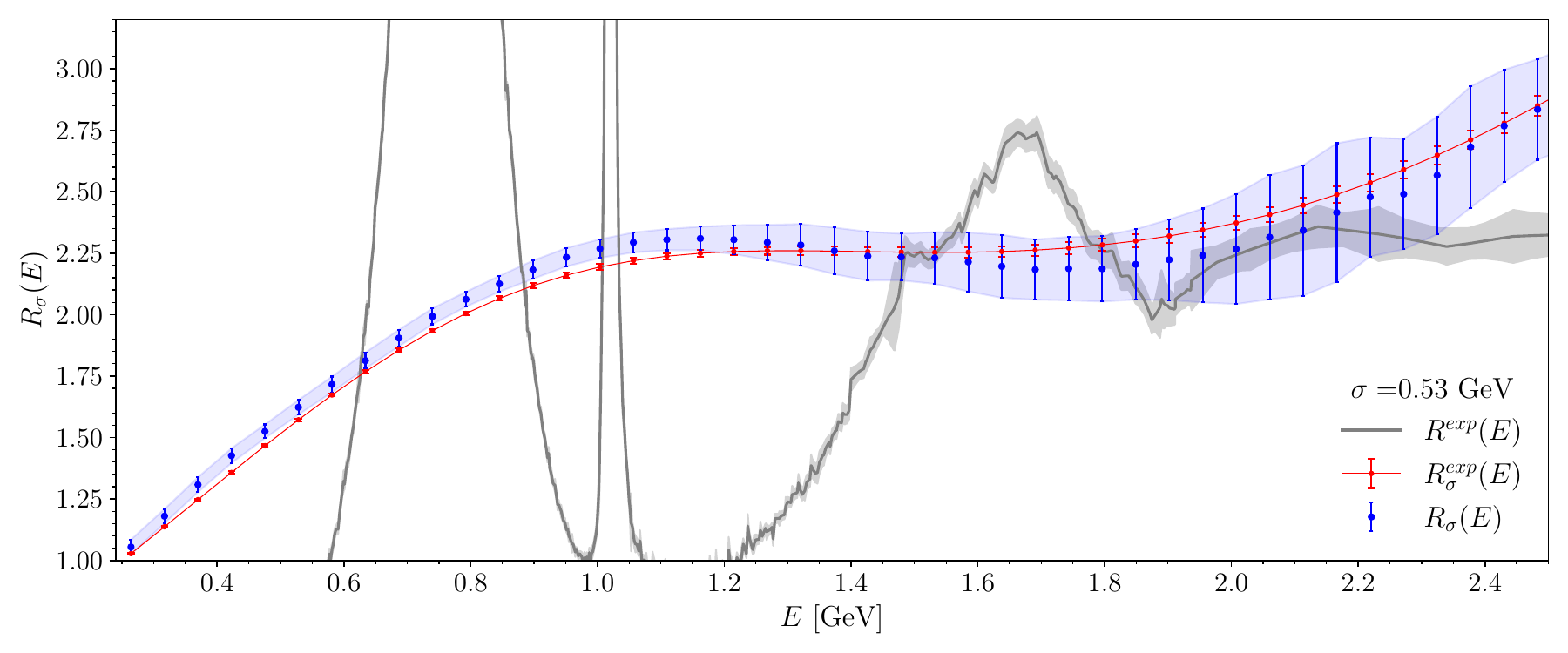}\\
\includegraphics[width=\columnwidth]{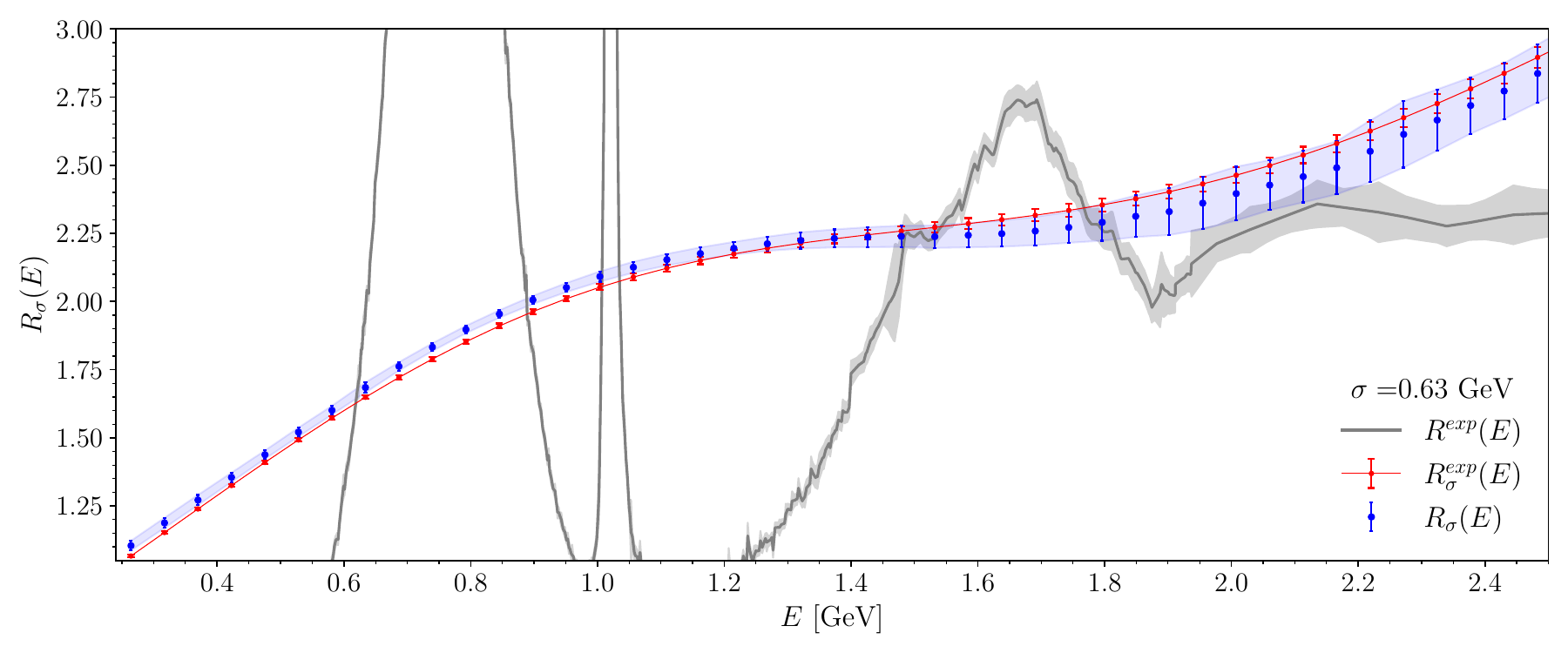}
\caption{\label{fig:comparison} Comparison of $R_\sigma(E)$ (blue points) and $R^\mathrm{exp}_\sigma(E)$ (red points) as functions of $E$ for $\sigma=0.44$~GeV (first row), $\sigma=0.53$~GeV (second row) and $\sigma=0.63$~GeV (third row).}
\end{figure}
\begin{figure}[t!]
	\includegraphics[width=\columnwidth]{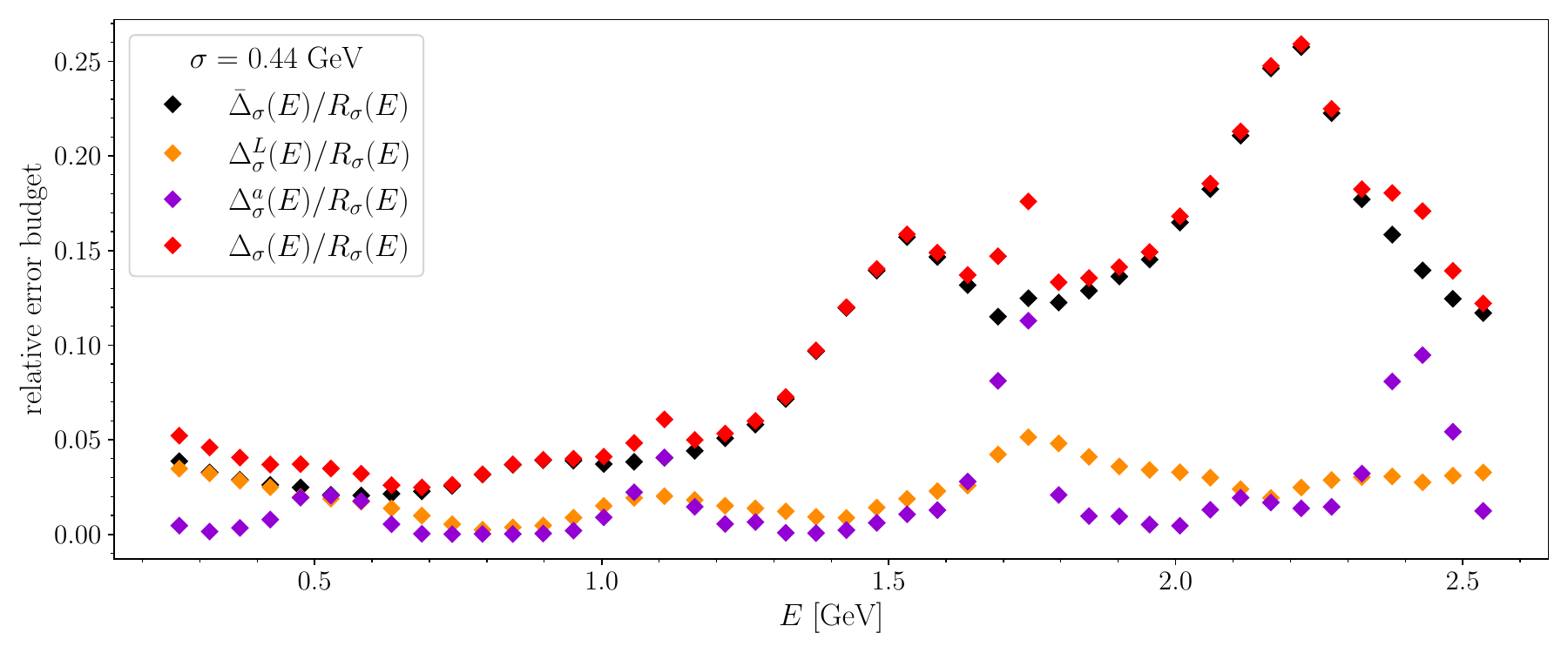}\\
	\includegraphics[width=\columnwidth]{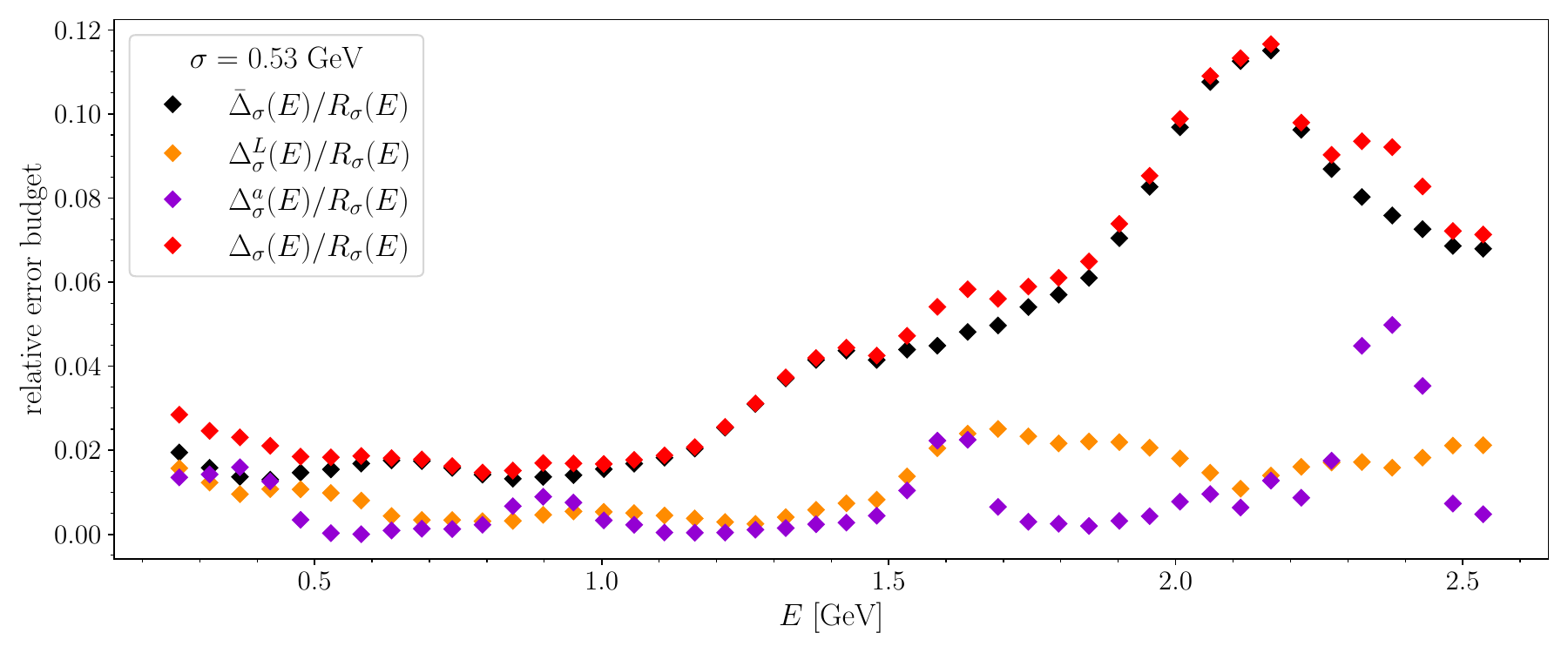}\\
	\includegraphics[width=\columnwidth]{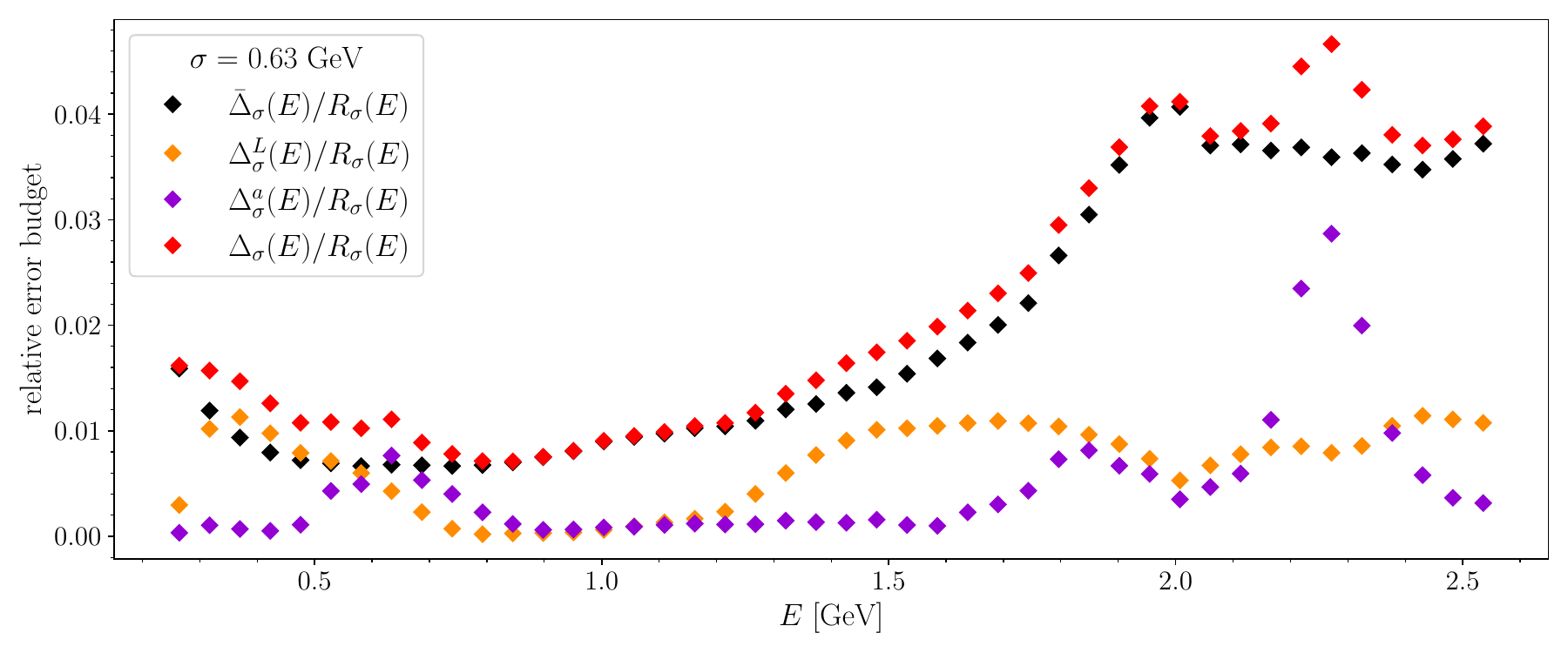}
	\caption{\label{fig:error_budget} Error budget for $R_\sigma(E)$ at $\sigma=0.44$~GeV (first row), $\sigma=0.53$~GeV (second row) and $\sigma=0.63$~GeV (third row). The red points correspond to the total relative error, $\Delta_\sigma(E)/R_\sigma(E)$. The black points are the statistical errors combined in quadrature with the systematics errors coming from the spectral reconstruction algorithm, $\bar \Delta_\sigma(E)$, divided by $R_\sigma(E)$. The violet and orange points are, respectively, our estimates of the relative systematics errors associated with the continuum extrapolations, $\Delta^a_\sigma(E)/R_\sigma(E)$, and finite volume effects, $\Delta^L_\sigma(E)/R_\sigma(E)$.}
\end{figure}
\begin{figure}[h!]
	\includegraphics[width=\columnwidth]{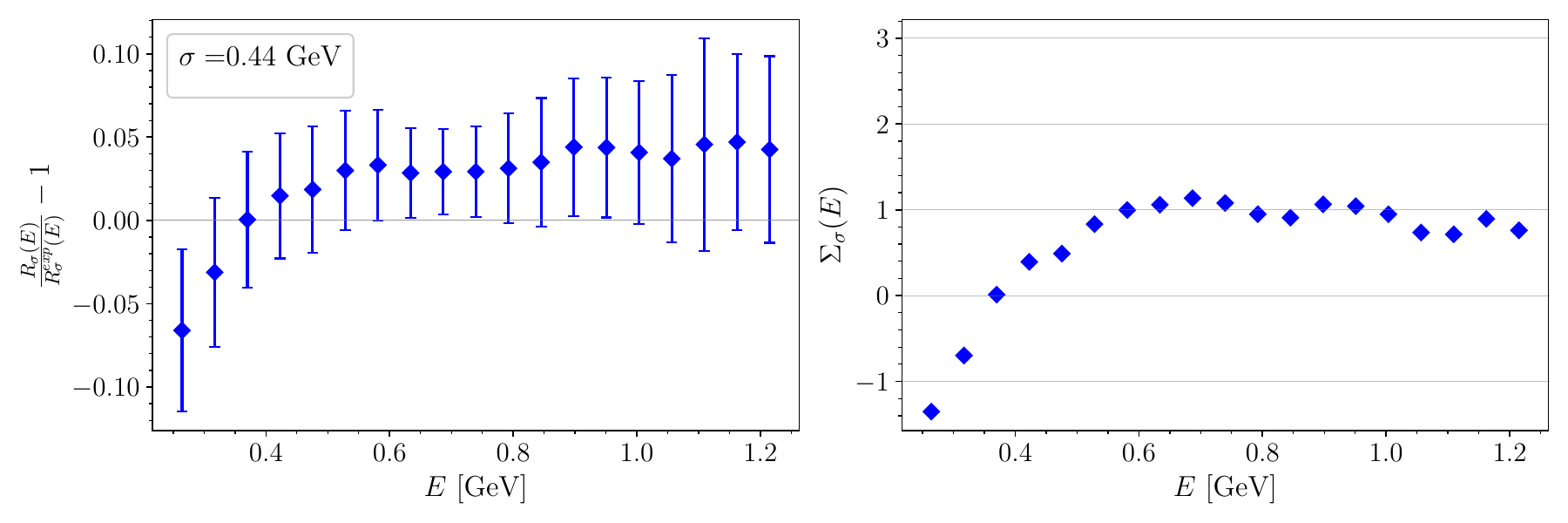}\\
	\includegraphics[width=\columnwidth]{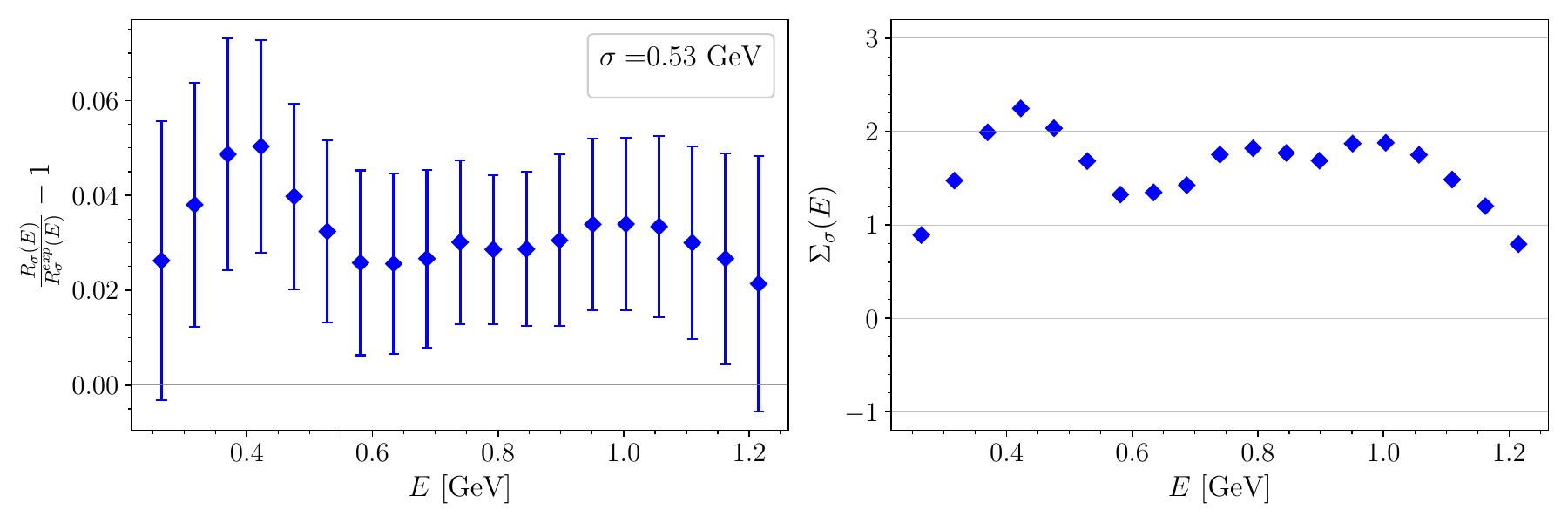}\\
	\includegraphics[width=\columnwidth]{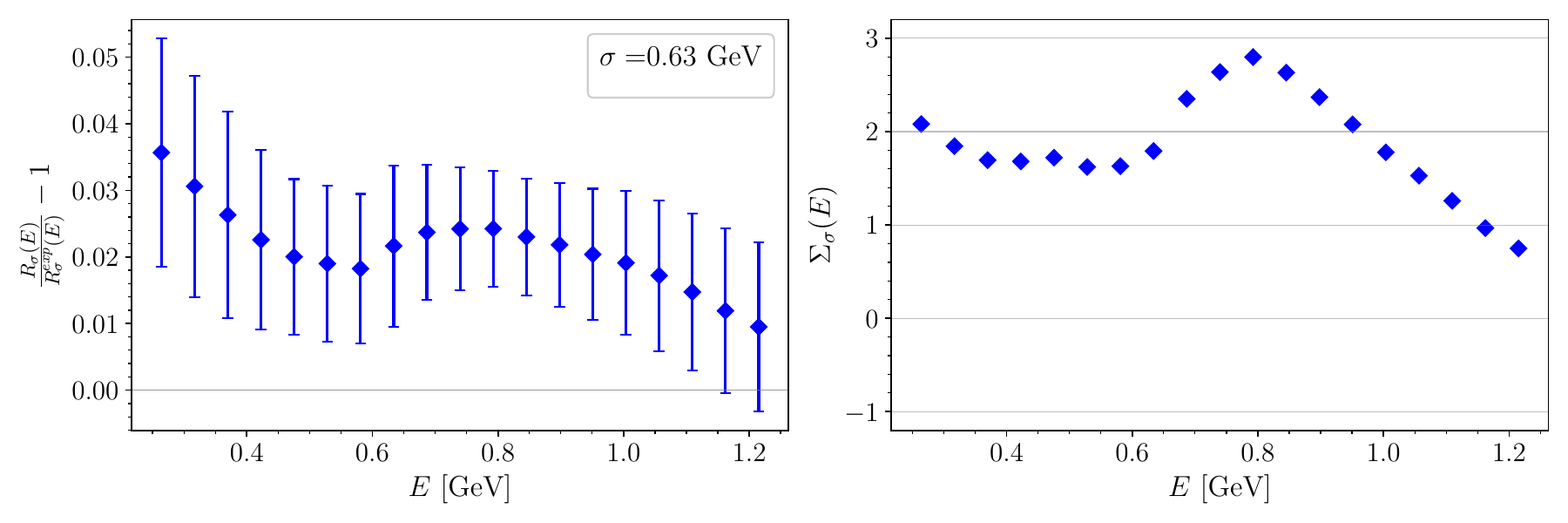}
	\caption{\label{fig:zoom} \emph{Left-plots}: Relative difference $R_\sigma(E)/R_\sigma^\text{exp}(E)-1$ as a function of the energy for $\sigma=0.44$~GeV (first row), $\sigma=0.53$~GeV (second row) and $\sigma=0.63$~GeV (third row) \emph{Right-plots}: The pull quantity $\Sigma_\sigma(E)$, see Eq.~(\ref{eq:finalpull}), as function of the energy for the three values of $\sigma$.}
\end{figure}
In our lattice calculation we considered three values for the smearing parameter, $\sigma=\{0.44,0.53,0.63\}$~GeV, and central energies in the range $E\in[0.21,2.54]$~GeV. A detailed discussion of the analysis procedure, including the break-down of $R_\sigma(E)$ into the contributions coming from the different flavours and from connected and disconnected fermionic Wick contractions, together with a careful study of the systematic uncertainties affecting each contribution, can be found in the supplementary material. Here, in FIG.~\ref{fig:contmain}, we show an example ($E=0.79$~GeV and $\sigma=0.63$~GeV) of the continuum extrapolations of the different contributions to $R_\sigma(E)$ and, in the following, concentrate on the comparison of our first-principles determination with the experimental results $R^\mathrm{exp}_\sigma(E)$.

This is done in FIG.~\ref{fig:comparison} where the plots show $R_\sigma(E)$ (blue points) and $R^\mathrm{exp}_\sigma(E)$ (red points) as functions of $E$ for $\sigma=0.44$~GeV (first row), $\sigma=0.53$~GeV (second row) and $\sigma=0.63$~GeV (third row). Our quoted final errors include the estimates of the systematics associated with continuum extrapolations, with finite-volume effects and also the ones coming from the spectral reconstruction algorithm, see FIG.~\ref{fig:error_budget}. In order to properly interpret FIG.~\ref{fig:comparison} it is very important to realize that the information contained into $R_\sigma(E)$ and $R_\sigma(E^\prime)$ for central energies such that $\vert E-E^\prime\vert \ll \sigma$ is essentially the same. Moreover, our theoretical results at different values of $E$ and $\sigma$ are obtained from the same correlators and, therefore, are correlated (a table with the numerical results and their correlation matrix is provided in the supplementary material). It is also very important to stress that our lattice simulations have been calibrated by using hadron masses to fix the quark masses and the lattice spacing and, therefore, $R_\sigma(E)$ is a theoretical prediction obtained without using any input coming from $R^\mathrm{exp}_\sigma(E)$. In view of these observations, and of the fact that the extraction of spectral densities from Euclidean correlators is a challenging numerical problem, we consider the overall agreement between the theoretical and experimental data quite remarkable.

Although our theoretical errors, $\Delta_\sigma(E)$, are still substantially larger than the experimental ones, $\Delta^\mathrm{exp}_\sigma(E)$, there is a tension between $R_\sigma(E)$ and $R^\mathrm{exp}_\sigma(E)$ when the smearing Gaussian is centred in the region around the $\rho$ resonance. This can be better appreciated in FIG.~\ref{fig:zoom} where, for $E<1.3$~GeV, the plots on the left show the relative difference $R_\sigma(E)/R^\mathrm{exp}_\sigma(E)-1$ while those on the right show the ``pull'' 
\begin{flalign}
\Sigma_\sigma(E) = \frac{R_\sigma(E)- R^\mathrm{exp}_\sigma(E)}{\sqrt{\left[\Delta_\sigma(E)\right]^2 + \left[\Delta^\mathrm{exp}_\sigma(E)\right]^2}}\;.
\label{eq:finalpull}
\end{flalign}
Before ascribing this tension, of about three standard deviations, to new physics or to underestimated experimental uncertainties a very important remark is in order. 

The calculation of $R_\sigma(E)$ that we have performed in this study is an iso-symmetric $n_f=2+1+1$ lattice QCD calculation and, therefore, we have not calculated yet, from first principles, the contributions to $R_\sigma(E)$ coming from $b$-quarks and from the QED and strong isospin breaking corrections. Concerning the $b$-quark contribution, if sizeable, this would represent a positive correction to $R_\sigma(E)$ and thus, given the fact that $R^\mathrm{exp}_\sigma(E)$ is below $R_\sigma(E)$ in the region in which these are in tension, it can only lead to an enhancement of the observed discrepancy. On the other hand, in the supplementary material we provide numerical evidence that even the charm contribution is negligible for $E<1.5$~GeV at the current level of the theoretical precision. This is evident at $E=0.79$~GeV and $\sigma=0.63$~GeV, where we observe the largest tension, from the comparison of the first and third panels in FIG.~\ref{fig:contmain}. We therefore exclude that the observed tension can be ascribed to the $b$-quark contribution.

\begin{figure}[t!]
	\begin{center}
		\includegraphics[width=\columnwidth]{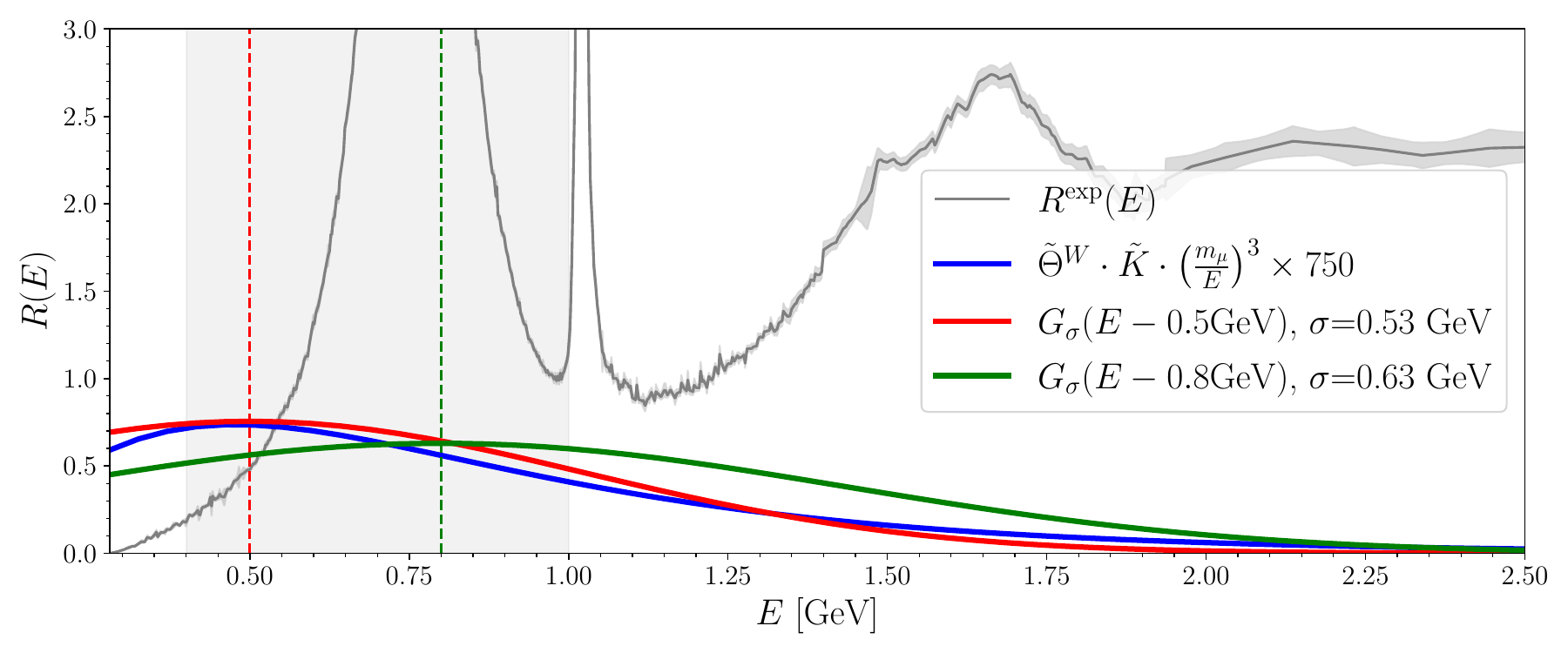}
		\caption{\label{fig:kernelW} 
		The Gaussian kernels with central energy 0.5 GeV and width 0.53 GeV (red) and central energy 0.8 GeV and width 0.63 GeV (green) are compared with the intermediate window kernel $\tilde{\Theta}^W\cdot \tilde{K}\cdot\left(\frac{E}{m_\mu}\right)^3$ (see e.g. Ref.~\cite{Alexandrou:2022amy} for the explicit expression). The red Gaussian is centred at the peak of the intermediate window kernel (vertical red line) that is shown in blue and normalized such that the heights of the two peaks coincide.  The green Gaussian is centred at the energy (vertical green line) where we observe the most significant tension (about $2.5$\% and $3$ standard deviations) between $R_\sigma(E)$ and $R_\sigma^\text{exp}(E)$. Using the red Gaussian we observe instead a 5\% tension corresponding to $2.2$ standard deviations, see Figure~\ref{fig:comparison}.}
	\end{center}
\end{figure}
\emph{Isospin breaking effects definitely have to be evaluated from first principles}. 
Indeed, for very small values of $\sigma$ very large isospin breaking effects have to be expected at certain values of $E$, e.g. at very low energy where the channel $\pi^0+\gamma$ opens in QCD$+$QED and also close to other thresholds (see Refs.~\cite{Colangelo:2022prz,Hoferichter:2022iqe}).  
Nevertheless, we notice that in order to explain the observed tension at $E\sim 0.8$~GeV and $\sigma\sim 0.6$~GeV an isospin breaking effect larger than $2\%$ would be needed and this is hard to reconcile with the first principle lattice calculation performed in Ref.~\cite{Borsanyi:2020mff} of the isospin breaking corrections on closely related quantities, in particular on $a_\mu^{\mathrm{HVP},W}$. Indeed, the smearing kernel that in energy space defines $a_\mu^{\mathrm{HVP},W}$ is very similar in shape to the Gaussian kernel with central energy $E=0.5$~GeV and width $\sigma=0.53$~GeV  (see Figure~\ref{fig:kernelW}) and the isospin breaking effect on $a_\mu^{\mathrm{HVP},W}$ is found to be at the two permille level. We also note that, when $R(E)$ is convoluted with the quite different (but always very much spread out in energy)  kernels that define the long and short distance contributions to $a_\mu^\text{HVP}$ (see Ref.~\cite{Alexandrou:2022jlc}), the isospin breaking corrections w.r.t.\ iso-symmetric QCD remain very small, namely of about one permille~\cite{Borsanyi:2020mff} and three permille~\cite{HARLANDER2003244} respectively.

\section{
\label{sec:conclusions}
Conclusions
}
We presented, for the first time, a non-perturbative theoretical study of the $e^+e^-$ cross-section into hadrons. We have calculated the $R$-ratio convoluted with Gaussian smearing kernels of widths between $440$~MeV and $630$~MeV and center energies up to $2.5$~GeV. We compared our first-principles theoretical results with the corresponding quantity obtained by using the KNT19 compilation~\cite{keshavarzi2020g} of $R$-ratio experimental data courteously provided by the authors. 

For central energies of the smearing Gaussian in the region around the $\rho$ resonance our results are sufficiently precise to let us observe a tension of about three standard deviations with experiments. A solid evidence of a significant discrepancy between theory and experiment already emerged also from the comparison of the lattice calculations~\cite{Borsanyi:2020mff,Alexandrou:2022amy,FermilabLattice:2022smb,Ce:2022kxy} of the (window) contributions to $a_\mu^\mathrm{HVP}$ and the corresponding dispersive determinations~\cite{keshavarzi2020g}. 
Our results corroborate this evidence and, being totally unrelated to the muon $g-2$ experiment, highlight the fact that the tension is between experimental measurements of the $e^+e^-$ inclusive hadronic cross-section and first-principles Standard Model theoretical calculations and are localized in a Gaussian energy bin of width $\sigma\sim 600$~MeV and center energy $E\sim 800$~MeV. 

Although we argued that an isospin breaking corrections larger than $2\%$ would be required to fully reconcile our lattice data with experiments, and that such a large correction is hardly conceivable in view of the few permille effects found in the related full and intermediate window contributions to $a_\mu^\mathrm{HVP}$ in ref.~\cite{Borsanyi:2020mff}, as a matter of fact, the phenomenological relevance of our theoretical results is partially reduced by the missing QED and strong isospin breaking corrections.

At the same time, from the methodological perspective, the observed tension provides a solid numerical evidence of the fact that it is possible to study the $R$-ratio in Gaussian energy bins on the lattice at the precision level required to perform precision tests of the Standard Model.

In future work on the subject we plan to substantially reduce the widths of the smearing Gaussians. Preliminary investigations make us confident on the possibility of studying $R_\sigma(E)$ with $\sigma\sim 200$~MeV by doubling the statistics on the iso-symmetric QCD correlators already considered in this study. Moreover, we plan to compute from first principles the missing QED and strong isospin breaking corrections to $R_\sigma(E)$.

\subsection{Acknowledgments}
\begin{acknowledgments}
We warmly thank A.~Keshavarzi, D.~Nomura and T.~Teubner, the authors of the KNT19 combination~\cite{keshavarzi2020g} of $R$-ratio experimental measurements, for kindly providing us their results. We thank all members of ETMC for the most enjoyable collaboration. 
N.T. warmly thanks L.~Del Debbio, A.~Lupo and M.~Panero for illuminating discussions on the Bayesian probabilistic interpretation of the method of Ref.~\cite{Hansen:2019idp}.
We thank the developers of the
QUDA~\cite{clark2010solving,babich2011scaling,clark2016accelerating}  library for their continued support, without which the calculations for this project would not have been possible.
S.B. and J.F. are supported by the H2020 project PRACE 6-IP (grant agreement No. 82376) and the EuroCC project (grant agreement No. 951740). We acknowledge support by the European Joint Doctorate program STIMULATE grant agreement No. 765048. P.D. acknowledges support
from the European Unions Horizon 2020 research and innovation programme under the Marie Sklodowska-Curie grant agreement No. 813942 (EuroPLEx) and also support from INFN under the research project INFN-QCDLAT. K.H. is supported by the Cyprus Research and Innovation
Foundation under contract number POST-DOC/0718/0100, under contract number CULTURE-AWARD-YR/0220/0012 and by the EuroCC project (grant agreement No. 951740). R.F. and N.T. acknowledge partial support from the University of Tor Vergata program ``Beyond Borders/ Strong Interactions: from Lattice QCD to Strings, Branes and Holography''. F.S., G.G. and S.S. are supported by the Italian Ministry of University and Research (MIUR) under grant PRIN20172LNEEZ.
F.S. and G.G. are supported by INFN under GRANT73/CALAT. This work is supported by the Deutsche Forschungsgemeinschaft (DFG, German Research Foundation) and the NSFC through the funds provided to the Sino-German Collaborative Research Center CRC 110 ``Symmetries and the Emergence of Structure in QCD'' (DFG Project-ID 196253076 - TRR 110, NSFC Grant No. 12070131001). The authors gratefully acknowledge the Gauss Centre for Supercomputing e.V. (www.gauss-centre.eu) for funding the project pr74yo by providing computing time on the GCS Supercomputer SuperMUC at Leibniz Supercomputing Centre (www.lrz.de), as well as computing time projects on the GCS supercomputers JUWELS Cluster and JUWELS Booster~\cite{Krause:2019pey} at the J\"ulich Supercomputing Centre (JSC) and time granted by the John von Neumann Institute for Computing (NIC) on the supercomputers JURECA and JURECA Booster~\cite{Krause:2018}, also at JSC. Part of the results were created within the EA program of JUWELS Booster also with the help of the JUWELS Booster Project Team (JSC, Atos, ParTec, NVIDIA). We further acknowledge computing time granted on Piz Daint at Centro Svizzero di Calcolo Scientifico (CSCS) via the
project with id s702. The authors acknowledge the Texas Advanced Computing Center (TACC) at The University of Texas at Austin for providing HPC resources that have contributed to the research results. The authors gratefully acknowledge PRACE for awarding access to HAWK at HLRS within the project with Id Acid 4886.
\end{acknowledgments}

\bibliography{rf}

\appendix

\newpage

\section{\bf \large SUPPLEMENTARY MATERIAL}

\subsection{Spectral reconstruction algorithm}
On a finite lattice, with periodic boundary conditions in time, 
Eq.~(3) of the main text becomes
\begin{flalign}
V(a\tau)=\frac{1}{12\pi^2}\int_{0}^\infty d\omega \, \omega^2 R_{LT}(\omega)\, b_\tau(\omega)\;,
\label{eq:fltcorr}
\end{flalign}
where
\begin{flalign}
&
b_\tau(\omega)=e^{-a\omega \tau}+e^{-a\omega(T-\tau)}\;.
\end{flalign}
The finite-volume distribution $R_{LT}(\omega)$ is radically different from its infinite-volume counterpart, mainly because of the quantization of the spectrum of the Hamiltonian on the finite volume $aL$ but also because of thermal effects at the finite temperature $1/aT$ (see Refs.~\citeSM{SMHansen:2019idp,SMBulava:2021fre}). The infinite-volume limit of the smeared distribution is a well defined quantity and, since we calculate here $R_\sigma(E)$ at $\sigma>0$, our task is that of estimating the systematics associated with the limits
\begin{flalign}
R_\sigma(E)=\lim_{L,T\mapsto \infty}\int_{0}^\infty d\omega\, G_\sigma(E-\omega)\, R_{LT}(\omega)\;. 
\end{flalign}
This will be done in the next section by using a data-driven approach relying on the ensembles B64 and B96 (see 
TABLE~I in the main text
) that have been generated with the same bare parameters but with different volumes. Therefore, in order to simplify the notation, we shall omit in the following the explicit dependence of $R_{LT}(\omega)$ on $L$ and $T$.

In the method of Ref.~\citeSM{SMHansen:2019idp} smearing kernels are represented as
\begin{flalign}
K(\omega;\vec g) =\sum_{\tau=1}^{\tau_\mathrm{max}} g_\tau\, \left\{
e^{-a\omega \tau}+e^{-a\omega(T-\tau)}
\right\}\;.
\label{eq:frepresentation}
\end{flalign}
In the present implementation of the method, the distance between the target kernel and its representations in terms of the coefficients $g_\tau$ is measured by the functionals
\begin{flalign}
A_\mathrm{n}[\vec g] = \int_{E_0}^\infty d\omega\, w_\mathrm{n}(\omega)\left\vert
K(\omega;\vec g) - \frac{12\pi^2G_\sigma(E-\omega)}{\omega^2}
\right\vert^2\;,
\label{eq:afunctionals}
\end{flalign}
that, for weight-functions $w_\mathrm{n}(\omega)>0$, correspond to a class of weighted $L_2$-norms in functional space. In the previous formula $E_0$ is an algorithmic parameter. By relying on the fact that $R(\omega)=0$ for $\omega<E_{th}$, where the threshold energy $E_{th}$ is $2m_\pi$ in iso-symmetric QCD and $m_{\pi^0}$ in QCD$+$QED (because of the opening of the $e^+e^-\mapsto \pi^0\gamma$ channel), $E_0$ can conveniently be optimized under the condition $E_0<E_{th}$. We have considered the following weight functions
\begin{flalign}
&w_\alpha(\omega) = e^{a\omega \alpha}\;,
\qquad
\alpha=\left\{0,\frac{1}{2},2^-\right\}\;,
\label{eq:walpha}
\\
&w_c(\omega) = \frac{1}{\sqrt{e^{a(\omega-E_0)}-1}}\;,
\label{eq:wcheby}
\end{flalign}
that we distinguish by using the tag $\mathrm{n}=\{0,1/2,2^-,c\}$. 
The parameter $\alpha$ had already be introduced in the original version of the algorithm, see appendix~A of Ref.~\citeSM{SMHansen:2019idp} where the fact that the condition $\alpha<2$ is required for convergence is explained (in practice $\alpha=2^-$ means for us $\alpha=1.99$). As we are going to argue at the end of the section, using a value $\alpha>0$ is particularly useful in order to reduce the systematic error due to the necessarily imperfect reconstruction of the smearing kernel. Before doing that, however, we explain the alternative choice made in Eq.~(\ref{eq:wcheby}). 

By making the change of variable
\begin{flalign}
x=2e^{a(E_0-\omega)}-1\;,
\qquad
x\in [-1,1]\;,
\end{flalign}
for a generic integrable function $f(\omega)$ one has
\begin{flalign}
&\int_{E_0}^\infty \frac{d\omega\, f(\omega)}{\sqrt{e^{a(\omega-E_0)}-1}}
\nonumber \\
&
\qquad\quad=
\frac{1}{a}\int_{-1}^1 \frac{dx}{\sqrt{1-x^2}}\, f\left(E_0-\frac{\log\left(\frac{x+1}{2}\right)}{a}\right)\;,
\end{flalign}
and this implies that in the $T\mapsto \infty$ limit, by minimizing $A_c[\vec g]$ w.r.t. the coefficients vector $\vec g$, one is actually searching the best polynomial approximation of the target kernel by using Chebyshev polynomials. The weights of Eq.~(\ref{eq:walpha}) correspond instead to other Jacobi polynomials (to Legendre ones for $\alpha=0$).

Chebyshev polynomials have been introduced in the spectral reconstruction game in the fundamental paper~\citeSM{SMBarata:1990rn} and, more recently, in Ref.~\citeSM{SMBailas:2020qmv}. As explained in the main text, the numerical problem of reconstructing the kernel $G_\sigma(E-\omega)$ becomes rapidly ill-posed for $E>E_{th}$ in the $\sigma\mapsto 0$ limit. Without a \emph{regularization mechanism} the coefficients $\vec g$ become huge in absolute value for all choices of the weighting function $w_\mathrm{n}(\omega)$, including the one corresponding to Chebyshev polynomials. The regularization method adopted in Ref.~\citeSM{SMBailas:2020qmv} consists in fitting the correlator on a Chebyshev polynomial basis and in filtering the noise that doesn't satisfy the theory constraints coming from the expected $\exp{(-tH)}$ behaviour w.r.t. time, with $H$ being the QCD Hamiltonian.

The regularization method proposed in Ref.~\citeSM{SMHansen:2019idp}, and adopted here, is the model-independent mechanism originally proposed by Backus and Gilbert~\citeSM{SMBackus} and does not require any pre-processing/filtering of the input correlator data. The coefficients $\vec g$ are obtained by minimizing a linear combination,
\begin{flalign}
W_\mathrm{n}[\vec g] =\frac{A_\mathrm{n}[\vec g]}{A_\mathrm{n}[\vec 0]} + \lambda\, B[\vec g]\;,
\label{eq:wexpression}
\end{flalign}
of the norm-functional $A_\mathrm{n}[\vec g]$ and of the error-functional  
\begin{flalign}
B[\vec g] =  
B_\mathrm{norm} \sum_{\tau_1,\tau_2=1}^{\tau_\mathrm{max}} g_{\tau_1} g_{\tau_2}\, \mathrm{Cov}(\tau_1,\tau_2)\;.
\label{eq:bdef}
\end{flalign}
The matrix $\mathrm{Cov}(\tau_1,\tau_2)$ appearing in the previous expression is the covariance matrix of the lattice correlator $V(a\tau)$. In this paper we set the relative normalization between the norm and error functionals by choosing
\begin{flalign}
B_\mathrm{norm}=\frac{E^6}{\left(V(a \tau_\mathrm{norm})\right)^2}
\label{eq:relative_normalization}
\end{flalign}
and, moreover, we use a slightly different expression for $W_\mathrm{n}[\vec g]$ w.r.t. Refs.~\citeSM{SMHansen:2019idp,SMBulava:2021fre}. In fact, the relative normalization of the two functionals can be reabsorbed into a redefinition of the unphysical algorithmic parameter $\lambda$. 
Once the relative normalization of the two functionals has been fixed, conditions such as $A_\mathrm{n}[\vec g]=A_\mathrm{n}[\vec 0] B[\vec g]$ (that we use in the search for the optimal approximation of $R_\sigma(E)$, see below) acquire a meaning regardless of the value of $\lambda$. With our choice of $B_\mathrm{norm}$ the error functional is dimensionless.

At fixed values of the algorithmic parameters
\begin{flalign}
\vec p=(\mathrm{n},\lambda, E_0,\tau_\mathrm{max},\tau_\mathrm{norm})
\end{flalign}
the linear minimization problem
\begin{flalign}
\left.\frac{\partial W_\mathrm{n}[\vec g]}{\partial g_\tau}\right\vert_{\vec g=\vec g^{\vec p}}=0
\end{flalign}
gives the coefficients $\vec g^{\vec p}$ and the corresponding approximation of $R_\sigma(E)$ according to
\begin{flalign}
R_\sigma(E;\vec g^{\vec p}) = \sum_{\tau=1}^{\tau_\mathrm{max}} g_\tau^{\vec p}\, V(a\tau)\;.
\end{flalign}
The name error functional comes from the fact that 
\begin{flalign}
\Delta_\sigma^\mathrm{stat}(E;\vec g^{\vec p}) = \sqrt{\frac{B[\vec g^{\vec p}]}{B_\mathrm{norm}}}
\label{eq:biserror}
\end{flalign}
is the statistical error
of $R_\sigma(E;\vec g^{\vec p})$. Therefore, the regularization of the problem induced by the presence of $B[\vec g]$ in Eq.~(\ref{eq:wexpression}) disappears in the ideal limit of infinitely precise input correlators.

In order to quantify the systematic error associated with the necessarily imperfect reconstruction of the smearing kernel we study $R_\sigma(E;\vec g^{\vec p})$ as a function of the normalized $L_2$--norm at $\alpha=0$ (also in the case where $\vec g^{\vec p}$ has been obtained with $\alpha \neq 0$ or with the Chebyshev weight), 
\begin{flalign}
d(\vec g^{\vec p})=\sqrt{\frac{A_0[\vec g^{\vec p}]}{A_0[\vec 0]}}\;.
\label{eq:dgdef}
\end{flalign}
We quote our best estimate for $R_\sigma(E)$ by selecting a result from the region of the statistically dominated regime, i.e. the region of small values of $d(\vec g^{\vec p})$ where the results are stable, within statistical errors, w.r.t. variations of the unphysical algorithmic parameters $\vec p$. In the following we refer to this procedure, introduced and validated in Ref.~\citeSM{SMBulava:2021fre}, as \emph{stability analysis}. 

The rationale behind the stability analysis procedure is contained in the following two simple observations. For large values of $d(\vec g^{\vec p})$ the results corresponding to the different weight functions and/or different values of $\lambda$ are substantially different, simply because the reconstructed kernels are very different from the target and among themselves. Conversely, for sufficiently small values of $d(\vec g^{\vec p})$ the results of $R_\sigma(E;\vec g^{\vec p})$ tend to agree within the statistical errors (see e.g. FIG.~\ref{fig:llcstability}) simply because in this regime $\Delta_\sigma^\mathrm{stat}(E;\vec g^{\vec p})$ tend to grow, for any choice of the weight function, because of the ill-posedness of the numerical problem.

In fact, in full compliance with Refs.~\citeSM{SMHansen:2019idp,SMBulava:2021fre}, we estimate the central-value of $R_\sigma(E)$ and the residual systematic error from the results for $R_\sigma(E;\vec g^{\vec p})$ corresponding to the conditions 
\begin{flalign}
\frac{A_{2^-}[\vec g^\star]}{A_{2^-}[\vec 0] }=10 B[\vec g^\star]\;,
\quad
\frac{A_{2^-}[\vec g^{\star\star}]}{A_{2^-}[\vec 0]}= B[\vec g^{\star\star}]\;.
\label{eq:stars}
\end{flalign}
Our choice of the relative normalization of the functionals is such that these two points are both inside the region of the statistically dominated regime in most of the cases. Even when this doesn't happen, the central value of $R_\sigma(E)$ is reliably estimated by $R_\sigma(E;\vec g^{\star})$ and the difference $R_\sigma(E;\vec g^{\star})-R_\sigma(E;\vec g^{\star\star})$ provides a conservative estimate of the residual systematic uncertainty (see next section for more details).

On the one hand, the search for the points $\vec g^{\star}$ and $\vec g^{\star\star}$ can be automated and this greatly simplifies the analysis. On the other hand, the actual meaning of the conditions of Eqs.~(\ref{eq:stars}) (that fix these points) depends on the choice for $B_\mathrm{norm}$. Moreover, in order to check that the estimated errors are reliable, it is extremely helpful to have results corresponding to different unphysical parameters that must agree in the statistically dominated regime. This explains our choice of considering different weight functions.

Concerning the choice of the weight functions we now provide the argument in favour of $w_\alpha(\omega)$ with $\alpha>0$. In the case of a generic spectral density $\rho(\omega)$ and a generic target kernel $K(\omega)$, the bias due to the imperfect reconstruction of the kernel is given by
\begin{flalign}
\int_{0}^\infty d\omega\, \left\{K(\omega;\vec g^{\vec p}) - K(\omega)\right\}\, \rho(\omega)\;. 
\end{flalign}
If $\rho(\omega)$ is sufficiently regular, different local variations of the difference $K(\omega;\vec g^{\vec p})-K(\omega)$ produce results that cannot be distinguished within the statistical errors for sufficiently small values of $d(\vec g^{\vec p})$. The contribution to the bias coming from the high energy region of the integration domain is particularly important. A generic spectral density, being a tempered distribution, is expected to grow as a power for high energy and this is the source of a potentially very large contribution to the bias. The faster the difference between the target and reconstructed kernels decays with energy, the smaller is this contribution and the simpler is the stability analysis.
It is thus very useful to realize that the high energy behaviour of the difference between the kernels strongly depends upon the choice of the weighting function. Indeed, at the end of the minimization procedure, one has a finite number for $A_\mathrm{n}[\vec g^{\vec p}]$ and this implies that the difference between the kernels has to decrease faster than $1/\sqrt{\omega w_\mathrm{n}(\omega)}$ in the $\omega\mapsto \infty$ limit (see Eq.~(\ref{eq:afunctionals})). This means faster than $\exp(-\alpha a\omega/2)$ for $\alpha>0$, faster than $1/\sqrt{\omega}$ in the $\alpha=0$ case or even a growth for $\alpha<0$ or in the Chebyshev case. This explains why the choice $\alpha>0$ is particularly convenient in order to stabilize the algorithm. Our numerical results, presented in the next section, confirm this observation.

\subsubsection{Probabilistic interpretation\\ in the language of Gaussian Processes}
The method of Ref.~\citeSM{SMHansen:2019idp} can be interpreted in the Bayesian probabilistic language of Gaussian Processes. Building on the results of Ref.~\citeSM{SMValentine2020}, the Gaussian Processes approach to the extraction of unsmeared spectral densities from noisy lattice correlators has been proposed in Ref.~\citeSM{SMHorak:2021syv} (see also Refs.~\citeSM{SMDelDebbio:2021whr,SMCandido:2023nnb}). In fact, the results of Ref.~\citeSM{SMValentine2020} (see in particular subsection~3.1.3 and Eqs.~(25)) can also be used to obtain smeared spectral densities and to establish a one-to-one correspondence with the method of Ref.~\citeSM{SMHansen:2019idp}.

In the language of our paper, the central value of the posterior Gaussian distribution of the smeared $R$-ratio, given the observations of the lattice correlator $V(a\tau)$ and its covariance $\mathrm{Cov}(\tau_1,\tau_2)$, is given by
\begin{flalign}
R_\sigma^\mathrm{GP}(E) = R_\sigma^\mathrm{mod}(E) + \sum_{\tau=1}^{\tau_\mathrm{max}} g^\mathrm{GP}_\tau\left\{ 
V(a\tau) -V^\mathrm{mod}(a\tau)
\right\}.
\label{eq:RGP}
\end{flalign}
The model-smeared $R$-ratio, $R_\sigma^\mathrm{mod}(E)$, and the model correlator, $V^\mathrm{mod}(a\tau)$, are obtained by smearing the central value $R^\mathrm{mod}(E)$ of the prior Gaussian distribution of the stochastic field $R(E)$, representing the $R$-ratio in this approach, according to
\begin{flalign}
&
R_\sigma^\mathrm{mod}(E) = \int_{E_0}^\infty d\omega\, \, G_\sigma(E-\omega)\, R^\mathrm{mod}(\omega)\;,
\nonumber \\
\nonumber \\
&
V^\mathrm{mod}(a\tau)=
\frac{1}{12\pi^2}\int_{E_0}^\infty d\omega \,\omega^2 R^\mathrm{mod}(\omega)\, b_\tau(\omega).
\end{flalign}
The model input distribution of $R(E)$ is a normalized Gaussian
\begin{flalign}
&\Pi[R-R^\mathrm{mod},S]
\nonumber \\
\nonumber \\
&=
\frac{e^{-\frac{1}{2}\int_{E_0}^{\infty} d\omega_1 d\omega_2 \left[R-R^\mathrm{mod}\right](\omega_1)S^{-1}(\omega_1,\omega_2) \left[R-R^\mathrm{mod}\right](\omega_2)}}{\mathcal N}
\end{flalign}
that, in addition to the central value $R^\mathrm{mod}(E)$, is fully specified once the model covariance $S(\omega_1,\omega_2)$ (the kernel of a positive definite, symmetric and invertible operator) is given. The coefficients $\vec g^\mathrm{GP}$ appearing in Eq.~(\ref{eq:RGP}) are given by
\begin{flalign}
\vec g^\mathrm{GP} = \frac{1}{\hat \Sigma + \hat B}\, \vec f\;,
\end{flalign}
where 
\begin{flalign}
&
\hat \Sigma(\tau_1,\tau_2)
=
\int_{E_0}^{\infty} d\omega_1 d\omega_2\, b_{\tau_1}(\omega_1) S(\omega_1,\omega_2)b_{\tau_2}(\omega_2) \;,
\nonumber \\
\nonumber \\
&
f_\tau
=
12\pi^2\int_{E_0}^{\infty} d\omega_1 d\omega_2\, b_{\tau}(\omega_1) S(\omega_1,\omega_2)  \frac{G_\sigma(E-\omega_2)}{\omega_2^2}\;,
\end{flalign}
and where
\begin{flalign}
\hat B(\tau_1,\tau_2)=\mathrm{Cov}(\tau_1,\tau_2)\;,
\end{flalign}
is the covariance of the lattice correlator appearing in Eq.~(\ref{eq:bdef}). 
The one-to-one correspondence with the method of Ref.~\citeSM{SMHansen:2019idp}, and with the formulae given in the rest of this paper, can now be established by making the following choice for the model central value and covariance
\begin{flalign}
R^\mathrm{mod}(\omega)=0\;,
\quad
S(\omega_1,\omega_2) = \frac{w_\mathrm{n}(\omega_1)}{A_\mathrm{n}[\vec 0]\, \lambda B_\mathrm{norm} }\, 
\delta(\omega_1-\omega_2)\;.
\label{eq:modelchoice}
\end{flalign}

Some important remarks are in order here. The choice of the input model given in the previous equations allows a probabilistic interpretation of the algorithm of Ref.~\citeSM{SMHansen:2019idp} and, therefore, of the stability analysis. Setting to zero the mean value of the input model, i.e.\ $R^\mathrm{mod}(\omega)=0$, is a rather common choice in the Gaussian Processes literature (see e.g.\ Ref.~\citeSM{SMValentine2020,SMHorak:2021syv}). Concerning the choice of the covariance, a common choice is
\begin{flalign}
S(\omega_1,\omega_2) = \frac{\gamma_1}{\sqrt{2\pi} \gamma_2}\,  e^{-\frac{(\omega_1-\omega_2)^2}{2\gamma_2^2}}
\end{flalign}
that reduces to our \emph{diagonal} choice in the $\gamma_2\mapsto 0$ limit. As discussed extensively in Ref.~\citeSM{SMValentine2020} (see in particular FIGs.~1 and~2), the smaller the value of $\gamma_2$ the less regular the input model is. In our problem, in order to explore the distributional space in which $R(E)$ lives on a finite volume, the choice $\gamma_2=0$ \emph{has} to be done.

The problem of optimizing the choice of $\gamma_1$ is much more delicate and, to our knowledge, has not been extensively discussed in the literature. By looking at the problem from the deterministic perspective in which the method of Ref.~\citeSM{SMHansen:2019idp} has originally been formulated, one has $\gamma_1\mapsto w_\mathrm{n}(\omega_1)/(A_\mathrm{n}[\vec 0]\, \lambda B_\mathrm{norm})$. In fact, the problem of optimizing the hyper-parameter $\gamma_1$ is the one that we address here with the stability analysis. In this respect, the fact that we are targeting the calculation of the smeared $R$-ratio $R_\sigma(E)$, and not of the unsmeared quantity $R(E)$, becomes crucial. Indeed, since the problem is linear, see Eq.~(\ref{eq:fltcorr}), and since our smearing kernel (being infinitely differentiable for $\omega>E_0>0$ and vanishing in the limit $\omega\mapsto \infty$) can \emph{exactly} be represented as
\begin{flalign}
12\pi^2 \frac{G_\sigma(E-\omega)}{\omega^2} =\lim_{\tau_\mathrm{max}\mapsto \infty}\sum_{\tau=1}^{\tau_\mathrm{max}} g_\tau(E,\sigma)\, b_\tau(\omega)\;,
\label{eq:exactrep}
\end{flalign}
the problem of extracting $R_\sigma(E)$ has a unique solution in the ideal limit of an infinite number of lattice points. Notice that in order to extract $R(E)$ one would need to represent a Dirac $\delta$-function as in the r.h.s.\ of Eq.~(\ref{eq:exactrep}). Once the sum in Eq.~(\ref{eq:exactrep}) is truncated and the problem is further regulated by setting $\lambda>0$, the solution acquires a dependence upon the trade-off parameter $\lambda$ and the weight function $w_\mathrm{n}(\omega)$. Indeed the dependence upon  the norm that defines the optimal representation of the smearing kernel at finite $\tau_\mathrm{max}$ (see Eq.~(\ref{eq:afunctionals})) disappears in the $\tau_\mathrm{max}\mapsto \infty$ limit. It follows that the systematic error induced by a finite $\tau_\mathrm{max}$ and $\lambda$ can be quantified by studying numerically the limits $\tau_\mathrm{max}\mapsto \infty$ and $\lambda\mapsto 0$. This is what we do in the stability analysis. When, within the statistical errors, the results are independent upon $\lambda$ and $w_\mathrm{n}(\omega)$, the onset of these limits has been reached. If this doesn't happen, we enlarge the statistical errors to estimate the residual systematic uncertainty.

In summary, by looking at the method of Ref.~\citeSM{SMHansen:2019idp} from the probabilistic perspective it is possible to clearly understand the prior assumptions (given in Eq.~(\ref{eq:modelchoice})) that lead to the solution. Conversely, by looking at the problem of finding an optimal choice for the hyper-parameters $\gamma_1$ and $\gamma_2$ from the deterministic perspective, it is possible to understand that the limits $\gamma_2\mapsto 0$ and $\gamma_1\mapsto \infty$ correspond to the exact solution for the \emph{smeared} $R$-ratio. In the light of these observations, the stability analysis of Ref.~\citeSM{SMBulava:2021fre} can now profitably be used within the Gaussian Processes approach.

\subsection{Data analysis}
In order to analyze our data we used a bootstrap procedure. The same number of bootstrap samples has been generated for each gauge ensemble. This allows to combine results obtained from different simulations and, at the same time, to take properly into account correlations when combining results extracted from the same set of gauge configurations. By varying the number of bootstrap samples (from $O(10^2)$ to $O(10^4)$) and by building bins of different sizes of the raw simulation data, i.e. by averaging data obtained on consecutive (w.r.t. Monte Carlo time) gauge configurations, we checked the reliability of our estimates of the statistical errors.

In the following, as customary, we shall consider separately the contributions corresponding to connected (C) and disconnected (D) fermionic Wick contractions to $V(t)$ and, in the case of the connected ones, also the contributions coming from the different flavours.  Moreover, since our connected lattice correlators have been computed in both the Twisted Mass (TM) and Osterwalder-Seiler (OS) regularizations~\citeSM{SMAlexandrou:2022amy}, we shall also distinguish these two cases. To this end, we shall use e.g. the notation $R_\sigma^{ss,C,\mathrm{TM}}(E)$ for the ``strange-strange connected'' contribution to $R_\sigma(E)$ obtained from the correlator $V(t)$ in which the electromagnetic currents, in the Twisted Mass regularization, are both given by $-\bar s\gamma_\mu s/3$ and only fermionic connected Wick contractions are considered. Analogously the connected contribution in the Osterwalder-Seiler regularization coming from the up and down (light) quarks will be denoted as $R_\sigma^{\ell\ell,C,\mathrm{OS}}(E)$, and so on for the other flavours. The disconnected contribution, computed only in the OS regularization and including all flavours, will be denoted as $R_\sigma^{D}(E)$. The same notation is adopted below for the parent correlators. 

We will discuss results obtained at three different values of $\sigma$, namely
\begin{flalign}
\sigma_1=0.44~\mathrm{GeV},
\
\sigma_2=0.53~\mathrm{GeV},
\
\sigma_3=0.63~\mathrm{GeV},
\end{flalign}
and at central energies $E$ in the range $[0.21,2.54]$~GeV. Although we have already produced a larger set of results, more statistics is needed (particularly in the case of the noisier but dominant light-light connected contribution) in order to be able to extract phenomenologically useful information at smaller values of $\sigma$ and/or at larger values of $E$. Therefore, in this work, we concentrate on the set of results specified above.  

All our results have been obtained by fixing $E_0=0.21$~GeV and $\tau_\mathrm{max}=T/2+1$, corresponding respectively to $65$, $97$, $81$ and $97$ on the B64, B96, C80 and D96 ensembles. A numerical investigation of the dependence of the results on $E_0$ and $\tau_\mathrm{max}$ revealed that choosing $E_0$ close to $E_{th}=2m_\pi$ and using the maximum number of lattice times available on each ensemble helps in reducing the size of the statistical errors (see also Ref.~\citeSM{SMBulava:2021fre}).

On any gauge ensemble we set $\tau_\mathrm{norm}=1$ in the case of the connected contributions and $\tau_\mathrm{norm}=0$ in the case of the disconnected contributions, see Eq.~(\ref{eq:relative_normalization}). Given these choices, providing a convenient relative normalization of the norm and error functionals in both cases (see previous section), the central values of our results are given by $R_\sigma(E)\equiv R_\sigma(E;\vec g^{\star})$ and the statistical errors are given by $\Delta_\sigma^\mathrm{stat}(E)\equiv\Delta_\sigma^\mathrm{stat}(E;\vec g^{\star})$ (see Eqs.~(\ref{eq:stars})). The systematics errors associated with the reconstruction, $\Delta^{\mathrm{rec}}_\sigma(E)$, are estimated by introducing the quantity
\begin{flalign}
P_\sigma(E)=\frac{R_\sigma(E;\vec g^{\star})-R_\sigma(E;\vec g^{\star\star})}{\Delta_\sigma^\mathrm{stat}(E;\vec g^{\star\star})}
\label{eq:syspull}
\end{flalign}
as a measure of the statistical compatibility with zero of the difference between the results obtained at $\vec g^{\star}$ and $\vec g^{\star\star}$ and by then evaluating
\begin{flalign}
\Delta^{\mathrm{rec}}_\sigma(E) = 
\left\vert R_\sigma(E;\vec g^{\star})-R_\sigma(E;\vec g^{\star\star}) \right\vert 
\mathrm{erf}\left(
\frac{\left\vert P_\sigma(E)\right\vert}{\sqrt{2}}
\right),
\label{eq:syserror}
\end{flalign}
i.e. the absolute value of this difference weighted with a (rough) estimate of the probability that its observed value is due to fluctuations within the associated error,
\begin{flalign}
\mathrm{erf}(x)=\frac{2}{\sqrt{\pi}}\int_0^x dt\, e^{-t^2}\;.
\end{flalign}
Our estimate of the total error, $\Delta_\sigma(E)$, is obtained by summing in quadrature $\Delta_\sigma^\mathrm{stat}(E)$, $\Delta_\sigma^{\mathrm{rec}}(E)$, the errors associated with the uncertainties on the renormalization constants and on the lattice spacing as well as an estimate of the systematic errors associated with finite-volume effects and continuum extrapolations. Our results at fixed cutoff are proportional to the square of the renormalization factors (different in the two regularizations) that have tiny errors (see Ref.~\citeSM{SMAlexandrou:2022amy} for more details). The systematic errors associated with the uncertainty ($\Delta a$) on the lattice spacing are estimated by repeating the analysis with $a\pm \Delta a$ and by taking the difference of the two results thus obtained. A detailed illustration of the procedures that we use to estimate the other systematics is given below.

\subsubsection{Light-light connected contribution}

\begin{figure}[t!]
\includegraphics[width=\columnwidth]{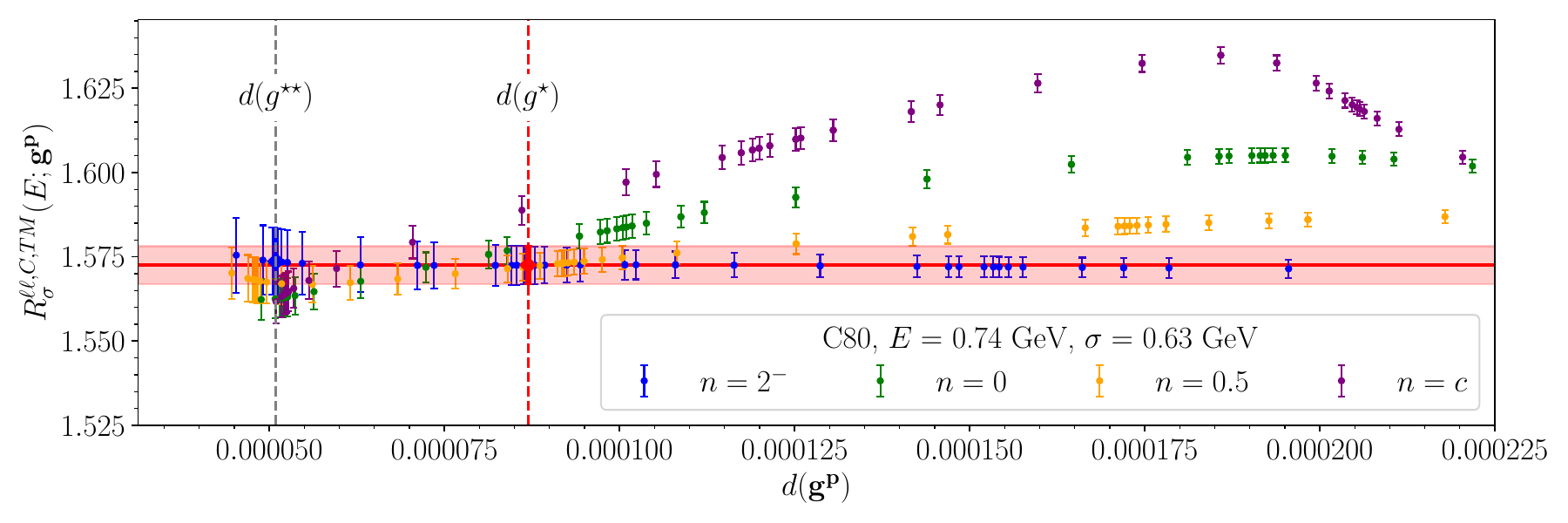}\\
\includegraphics[width=\columnwidth]{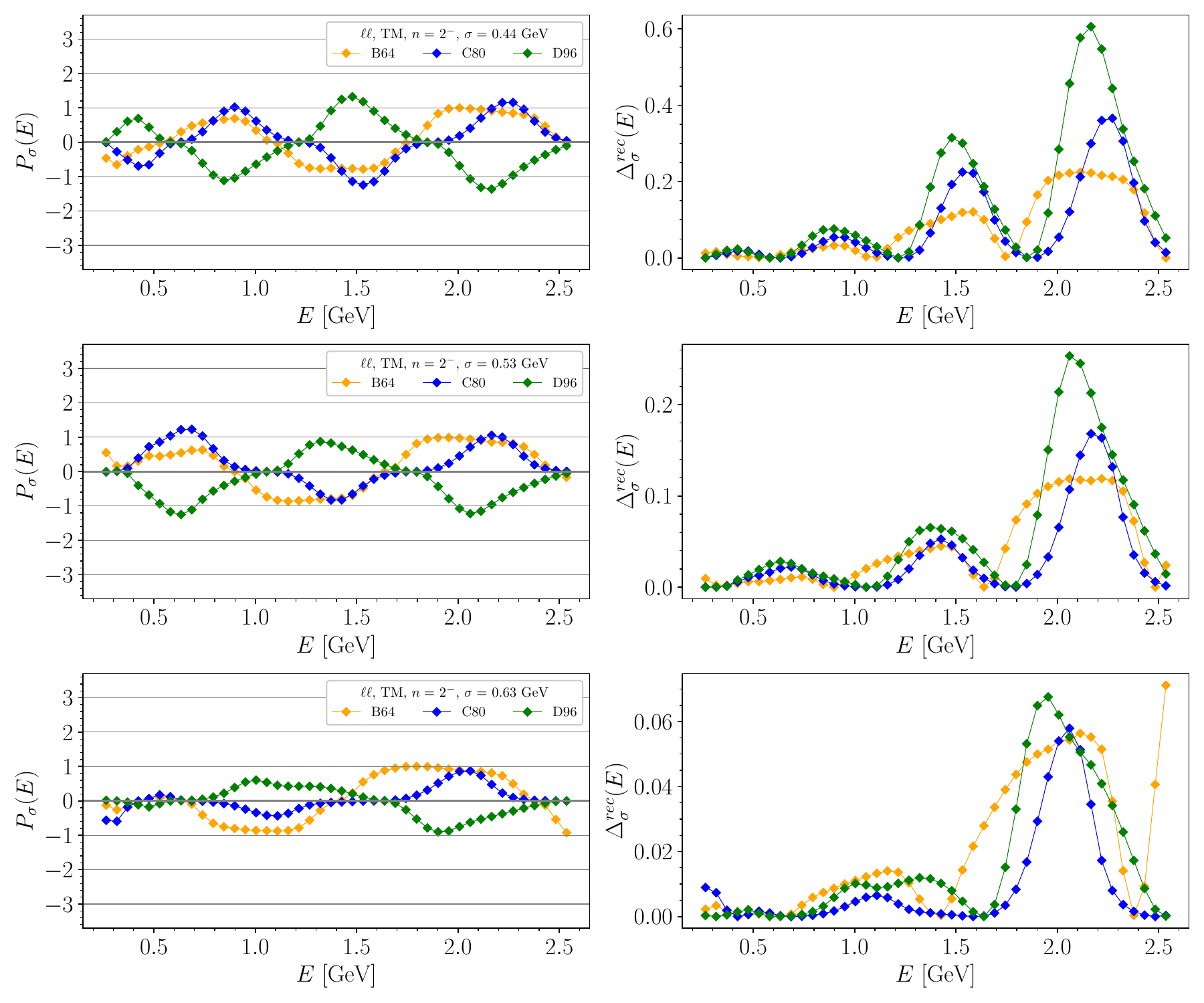}
\caption{\label{fig:llcstability} \emph{Top-panel}: Example of the stability analysis procedure in the case of the light-light connected contribution to $R_\sigma(E)$. The $\mathrm{n}=2^-$ data (blue points), that are remarkably stable in all cases analyzed in this work, have been used to estimate the central values of $R_\sigma(E)$ and the systematic errors $\Delta^{\mathrm{rec}}_\sigma(E)$. \emph{Other panels}: the plots on the left show $P_\sigma(E)$ while those on the right show $\Delta^{\mathrm{rec}}_\sigma(E)$ on the different ensembles at $aL\sim 5$~fm 
for $\sigma_1$ (second panel), $\sigma_2$ (third panel) and $\sigma_3$ (bottom panel). Most of the points are in the statistically dominated regime ($\vert P_\sigma(E)\vert<1$) and none in the systematics dominated regime ($\vert P_\sigma(E)\vert >2$).}
\end{figure}

\emph{Stability analysis.} In the top-panel of FIG.~\ref{fig:llcstability} we show an example of the stability analysis procedure in the case of $R^{\ell\ell,C,\mathrm{TM}}_\sigma(E)$. The data have been obtained on the C80 ensemble and correspond to $\sigma_3$ and $E=0.74$~GeV. The datasets corresponding to the different weighting functions have different colors and the errors on the points are statistical. Within each dataset the different points correspond to different values of $\lambda$ and, consequently, of $d(\vec g^{\vec p})$. As it can be seen, the behaviour of $R_\sigma(E;\vec g^{\vec p})$ as a function of $d(\vec g^{\vec p})$ is that expected according to the observations of the previous section. For large values of $d(\vec g^{\vec p})$ the results obtained at different values of the algorithmic parameters are significantly different and have small statistical errors. In the region of very small values of $d(\vec g^{\vec p})$ the statistical errors tend to increase and no significant differences are observed. The region from which we extract the results that we use for the central values of $R_\sigma(E;\vec g^{\vec p})$ is the intermediate one, where the results at different $\vec g^{\vec p}$ agree within the statistical errors and these are still under control. The dotted vertical lines correspond to $d(\vec g^\star)$ (red) and $d(\vec g^{\star\star})$ (black) and the point corresponding to $R_\sigma(E;\vec g^{\star})$ is marked in red (see Eq.~(\ref{eq:dgdef}) and Eqs.~(\ref{eq:stars}). The red horizontal band corresponds to our estimate of the error 
\begin{flalign}
\bar{\Delta}_\sigma(E) = \sqrt{\left[\Delta^\mathrm{stat}_\sigma(E)\right]^2+\left[\Delta^\mathrm{rec}_\sigma(E)\right]^2}
\end{flalign}
that, in this case, is slightly larger than the statistical one. The red band is always statistically compatible with the points at very small values of $d(\vec g^{\vec p})$ and with all points in the case of $\mathrm{n}=\alpha=2^-$ (blue points), thus representing a reliable estimate of the error. The other three panels in FIG.~\ref{fig:llcstability} show a quantitative summary of the results of the stability analyses on $R^{\ell\ell,C}_\sigma(E)$ by showing the quantity $P_\sigma(E)$ (see Eq.~(\ref{eq:syspull})) for all values of $\sigma$ and $E$ and for all the ensembles at $aL\sim 5$~fm. As it can be seen, none of our results has $\vert P_\sigma(E)\vert>2$ and a very large fraction of them is in the statistically dominated regime ($\vert P_\sigma(E)\vert<1$).

\begin{figure}[t!]
\includegraphics[width=\columnwidth]{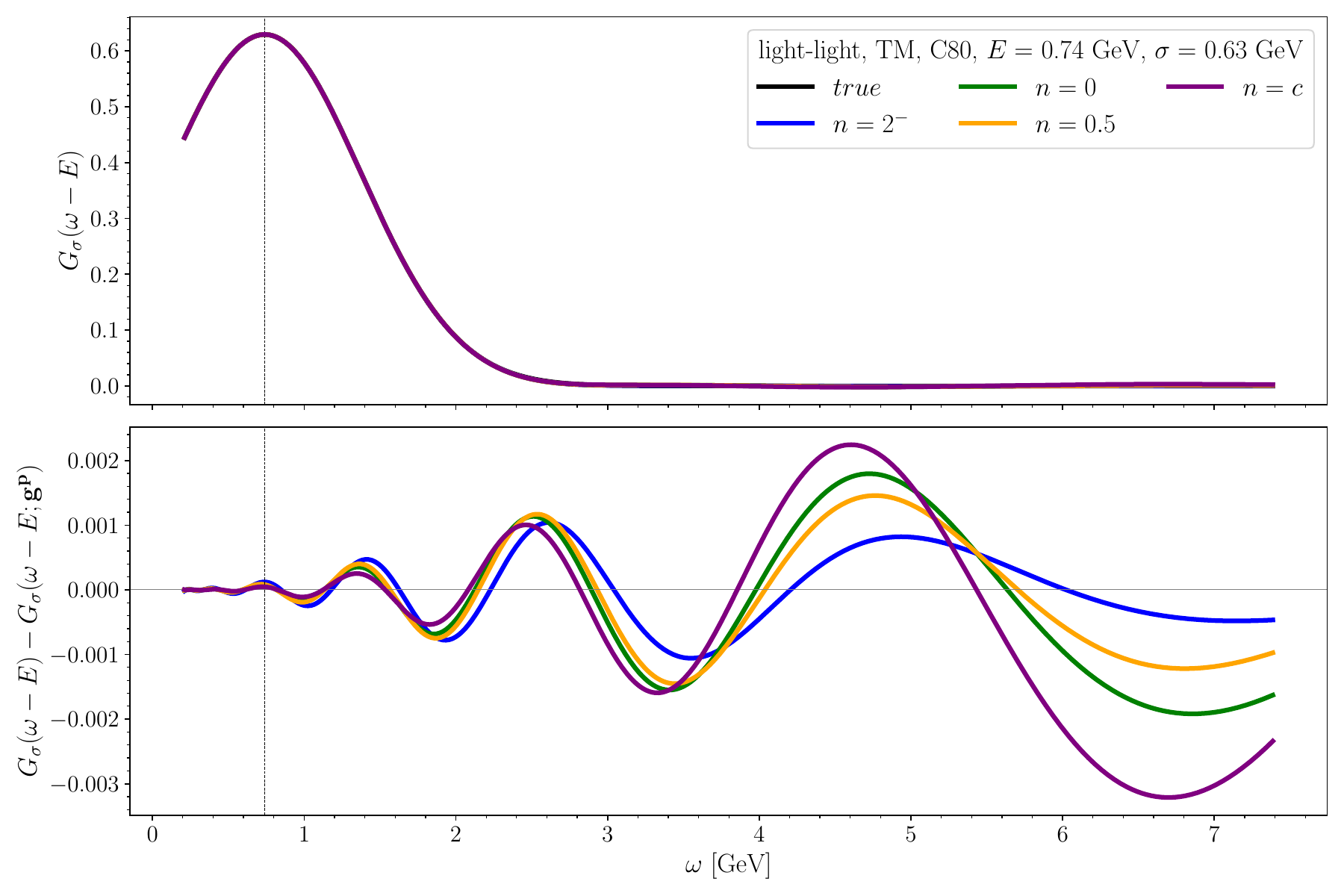}\\
\caption{\label{fig:gaussians} Reconstructed kernels at $d(\vec g^{\star})$ on the C80 ensemble at $\sigma_3$ and $E=0.74$~GeV. In both plots the results are shown for $\omega>E_0$ and the vertical lines mark the location of the peak of the target Gaussian.}
\end{figure}
The results corresponding to $\mathrm{n}=2^-$ are remarkably stable in all cases that we have been analyzing. This is evident in the top-panel of FIG.~\ref{fig:llcstability} and is also expected in light of the observations of the previous section concerning the importance of controlling the contributions to the reconstruction bias coming from the high energies. 

In order to better illustrate this point we show, in FIG.~\ref{fig:gaussians}, the kernels reconstructed at $d(\vec g^\star)$ by using the different weighting functions at $\sigma_3$ and $E=0.74$~GeV on the C80 ensemble, i.e. the same case considered in the top-panel of FIG.~\ref{fig:llcstability}. The quality of the reconstruction is excellent in all cases, see the top-plot of FIG.~\ref{fig:gaussians} where it is almost impossible to distinguish the different reconstructions. The bottom-plot shows the difference between the target and reconstructed kernels for the different choices of the weighting functions. At large energies, the $\mathrm{n}=\{0,1/2\}$ cases smoothly interpolate between the $\mathrm{n}=2^-$ case (blue), where the difference decreases in magnitude (oscillating in sign), and the Chebyshev case (violet), where the absolute value of the difference increases. This explains our choice of extracting the central values and errors for $R_\sigma(E)$ from the $\mathrm{n}=2^-$ datasets. In our experience there is no particular advantage in using the weighting function corresponding to Chebyshev polynomials.

\begin{figure}[t!]
\includegraphics[width=\columnwidth]{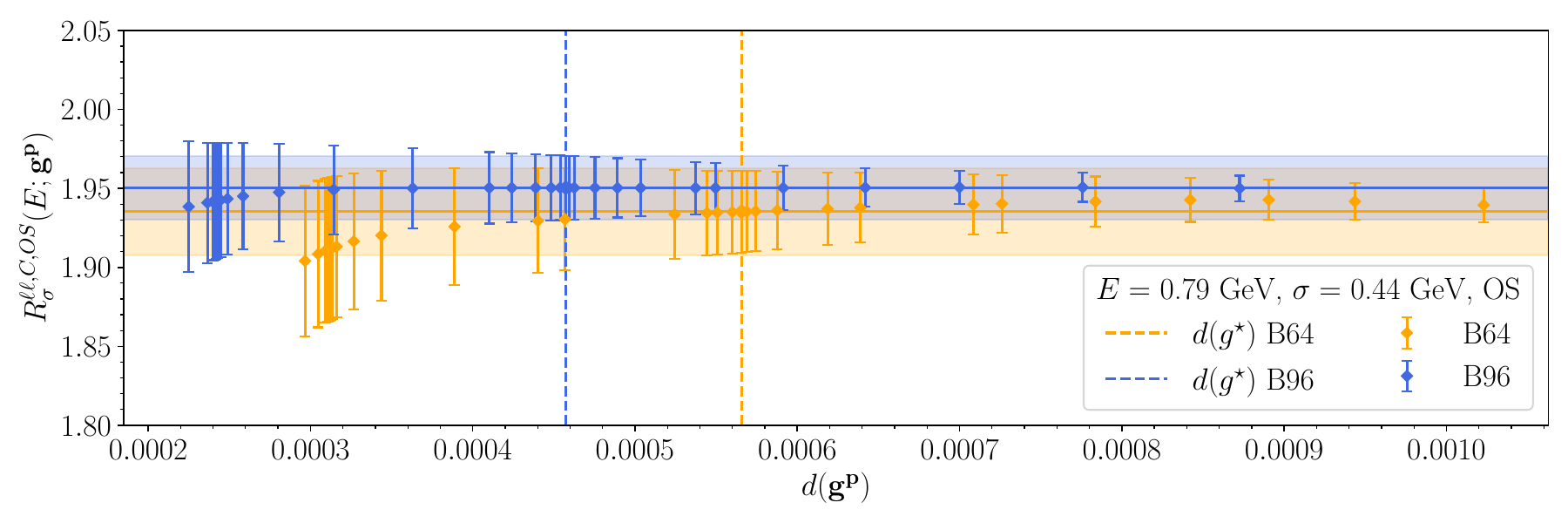}\\
\includegraphics[width=\columnwidth]{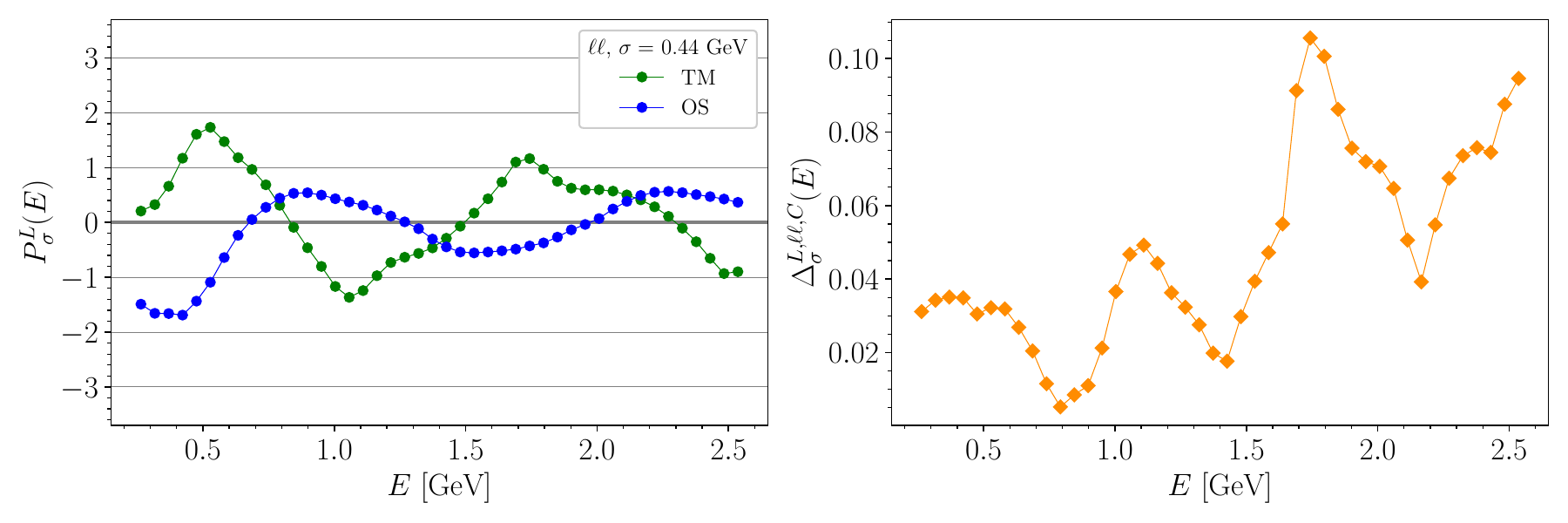}\\
\includegraphics[width=\columnwidth]{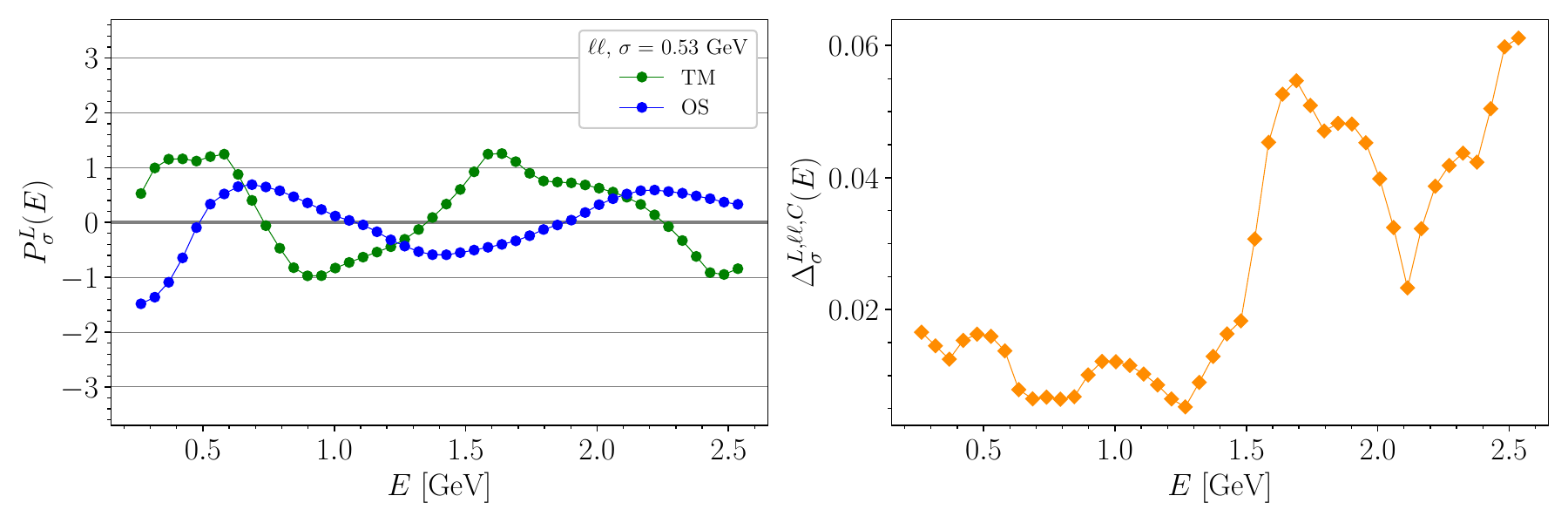}\\
\includegraphics[width=\columnwidth]{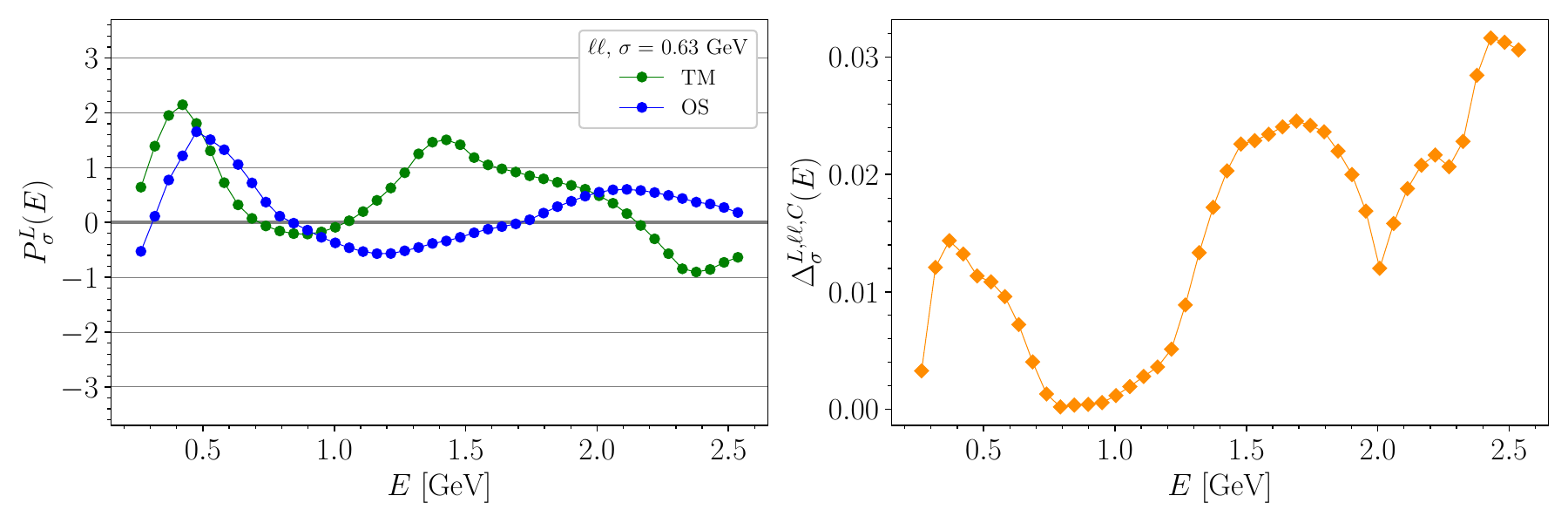}
\caption{\label{fig:llcvolume} \emph{Top-panel}: Example of the comparison of $R^{\ell\ell,C}_\sigma(E)$ on the B64 and B96 ensembles corresponding to volumes $aL\sim 5$~fm and $3 aL/2\sim 7.5$~fm. \emph{Other panels}: on the left we show $P_\sigma^L(E)$ for $\sigma_1$ (second panel), $\sigma_2$ (third panel) and $\sigma_3$ (bottom panel). On the right we show $\Delta^{L,\ell\ell, C}_\sigma(E)$, our estimate of the finite-volume systematic errors.}
\end{figure}
\emph{Volume dependence.} In the top-panel of FIG.~\ref{fig:llcvolume} we show an example of the comparison of $R^{\ell\ell,C}_\sigma(E)$ on the two ensembles B64 and B96 differing only for the spatial volume and time extension of the lattice. The data correspond to $\sigma_1$, $E=0.79$~GeV and to the OS regularization. The blue and orange bands are the results of the stability analysis performed independently on the two ensembles. The other three panels of FIG.~\ref{fig:llcvolume} show a quantitative summary of the comparison of $R^{\ell\ell,C}_\sigma(E)$ on the two volumes for all values of $\sigma$ and all energies. In these panels the plots on the left show, for both regularizations, the quantity
\begin{flalign}
P^{L}_\sigma(E)=
\frac{R_\sigma\left(E;\frac{3L}{2}\right) - R_\sigma(E;L)}{
\sqrt{\left[\bar{\Delta}_\sigma\left(E;\frac{3L}{2}\right)\right]^2 + \left[\bar{\Delta}_\sigma(E;L)\right]^2}}\;,
\label{eq:volumepull}
\end{flalign}
where $\bar{\Delta}_\sigma(E;L)$ is the error on $R_\sigma(E;L)$ extracted from the stability analysis performed on the B64 ensemble ($aL\sim 5$~fm) while $\bar{\Delta}_\sigma(E;3L/2)$ and $R_\sigma(E;3L/2)$ are the corresponding quantities extracted from the B96 ensemble. As it can be seen, although the light-light contribution to $R_\sigma(E)$ is the one on which we expect larger finite volume effects, particularly at small energies and small values of $\sigma$, we don't observe significant differences between the B64 and B96 data within the errors resulting from the stability analyses. All our points have $\vert P^L_\sigma(E)\vert <2.2$, most of them $\vert P^L_\sigma(E)\vert <1$ and $P^L_\sigma(E)$ oscillates quite regularly around zero as a function of the energy for all values of $\sigma$ and for both regularizations. This is presumably due to the fact that our data are not yet sufficiently precise to observe significant finite volume and finite temperature effects. 
Given the fact that the results on the two volumes are compatible we include both the B64 and B96 ensemble in our continuum extrapolations. Nevertheless, in order to provide an estimate of the systematics associated with possible residual finite-volume effects we consider the quantity 
\begin{flalign}
&
\Delta^{L}_\sigma(E)=
\max_{\mathrm{reg}=\{\mathrm{OS},\mathrm{TM}\}}\Bigg\{
\nonumber \\
&
\left\vert R^\mathrm{reg}_\sigma\left(E;\frac{3L}{2}\right) - R^\mathrm{reg}_\sigma(E;L)\right\vert
\mathrm{erf}\left(
\frac{\left\vert P^{L,\mathrm{reg}}_\sigma(E) \right\vert}{\sqrt{2}}
\right)
\Bigg\}.
\label{eq:volumesys}
\end{flalign}
The plots on the right in the last three panels of FIG.~\ref{fig:llcvolume} show $\Delta^{L,\ell\ell, C}_\sigma(E)$.

\begin{figure}[t!]
\includegraphics[width=\columnwidth]{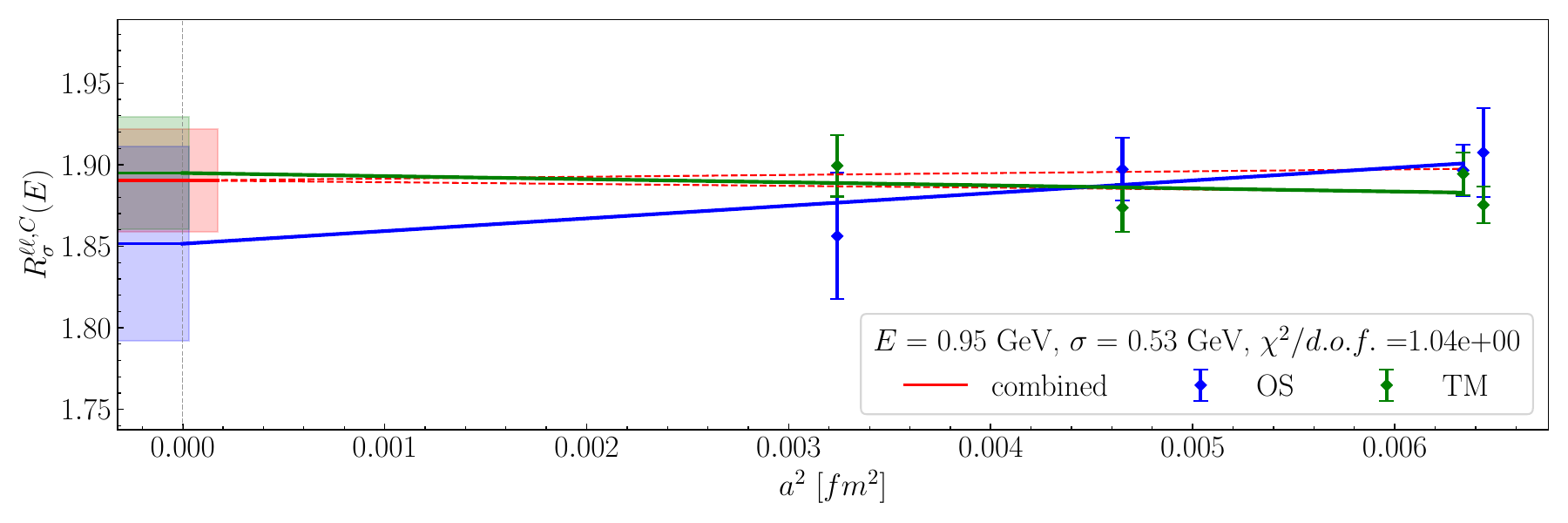}\\
\includegraphics[width=\columnwidth]{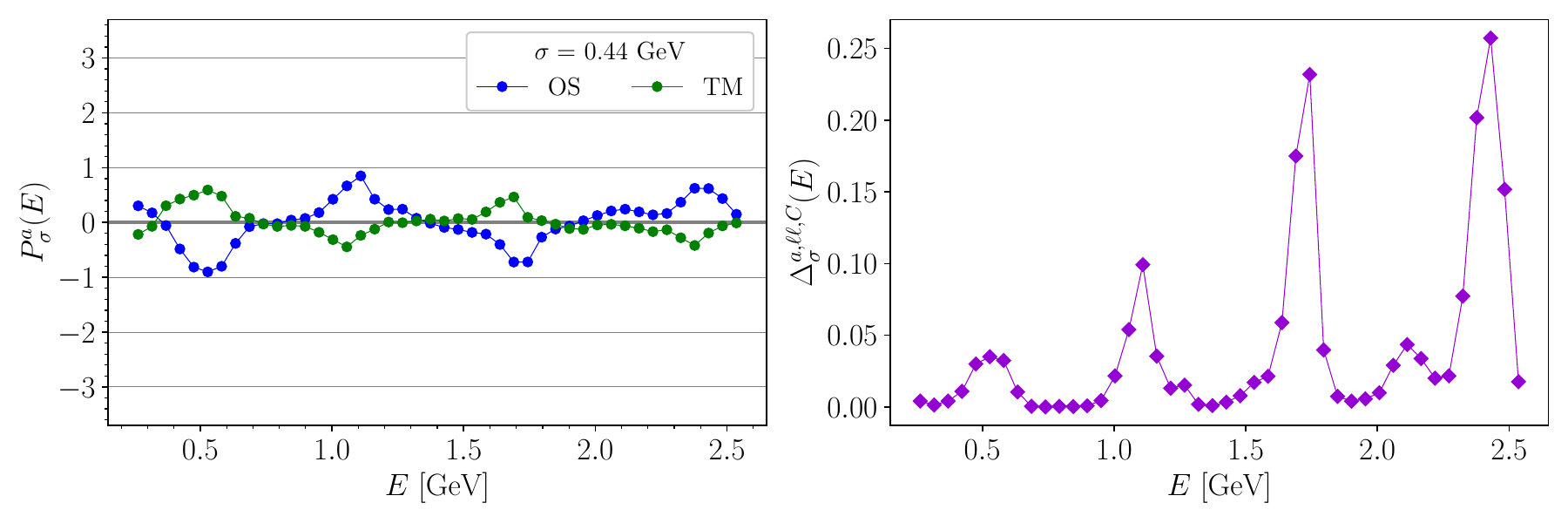}\\
\includegraphics[width=\columnwidth]{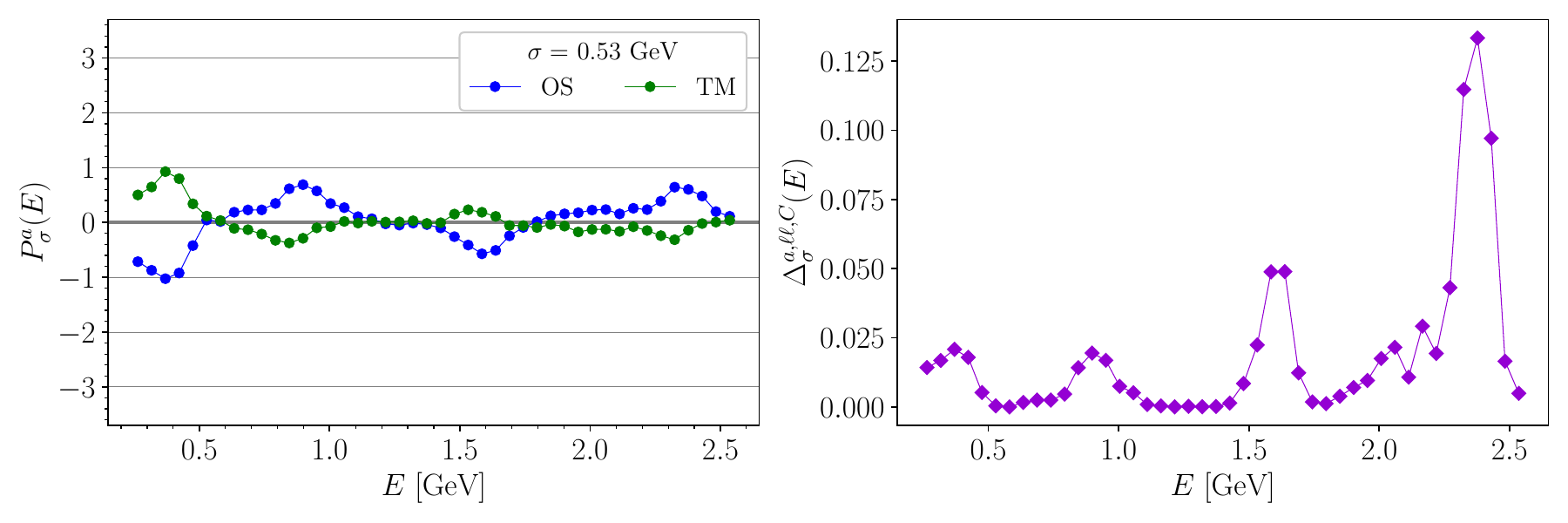}\\
\includegraphics[width=\columnwidth]{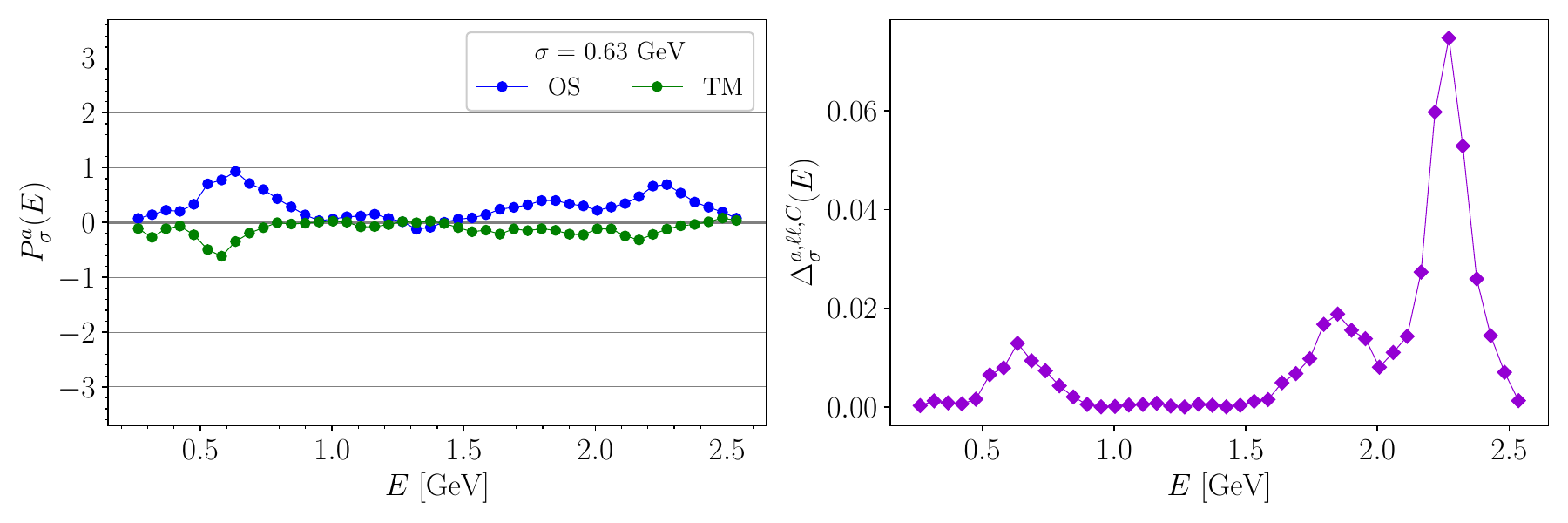}
\caption{\label{fig:llccontinuum} \emph{Top-panel}: Example of the continuum extrapolation of $R^{\ell\ell,C}_\sigma(E)$. The blue points correspond to the OS regularization while the green ones to the TM regularization. Although difficult to distinguish on the scale of the plots, at the coarsest value of the lattice spacing there are two points for each regularization that have been obtained on the B64 and B96 ensembles having different physical volumes. The red shaded area is the result of the combined continuum extrapolation of all data. The green and blue shaded areas are the results of the unconstrained extrapolations of respectively the TM and OS data. \emph{Other panels}: on the left we show $P_\sigma^a(E)$ for $\sigma_1$ (second panel), $\sigma_2$ (third panel) and $\sigma_3$ (bottom panel). On the right we show $\Delta^{a,\ell\ell, C}_\sigma(E)$, our estimate of the systematic errors associated with the continuum extrapolations.}
\end{figure}
\emph{Continuum extrapolations.} The top-panel of FIG.~\ref{fig:llccontinuum} shows an example of our continuum extrapolations, corresponding to $\sigma_2$ and $E=0.95$~GeV. The green points correspond to the TM regularization and the blue ones to the OS regularization. For each regularization at the coarsest value of the lattice spacing there are two points corresponding to the two ensembles B64 and B96 and, therefore, to different volumes. These would have been barely distinguishable on the scale of the plot, given the fact that finite volume effects are negligible within the quoted errors, and the B96 points have been slightly displaced on the $x$-axis to help the eye. 

\begin{figure}[t!]
\includegraphics[width=\columnwidth]{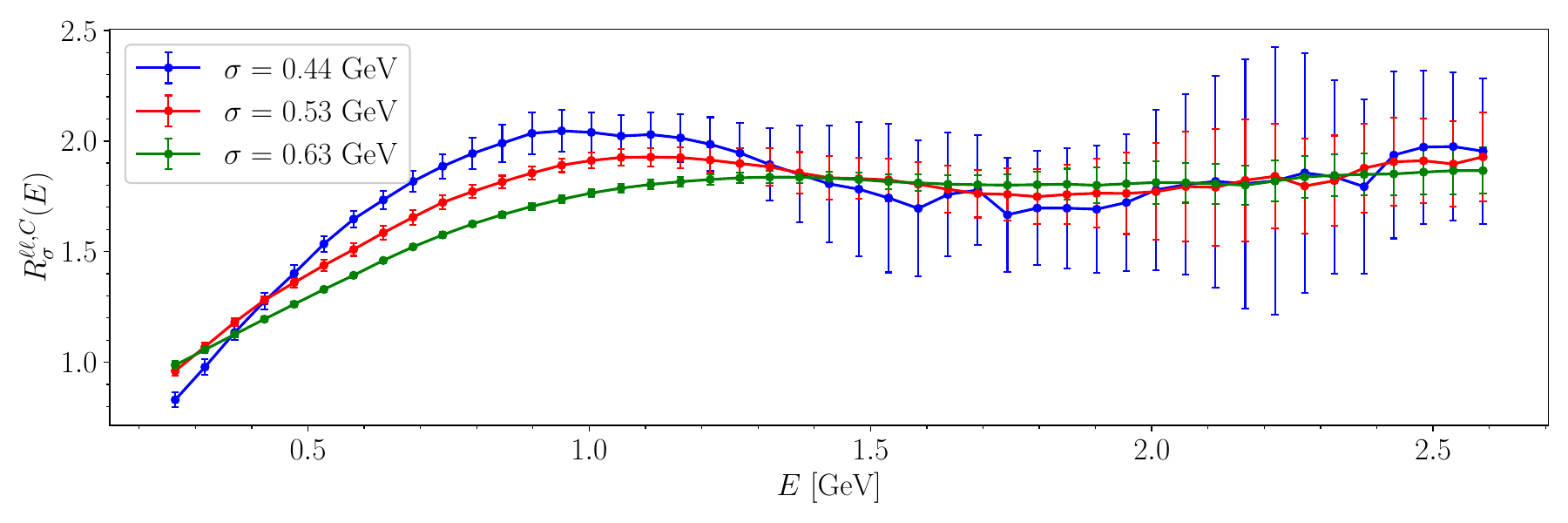}
\caption{\label{fig:llfinal} The figure show our final results for $R^{\ell\ell,C}_\sigma(E)$. }
\end{figure}
We perform both constrained and unconstrained continuum extrapolations. In the constrained extrapolation we fit OS and TM data by performing a correlated $\chi^2$-minimization that, at fixed $E$ and $\sigma$, fits the data of $R^{\ell\ell,C}_\sigma(E)$ with functions
\begin{flalign}
f^\mathrm{OS}(a) = A + B^\mathrm{OS} a^2\;,
\quad
f^\mathrm{TM}(a) = A + B^\mathrm{TM} a^2\;,
\end{flalign}
where the dependence w.r.t. the lattice spacing is assumed to be linear in $a^2$ with different slopes ($B^\mathrm{OS}$ and $B^\mathrm{TM}$) for the different regularizations and a common continuum limit ($A$) is enforced.  The correlation matrix of the data is block diagonal since the results corresponding to different ensembles are fully uncorrelated while the two points on the same ensemble, corresponding to the different regularizations, are obtained from the same gauge configurations. The result of this extrapolation is shown in red in the top-plot of FIG.~\ref{fig:llccontinuum}. In the unconstrained extrapolations, in which data are totally uncorrelated, we use the same fitting functions but we allow for different continuum limits, $A^\mathrm{OS}$ (blue band) and $A^\mathrm{TM}$ (green band). The other three panels in FIG.~\ref{fig:llccontinuum} show a quantitative summary of the comparison of the constrained and unconstrained extrapolations. The plots on the left show the quantity
\begin{flalign}
P^{a,\mathrm{reg}}_\sigma(E)=\frac{A - A^\mathrm{reg}}{\sqrt{\left[\Delta A\right]^2 + \left[\Delta A^\mathrm{reg}\right]^2}},
\label{eq:apull}
\end{flalign}
where $A$ is the result of the combined extrapolation at the given values of $\sigma$ and $E$, $\Delta A$ its error while $A^\mathrm{reg}$ and $\Delta A^\mathrm{reg}$ are the results and errors of the unconstrained extrapolations. As it can be seen, at all analyzed values of $\sigma$ and $E$ there is full compatibility between the constrained and unconstrained extrapolations.
In the case of $R^{\ell\ell,C}_\sigma(E)$, at all quoted values of $E$ and $\sigma$, we observe small cutoff effects. This, again, is presumably due to the fact that our data at fixed cutoff have large statistical errors and/or to a rather conservative estimate of the kernel reconstruction systematics that, in addition to finite volume effects, also masks cutoff effects. In light of this observation and of the compatibility of the constrained and unconstrained extrapolations, we decided to quote the central values and errors of our final results from the combined fits and to estimate the systematic errors associated with the continuum extrapolations by
\begin{flalign}
\Delta^{a}_\sigma(E)=\max_{\mathrm{reg}=\{\mathrm{OS},\mathrm{TM}\}}\left\{
\left\vert A - A^\mathrm{reg}\right\vert
\mathrm{erf}\left(
\frac{\left\vert P^{a,\mathrm{reg}}_\sigma(E) \right\vert}{\sqrt{2}}
\right)
\right\}.
\label{eq:asys}
\end{flalign}
The plots on the right in the last three panels of FIG.~\ref{fig:llccontinuum} show $\Delta^{a,\ell\ell,C}_\sigma(E)$ that we add in quadrature to the other errors on our final results for $R_\sigma(E)$. 
The final result for the connected light-light contribution $R^{\ell\ell,C}_\sigma(E)$ are shown in FIG~\ref{fig:llfinal}.

\subsubsection{Strange-strange connected contribution}
%
\begin{figure}[t!]
	\includegraphics[width=\columnwidth]{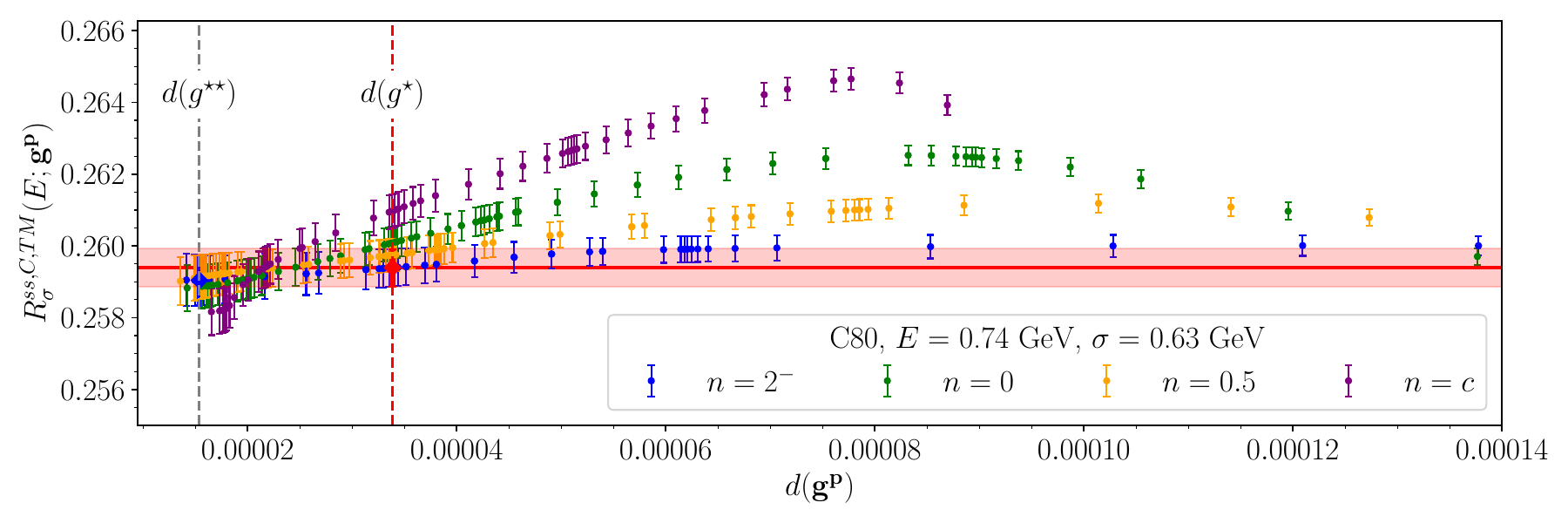}\\
	\includegraphics[width=\columnwidth]{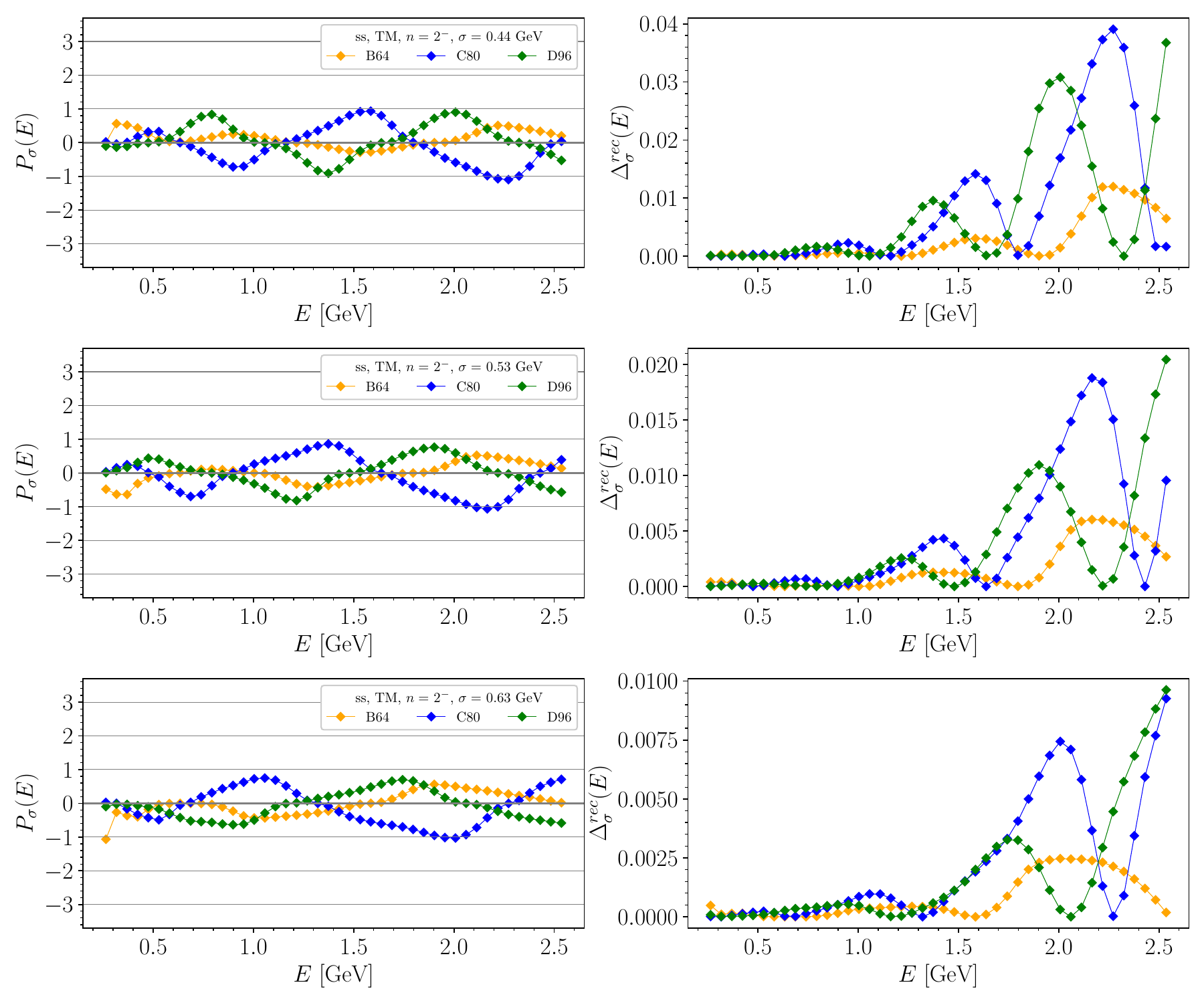}
	\caption{\label{fig:sscstability} \emph{Top-panel}: Example of the stability analysis procedure in the case of the strange-strange connected contribution to $R_\sigma(E)$. \emph{Other panels}: See FIG.~\ref{fig:llcstability}. }
\end{figure}
\begin{figure}[t!]
	\includegraphics[width=\columnwidth]{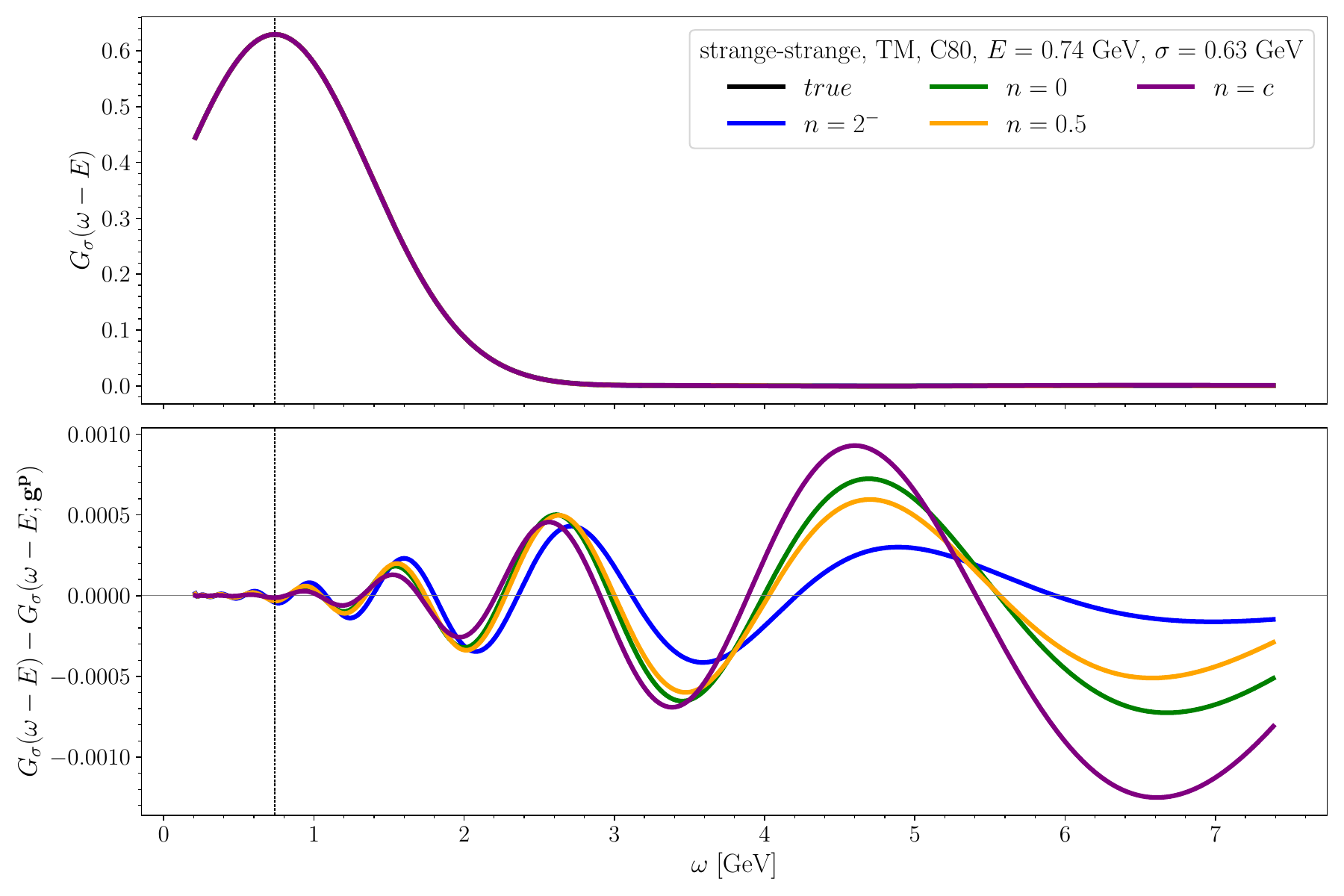}\\
	\caption{\label{fig:ssgaussians} Reconstructed kernels at $d(\vec g^{\star})$ on the C80 ensemble at $\sigma_3$ and $E=0.74$~GeV. In both plots the results are shown for $\omega>E_0$ and the vertical lines mark the location of the peak of the target Gaussian.}
\end{figure}

The following discussion of the stability analysis, volume dependence and continuum extrapolations is analogous to the one presented in the light-light case. Since, however, strange-strange connected correlators have been computed on each ensemble at two close-to-physical values of bare strange quark masses, an interpolation to the physical strange mass is required and, therefore, a detailed discussion of this additional step of the analysis will also be presented.

\emph{Stability analysis.} In the top-panel of FIG.~\ref{fig:sscstability} we show an example of the stability analysis procedure for  $R^{ss,C,\mathrm{TM}}_\sigma(E)$ in the same case as the one shown in FIG.~\ref{fig:llcstability}, that is, $E=0.74$ GeV, $\sigma_3$ and C80 ensemble.  The behaviour of the results at varying $d(\vec g^{\vec p})$ and weighting functions is totally analogous w.r.t.\ the corresponding light-light case and, again, the choice n$=2^-$ (blue points) is the most stable. The main difference is that now the $L_2$-norm $d(\vec g^\star)$ is reduced roughly by a factor 3. This is in agreement with the fact that the strange-strange connected correlator is more precise than the light-light one, thus allowing for a better reconstruction of the smearing kernel. Indeed, the systematic error associated with the imperfect reconstruction of the kernel never dominates compared to the statistical one since, as shown in the three bottom panels of FIG.~\ref{fig:sscstability}, $\vert P_\sigma(E)\vert \le 1$  in almost all cases (see Eq.~(\ref{eq:syspull})).
The excellent reconstruction of the smearing kernel can be appreciated in FIG.~\ref{fig:ssgaussians} where the difference between the target and  reconstructed kernels is shown ($E=0.74$ GeV, $\sigma_3$ and C80 ensemble).  Again, the difference is suppressed more rapidly in the n=$2^-$ case.

\begin{figure}[t!]
	\includegraphics[width=\columnwidth]{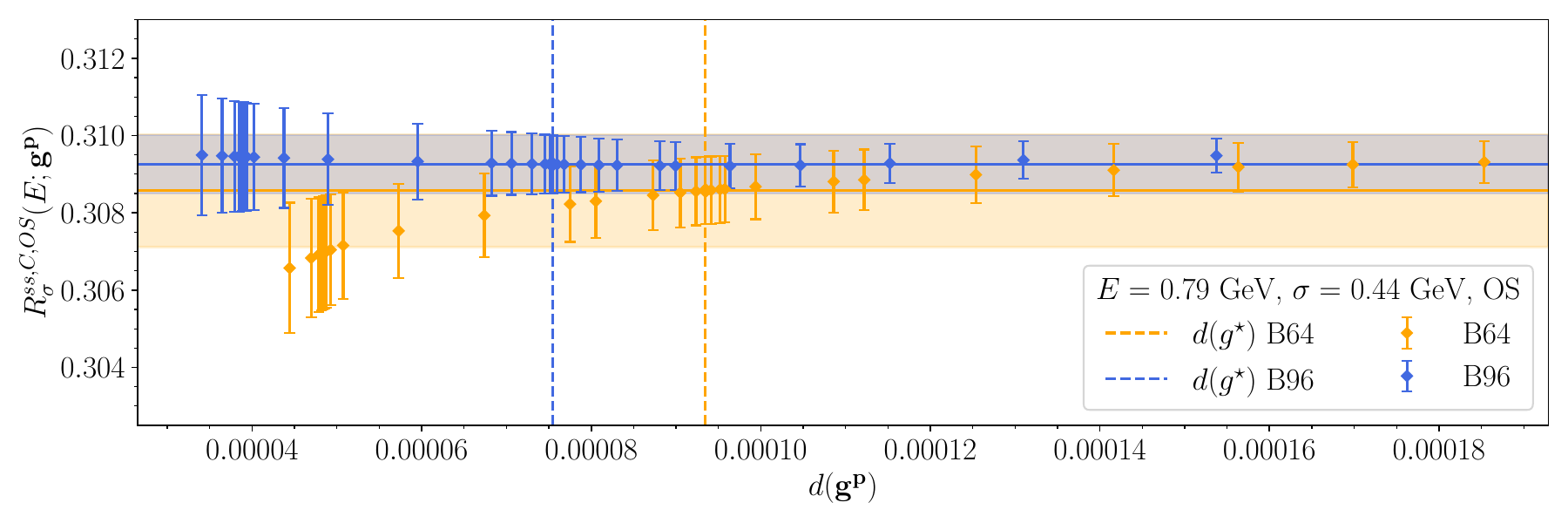}\\
	\includegraphics[width=\columnwidth]{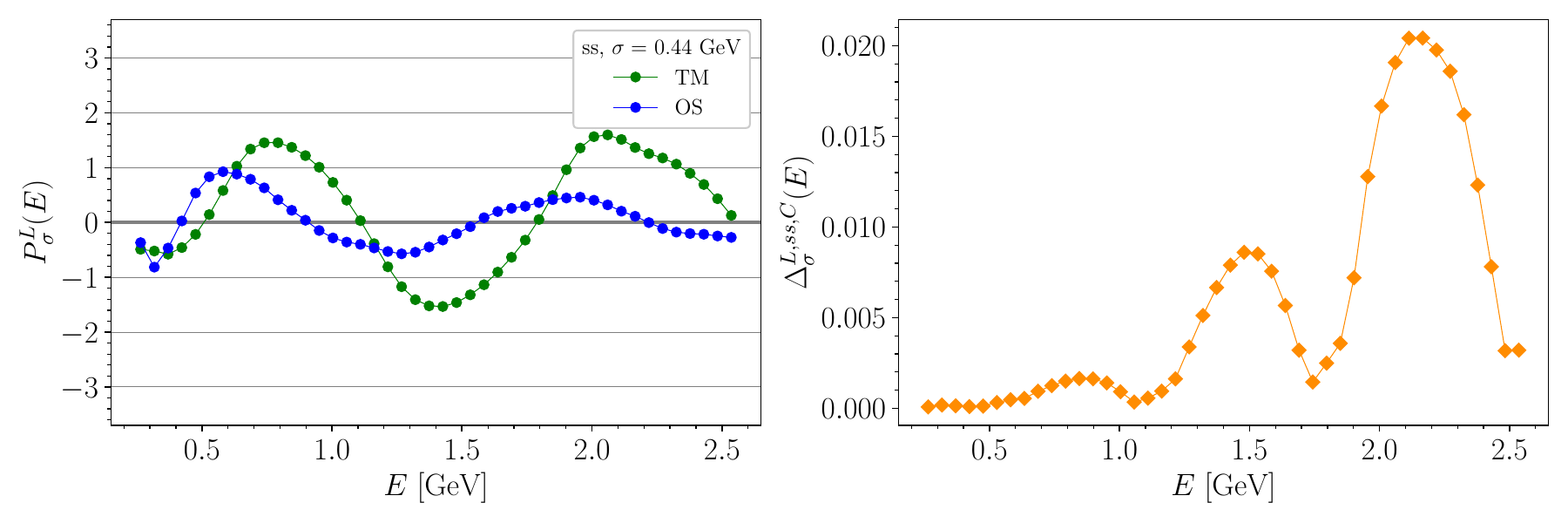}\\
	\includegraphics[width=\columnwidth]{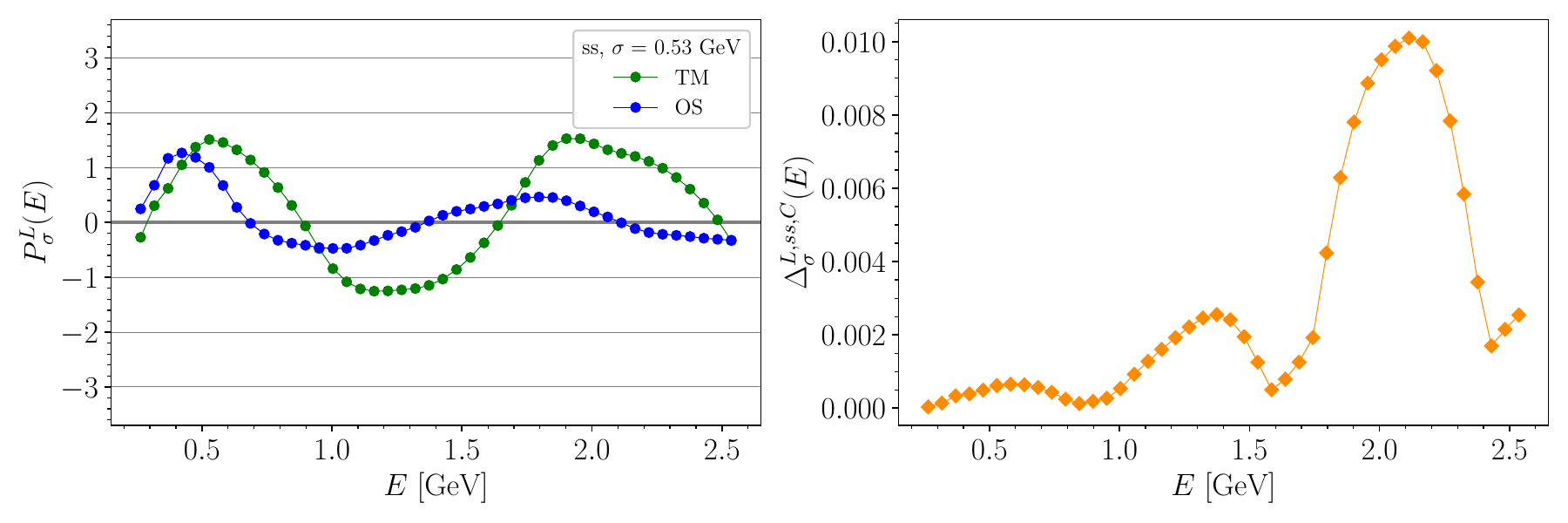}\\
	\includegraphics[width=\columnwidth]{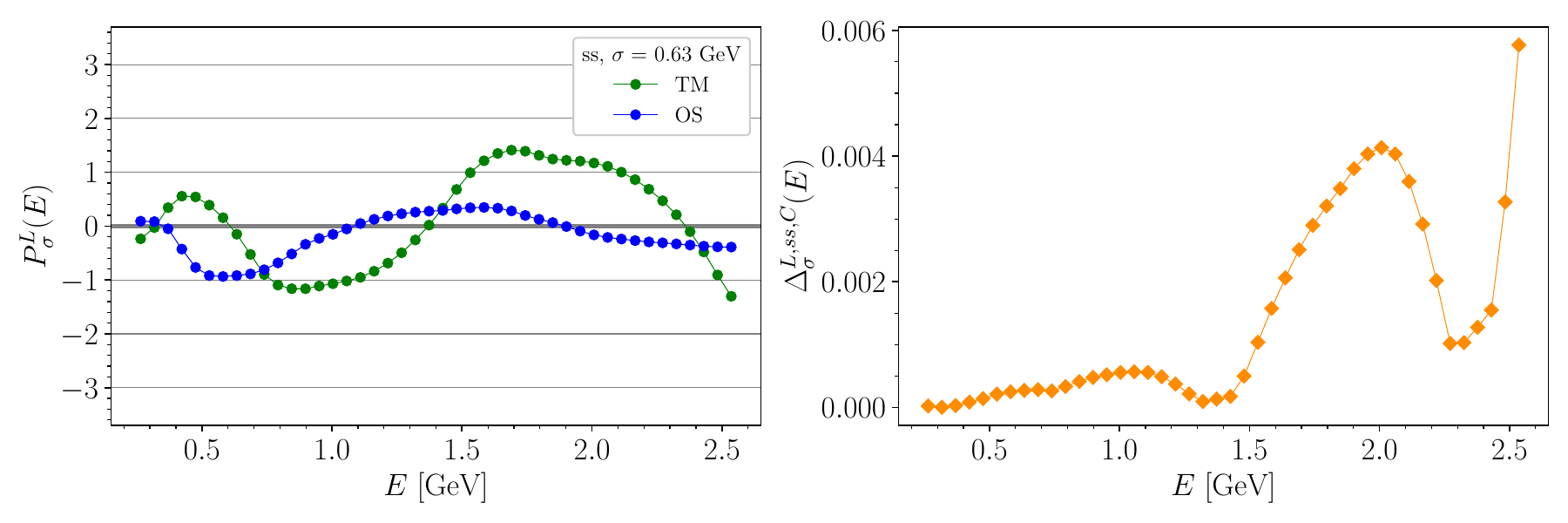}
	\caption{\label{fig:sscvolume} \emph{Top-panel}: Example of the comparison of of $R^{ss,C}_\sigma(E)$ on the B64 and B96 ensembles corresponding to volumes $aL\sim 5$~fm and $3 aL/2\sim 7.5$~fm. \emph{Other panels}: see FIG.~\ref{fig:llcvolume}.}
\end{figure}
\emph{Volume dependence.}
Even though the finite volume effects are expected to be slightly less important than in the light-light case, the smaller errors on \(R_\sigma^{ss,C,\text{reg}}(E)\) might enhance their significance in this case. The top-panel of FIG.~\ref{fig:sscvolume}
shows an example of the comparison of the values $R_\sigma^{ss,C}(E)$ obtained from the two ensembles B64 and B96  at $E=0.79$ GeV and $\sigma_1$ while a  summary of all the other cases is shown in the three bottom panels of the same figure. Also in this case $P_\sigma^L(E)$ oscillates around zero quite regularly as a function of $E$ and  $\vert P_\sigma^L(E)\vert <2$ in all cases. Therefore, despite the better accuracy of the results, the finite volume effects are not significant within the quoted statistical and systematic errors also for the strange-strange connected contribution. Nevertheless, we provide estimates for $\Delta^{L,ss, C}_\sigma(E)$, also shown in FIG.~\ref{fig:sscvolume}, that will be added in quadrature to the other errors on our final results. 

\begin{figure}[t!]
	\includegraphics[width=\columnwidth]{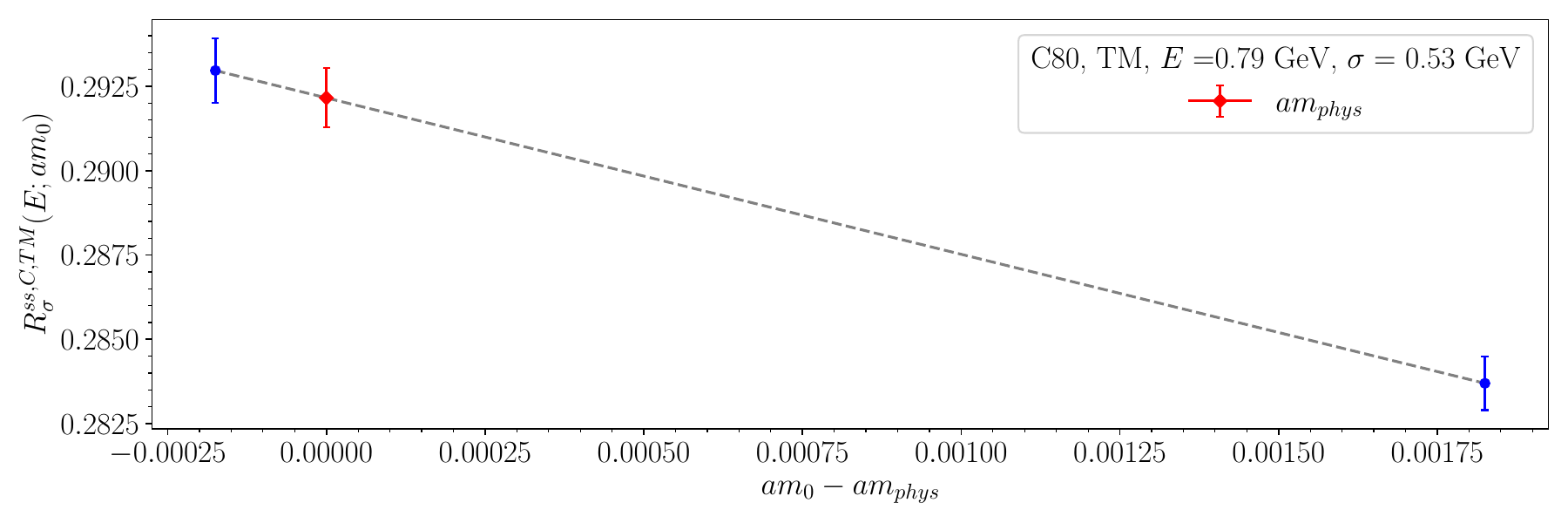}\\
	\includegraphics[width=\columnwidth]{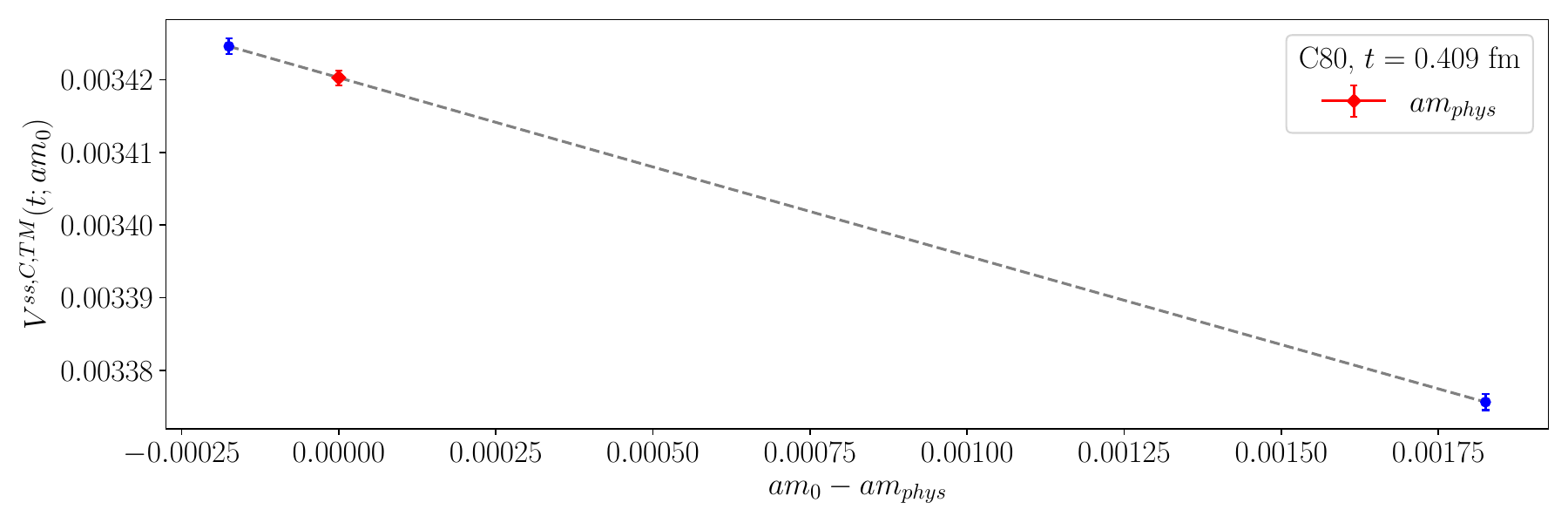}\\
	\includegraphics[width=\columnwidth]{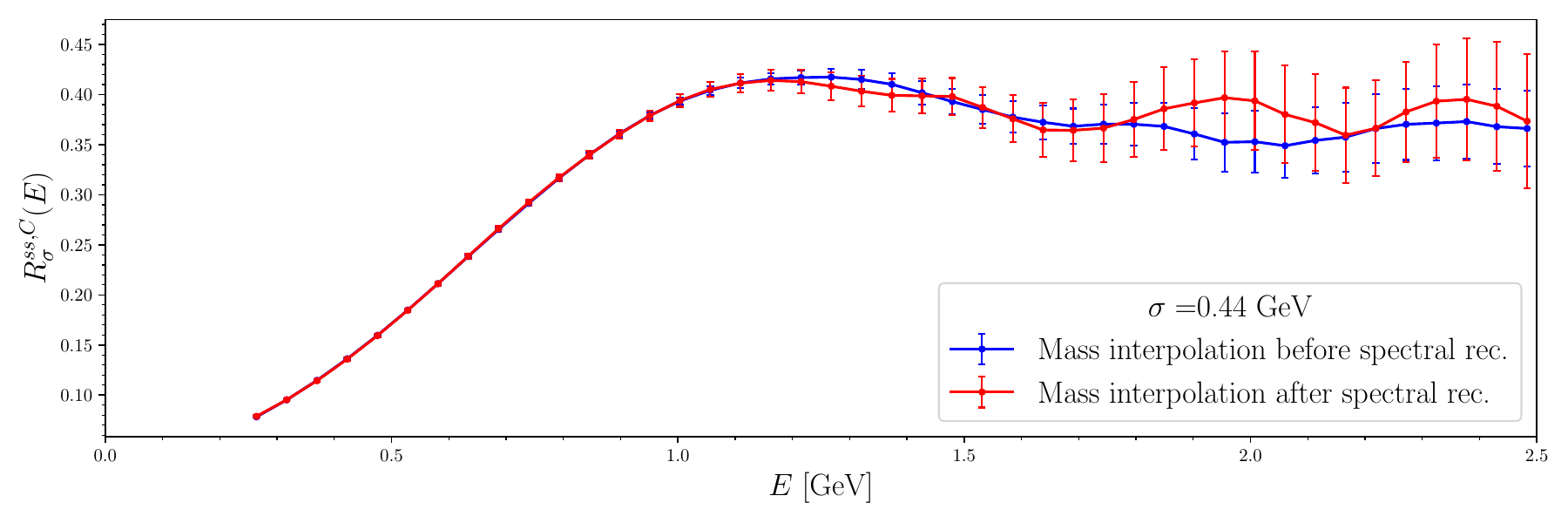}\\
	\includegraphics[width=\columnwidth]{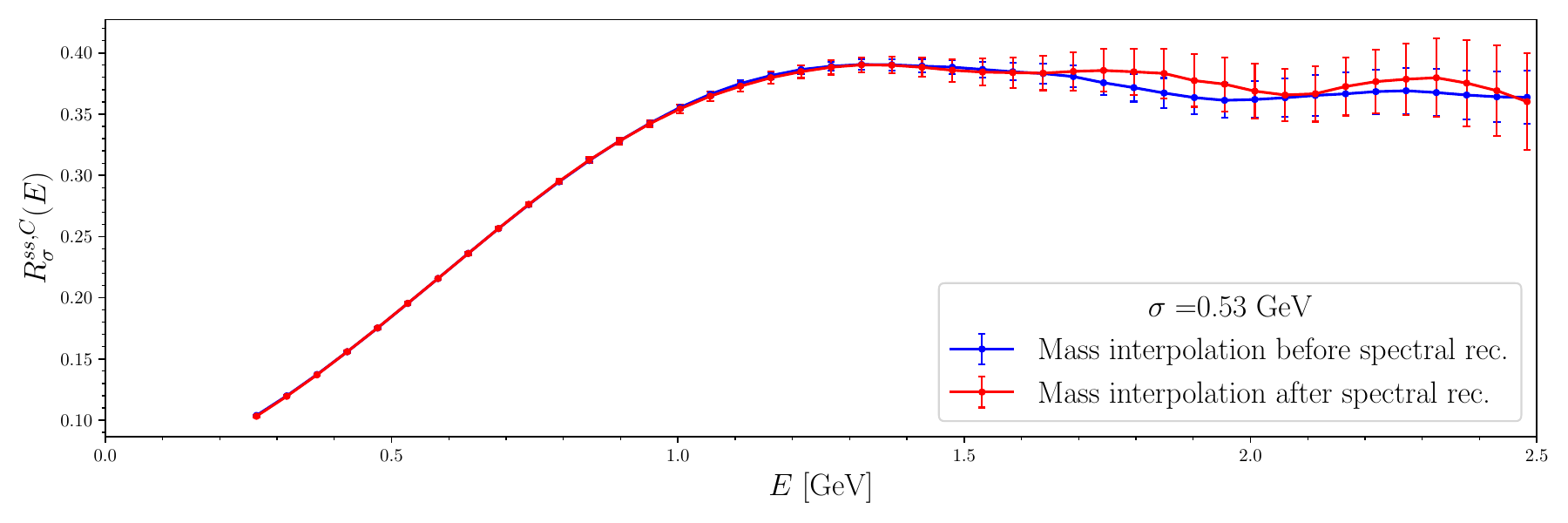}\\
	\includegraphics[width=\columnwidth]{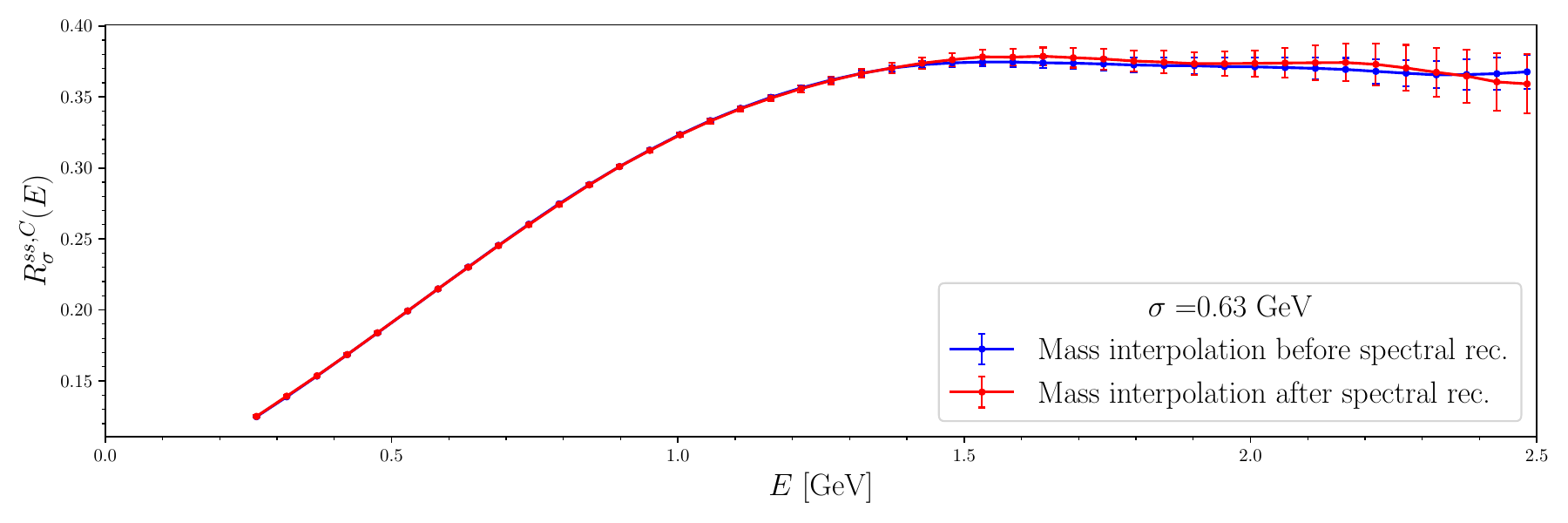}
	\caption{\label{fig:sscmass} \emph{Top-panel}: Example of the interpolation at the physical mass in the case of $R_\sigma^{ss,C,\mathrm{TM}}(E;am_0)$. \emph{Second panel}: Example of the interpolation at the physical mass for the correlator $V^{ss,C,\mathrm{TM}}(t;am_0)$. \emph{Other panels}: Comparison between the reconstructed $R_\sigma^{ss,C}(E)$ obtained by interpolating the correlators at the physical mass before applying the spectral reconstruction algorithm (blue points) and the one obtained by interpolating the results of the spectral reconstruction obtained from the correlators corresponding to the two bare masses (red points) for $\sigma_1$, $\sigma_2$ and $\sigma_3$. All the points are already extrapolated to the continuum.}
\end{figure}
\emph{Physical mass interpolation}.
As anticipated, the strange-strange correlator has been computed in correspondence of two different bare masses, very close to the physical strange quark mass, for all the ensembles (see Ref.~\citeSM{SMAlexandrou:2022amy} for more details) and an interpolation at the physical mass is then required as a further step of the analysis. 
We have two different, but physically equivalent, ways to proceed and we use both of them in order to assess the systematics associated with this analysis step.

The first route consists in applying the reconstruction algorithm separately to the correlators computed for the two bare masses. The quantity  $R_\sigma^{ss,C,\text{reg}}(E;am_0)$ thus obtained is therefore dependent upon the bare mass $am_0$. The physically relevant quantity $R_\sigma^{ss,C,\text{reg}}(E;am_\text{phys}^s)$ is then given by the interpolation at the physical strange mass $am^s_\text{phys}$ that we perform by using the linear ansatz
 \begin{equation}\label{eq:massRlinear}
 	R(am_0) = A + B\, am_0\;,
 \end{equation}
where the dependence upon $E$, $\sigma$ and the regularization has been omitted.
The top panel of FIG.~\ref{fig:sscmass} shows an example of such an interpolation for $R_\sigma^{ss,C,\mathrm{TM}}(E;am_0)$ at $E=0.79$ GeV, $\sigma_2$ on the C80 ensemble.  The interpolated value is the red point. The interpolation is repeated for all the ensembles, energies and values of $\sigma$.

In the second strategy we interpolate directly the correlator to the physical mass and then apply the spectral reconstruction algorithm to it. This procedure gives directly $R_\sigma^{ss,C,\text{reg}}(E;am^s_\text{phys})$. In this case we used a linear ansatz for the logarithm of the correlator,
\begin{equation}\label{eq:massVlinear}
\log V(am_0)= A +B\, am_0\;,
\end{equation}
where the dependence w.r.t. time and regularization has been omitted. An example of interpolation at the strange physical mass for $V^{ss,C,\mathrm{TM}}(t;am_0)$ at $t=0.409$ fm  is shown in the second panel of FIG.~\ref{fig:sscmass} in the case of the C80 ensemble. The interpolated value is the red point and notice that we are plotting directly the correlator and not its logarithm.

If the systematics induced by the mass interpolation are negligible within the quoted errors the two procedures have to give consistent results. This is what we indeed observe at fixed cutoff and also in the continuum. In the three bottom panels of FIG.~\ref{fig:sscmass} we show the comparison, at the three values of $\sigma$, of the results for $R_\sigma^{ss,C}(E)$ obtained by following the first strategy (red points) and the second one (blue points) after having performed the continuum extrapolations separately in the two cases (see next paragraph). The results are in perfect agreement within the errors at all energies and for all values of $\sigma$. 

\begin{figure}[t!]
	\includegraphics[width=\columnwidth]{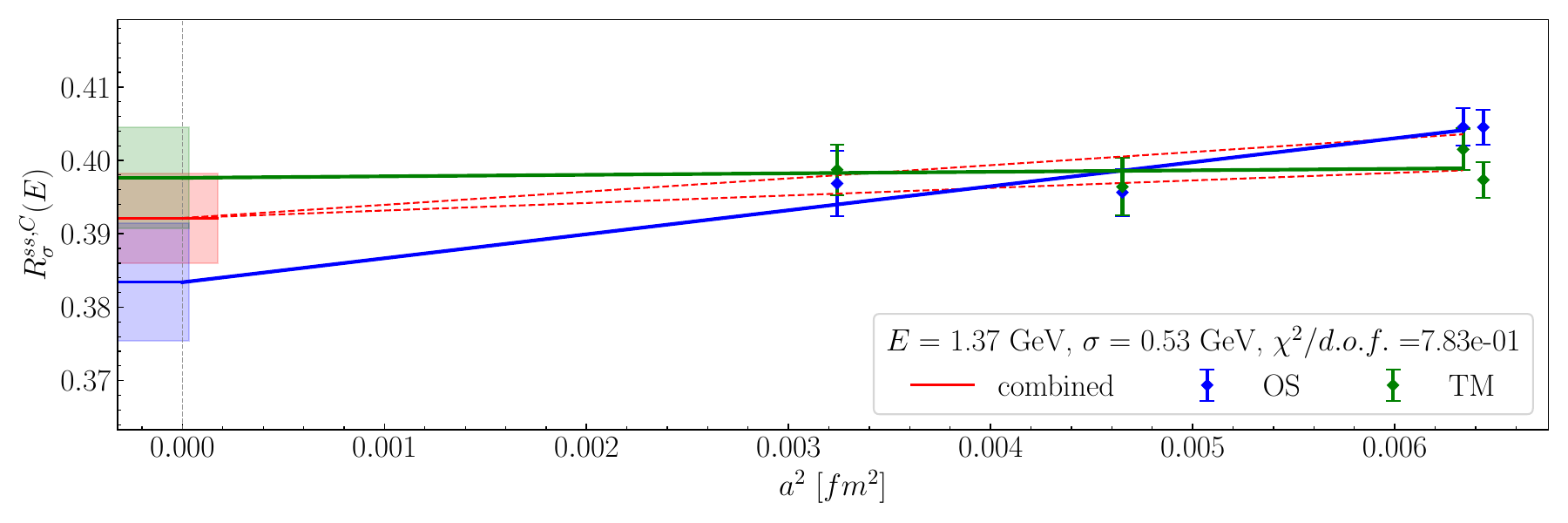}\\
	\includegraphics[width=\columnwidth]{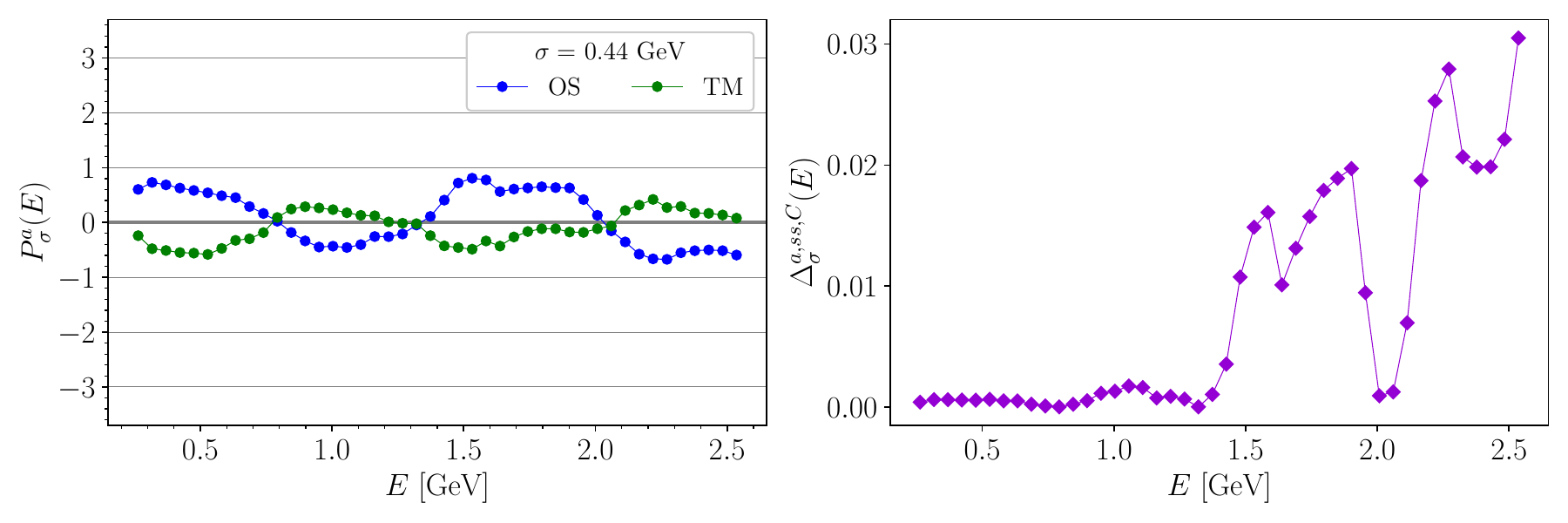}\\
	\includegraphics[width=\columnwidth]{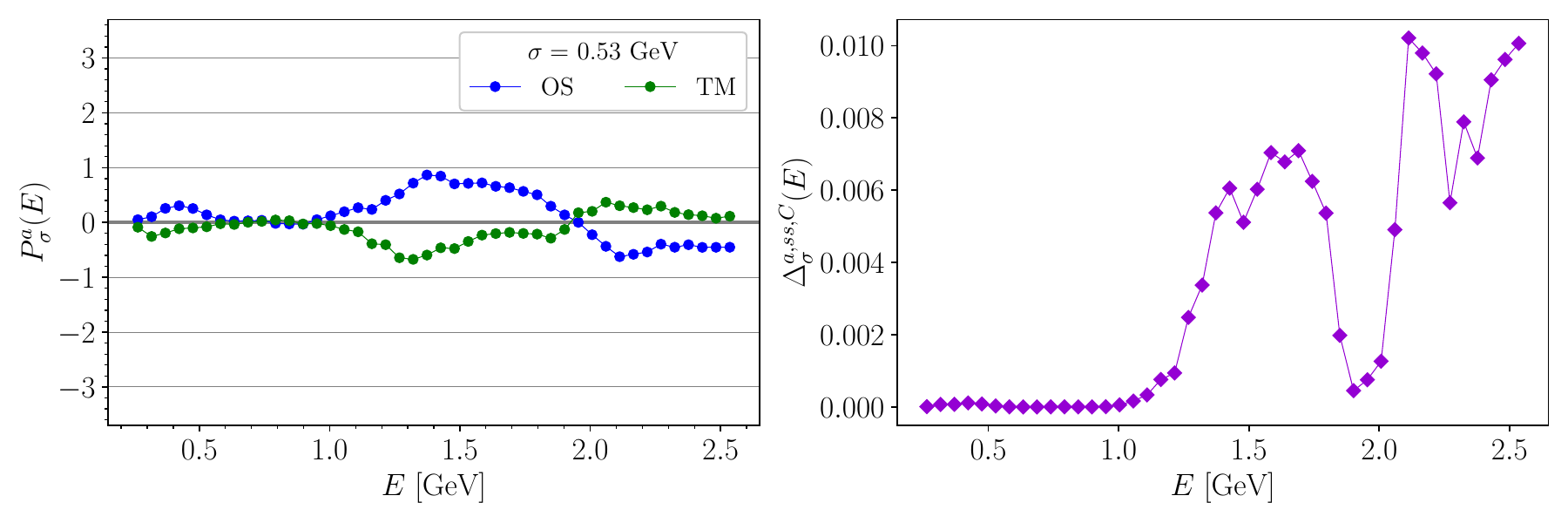}\\
	\includegraphics[width=\columnwidth]{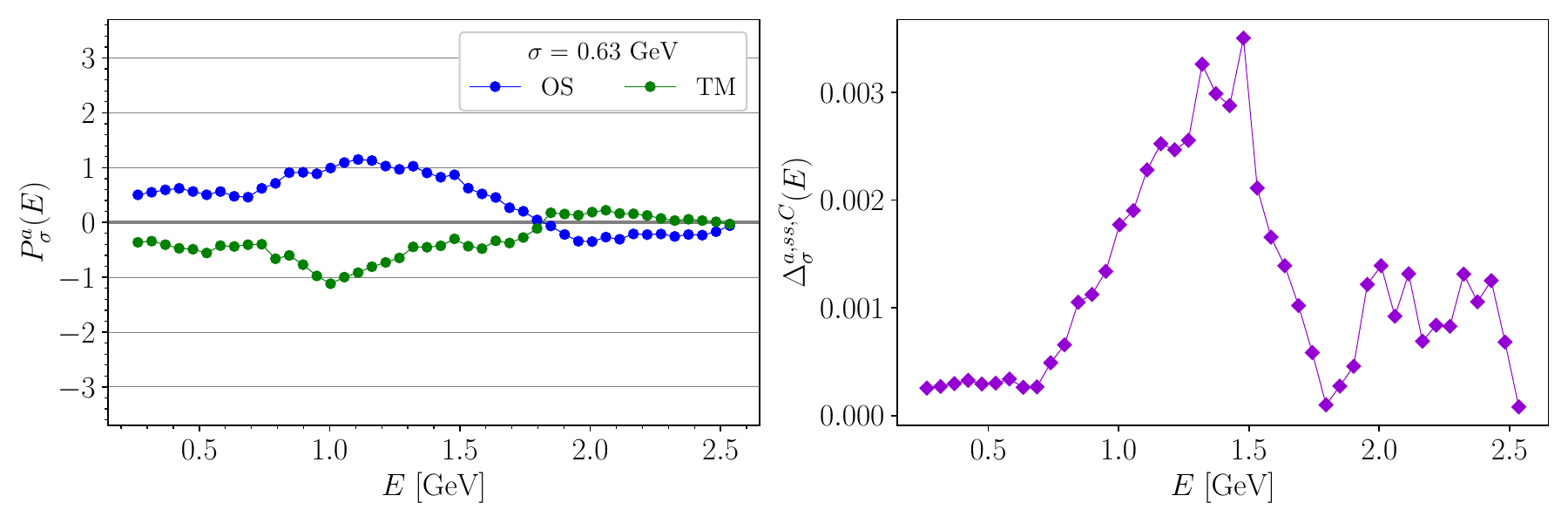}
	\caption{\label{fig:ssccontinuum} \emph{Top-panel}: Example of the continuum extrapolation of $R^{ss,C}_\sigma(E)$.\emph{Other panels}: see FIG.~\ref{fig:llccontinuum}.}
\end{figure}
\begin{figure}[t!]
	\includegraphics[width=\columnwidth]{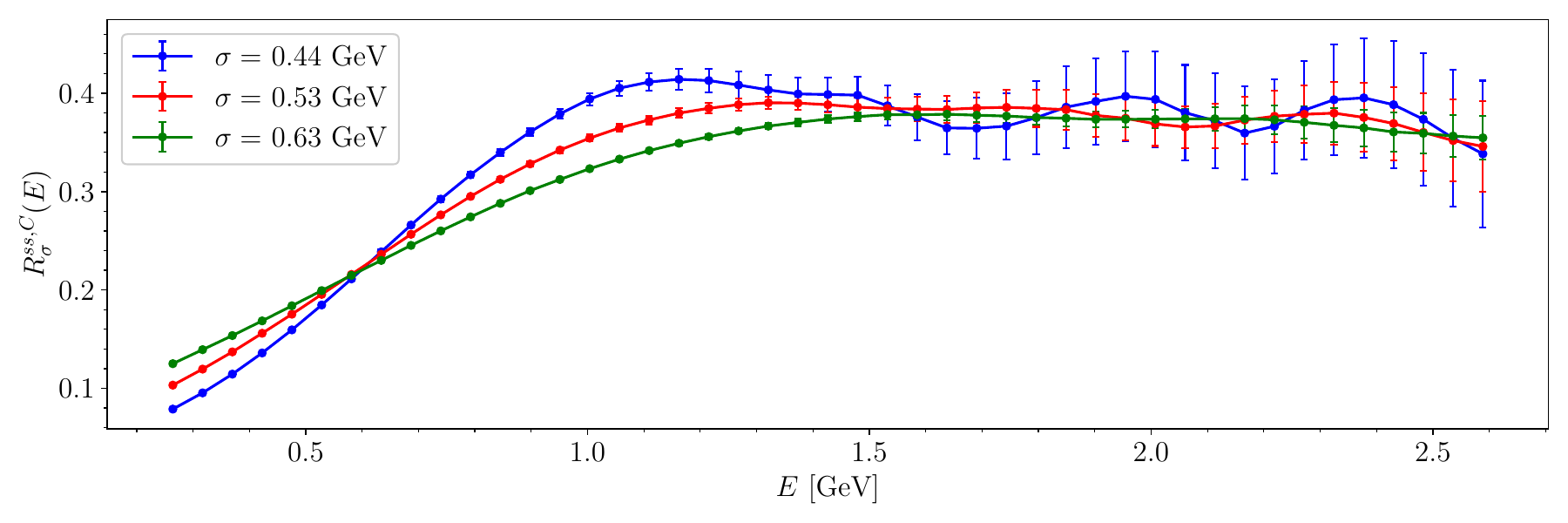}
	\caption{\label{fig:ssfinal} The figure show our final results for $R^{ss,C}_\sigma(E)$. }
\end{figure}
\emph{Continuum extrapolations.}  
All the results presented in this section correspond to the situation where the physical mass interpolation follows the application of the spectral reconstruction algorithm.
As in the light-light case, we perform both constrained and unconstrained linear continuum extrapolations  in $a^2$ and by their difference we estimate the systematics induced by this step of the analysis. The top-panel of FIG.~\ref{fig:ssccontinuum} shows an example of such continuum extrapolations, corresponding to $\sigma_2$ and $E=1.37$~GeV. At the coarsest lattice spacing we again include both the B64 and B96 ensembles. Despite the smaller errors w.r.t. the light-light case, the other panels in FIG.~\ref{fig:ssccontinuum} show full compatibility between the constrained and unconstrained extrapolations for all the values of $E$ and $\sigma$ since all points are such that $\vert P_\sigma^a(E)\vert <1.2$ (see Eq.~\ref{eq:apull}).
The final results, obtained from the combined fits, are shown in FIG.~\ref{fig:ssfinal} at all values of $E$ and $\sigma$.

\subsubsection{Charm-charm connected contribution}
%
\begin{figure}[t!]
	\includegraphics[width=\columnwidth]{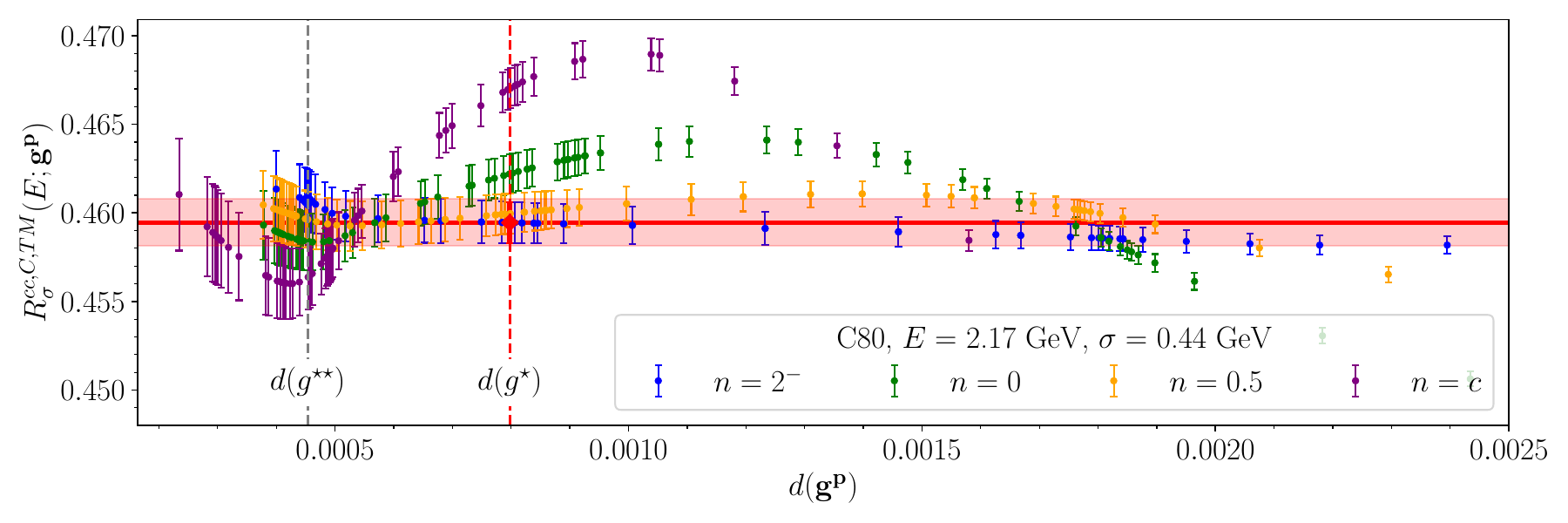}\\
	\includegraphics[width=\columnwidth]{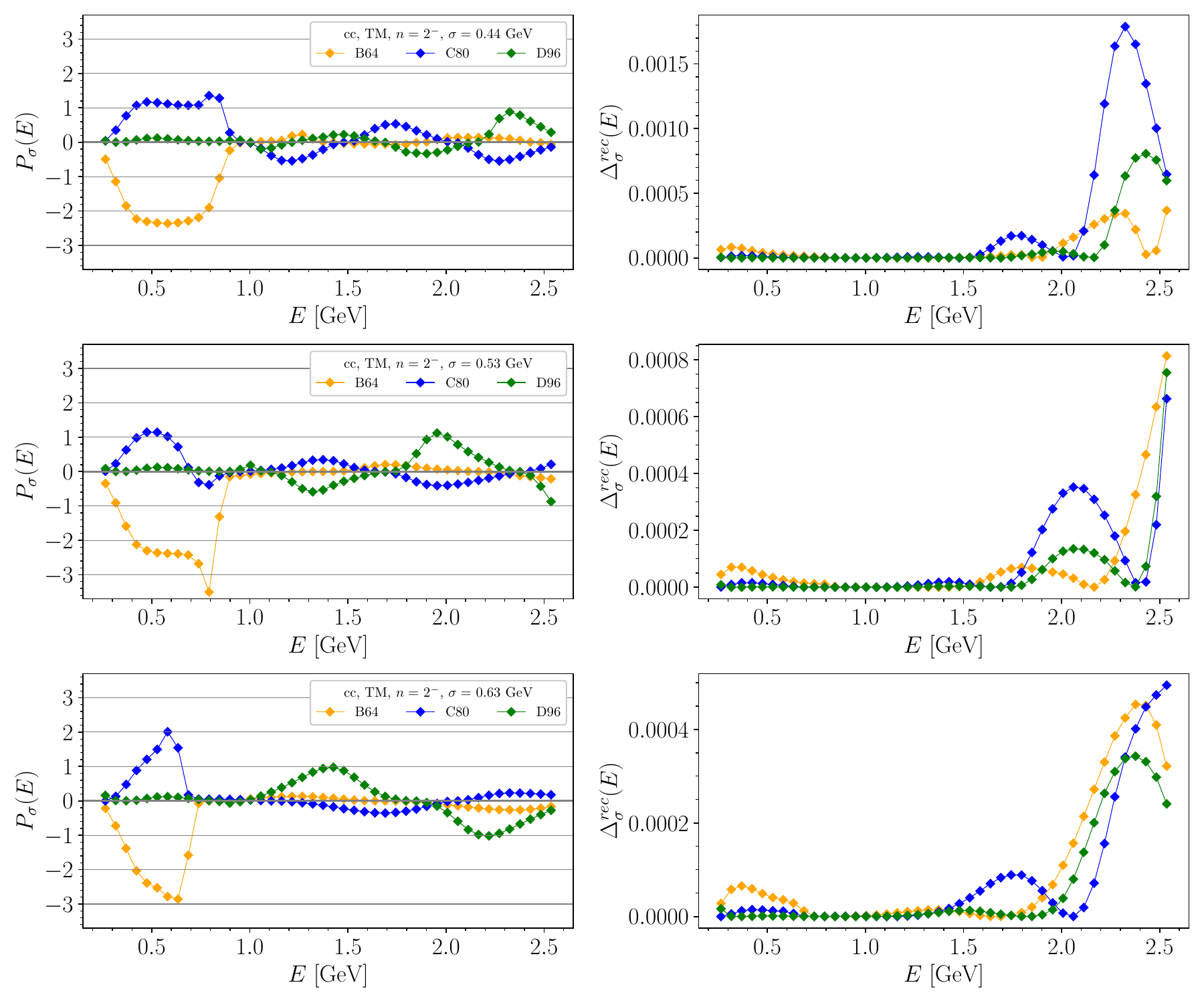}
	\caption{\label{fig:cccstability} \emph{Top-panel}: Example of the stability analysis procedure in the case of the charm-charm connected contribution to $R_\sigma(E)$. \emph{Other panels}: See FIG.~\ref{fig:llcstability}. }
\end{figure}
\begin{figure}[t!]
	\includegraphics[width=\columnwidth]{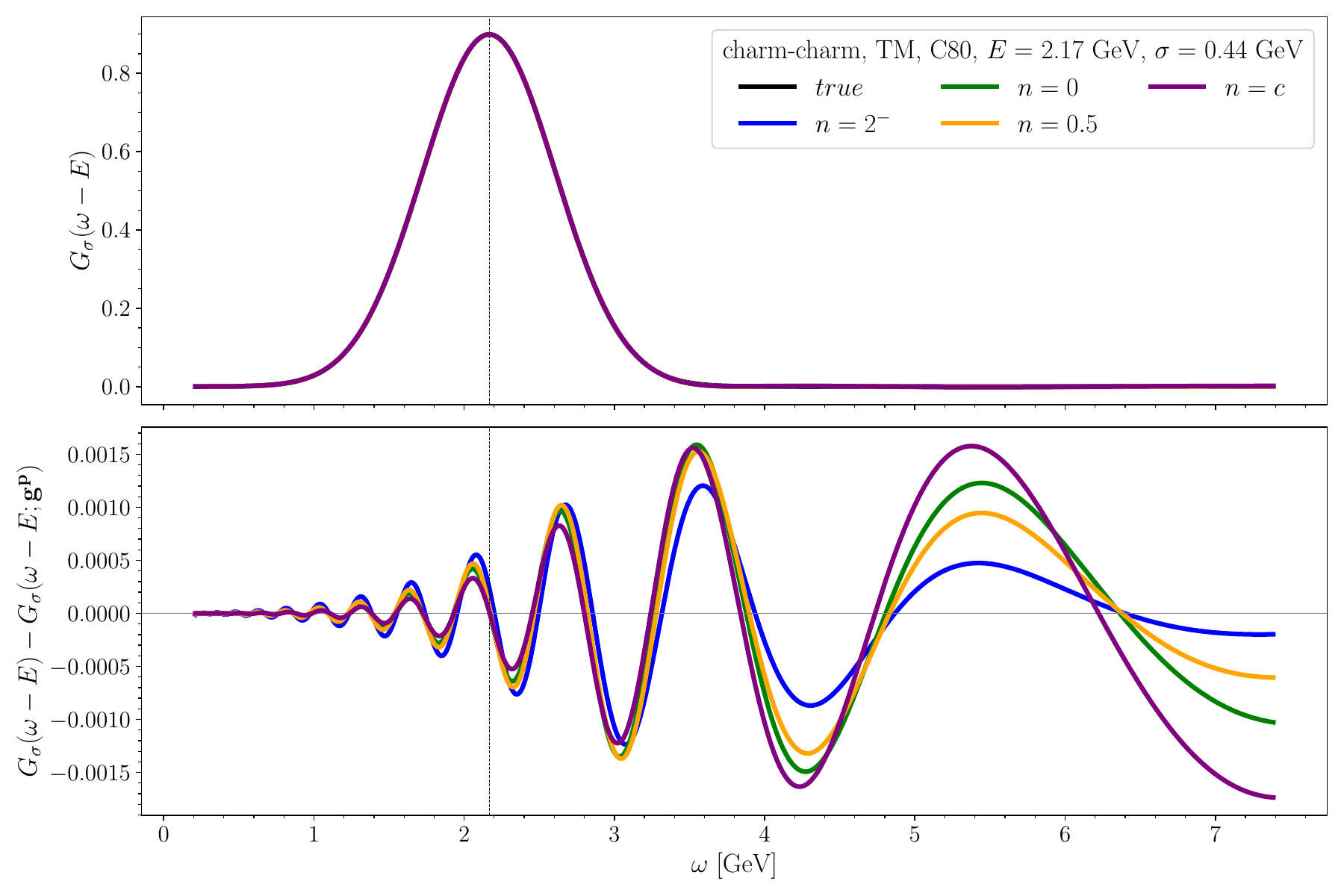}\\
	\caption{\label{fig:ccgaussians} Reconstructed kernels at $d(\vec g^{\star})$ on the C80 ensemble at $\sigma_1$ and $E=2.17$~GeV. In both plots the results are shown for $\omega>E_0$ and the vertical lines mark the location of the peak of the target Gaussian.}
\end{figure}
\begin{figure}[t!]
	\includegraphics[width=\columnwidth]{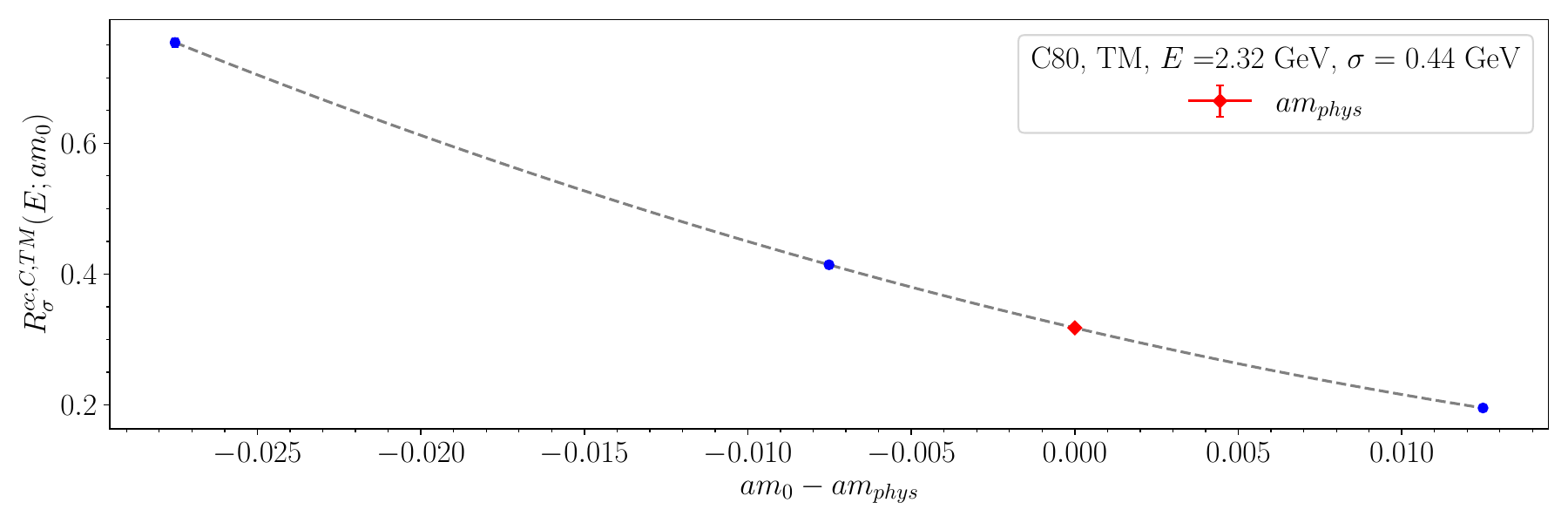}\\
	\includegraphics[width=\columnwidth]{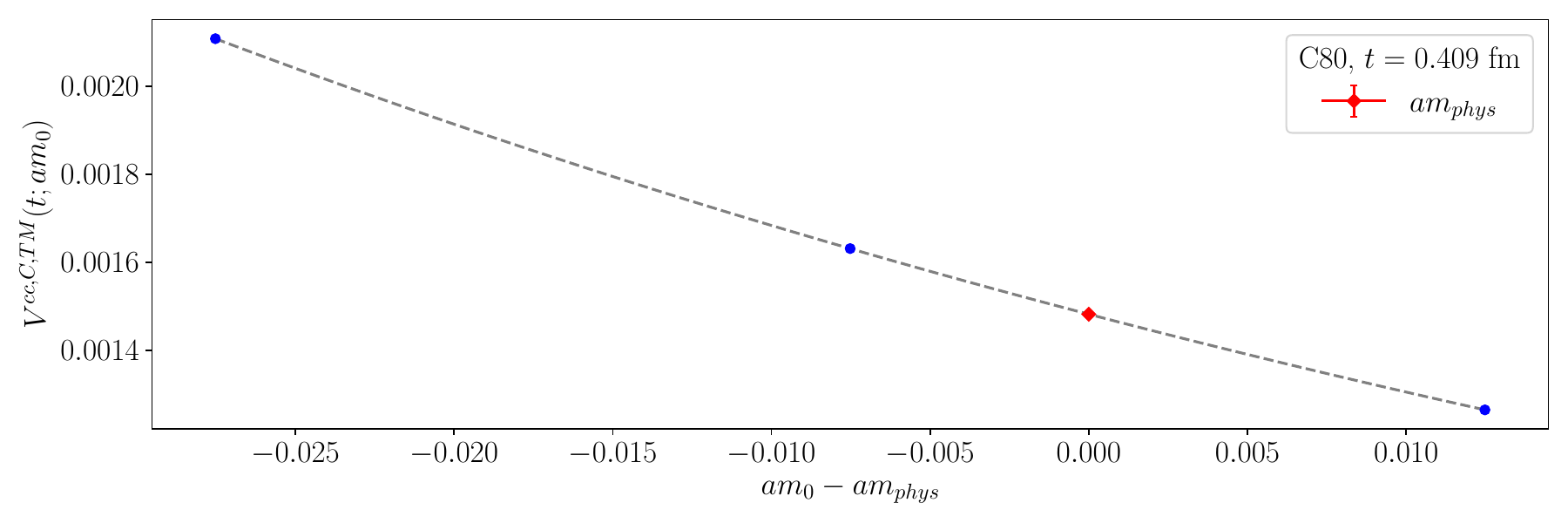}\\
	\includegraphics[width=\columnwidth]{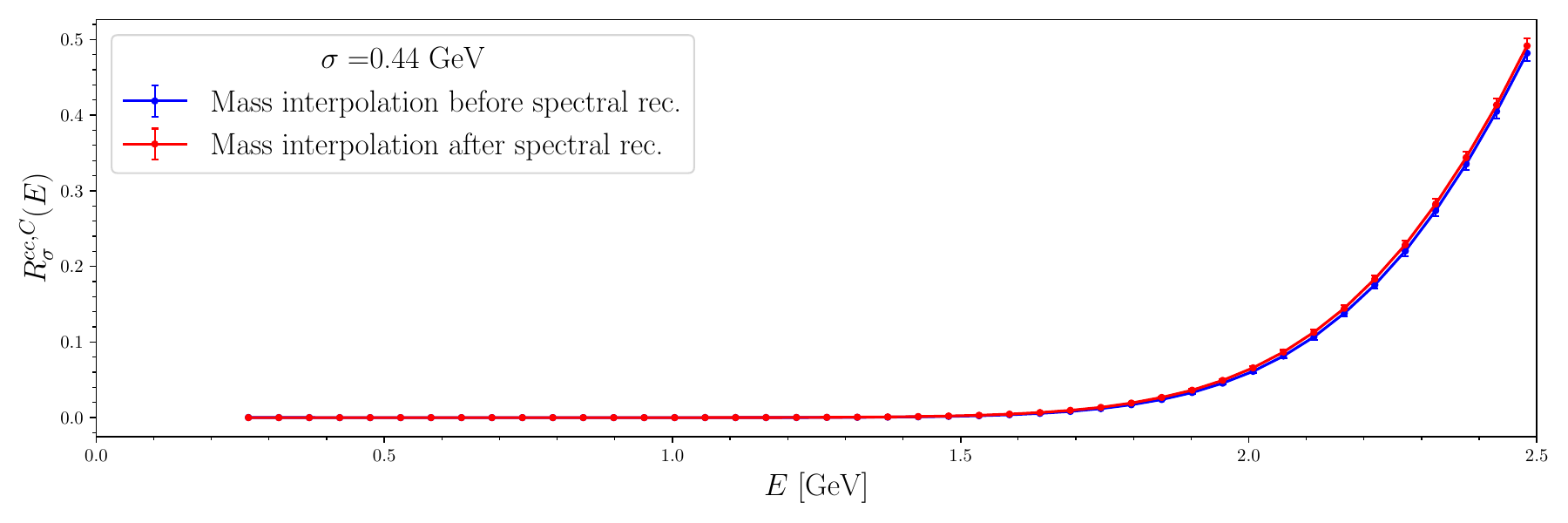}\\
	\includegraphics[width=\columnwidth]{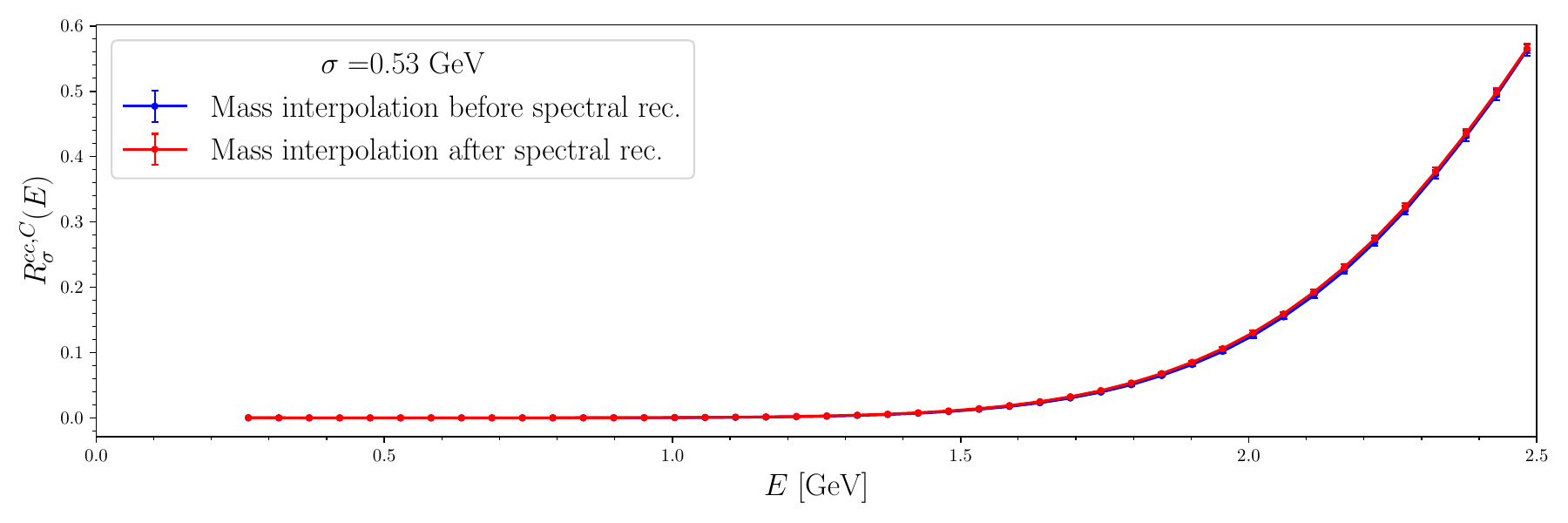}\\
	\includegraphics[width=\columnwidth]{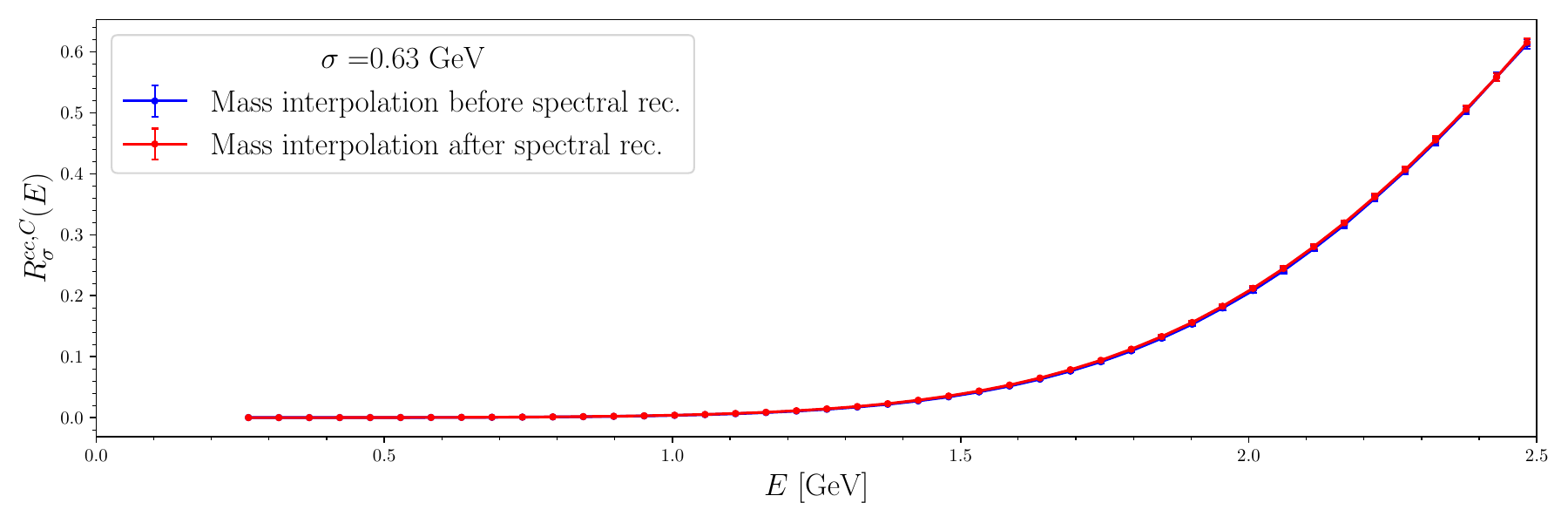}
	\caption{\label{fig:cccmass} \emph{Top-panel}: Example of the interpolation at the physical mass value in the case of $R_\sigma^{cc,C,\mathrm{TM}}(E;am_0)$. \emph{Second panel}: Example of the interpolation at the physical mass value for the correlator $V^{cc,C,\mathrm{TM}}(t;am_0)$. \emph{Other panels}: see FIG.~\ref{fig:sscmass}. }
\end{figure}
\begin{figure}[t!]
	\includegraphics[width=\columnwidth]{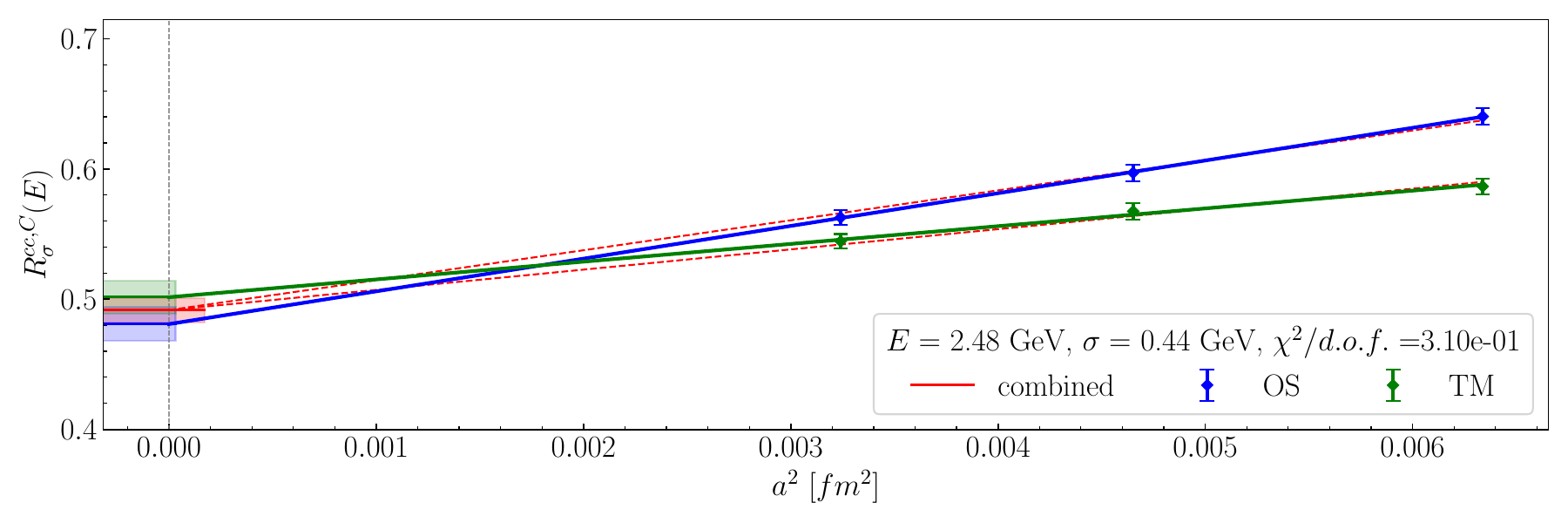}\\
	\includegraphics[width=\columnwidth]{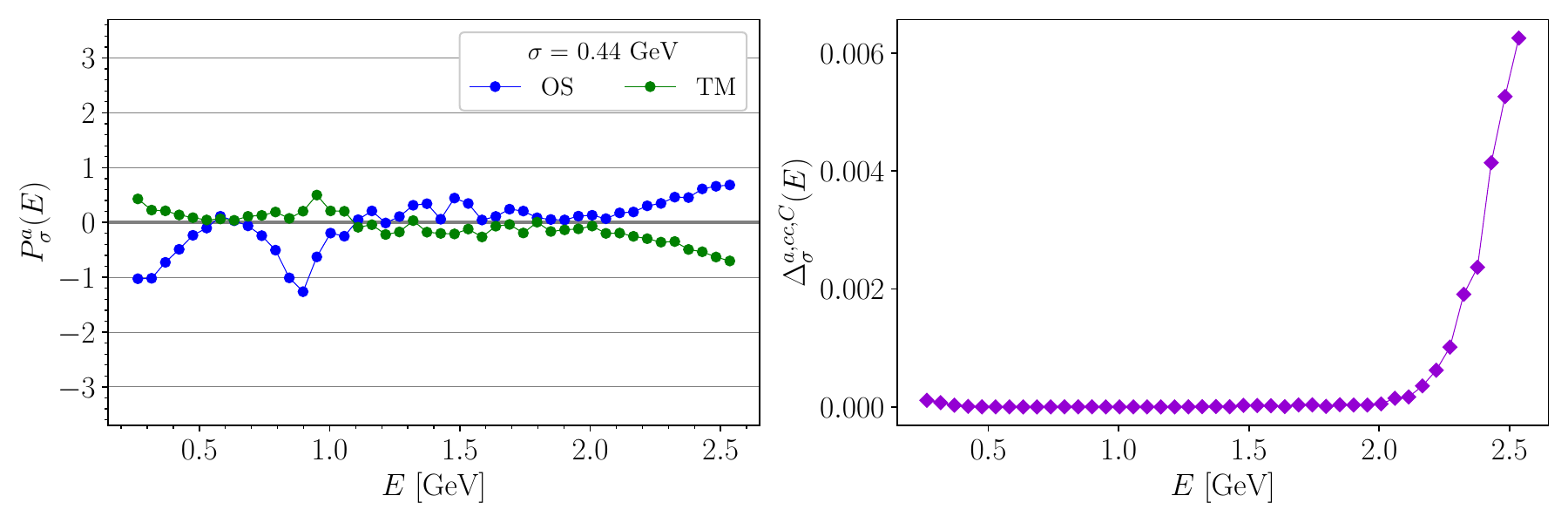}\\
	\includegraphics[width=\columnwidth]{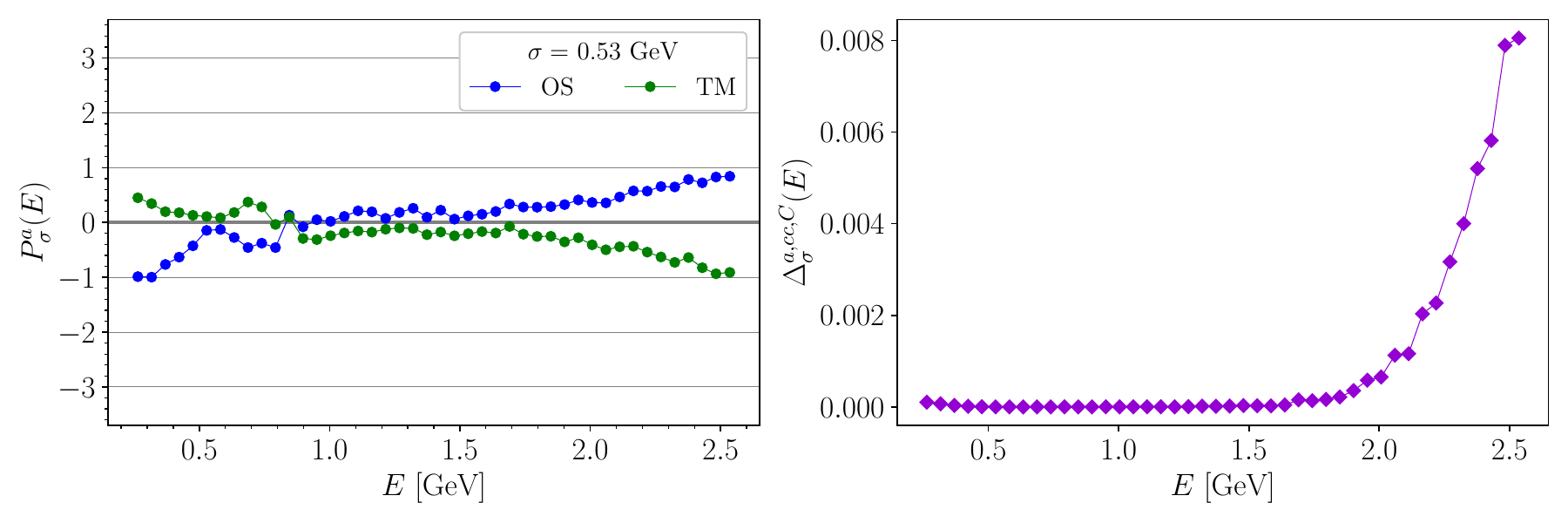}\\
	\includegraphics[width=\columnwidth]{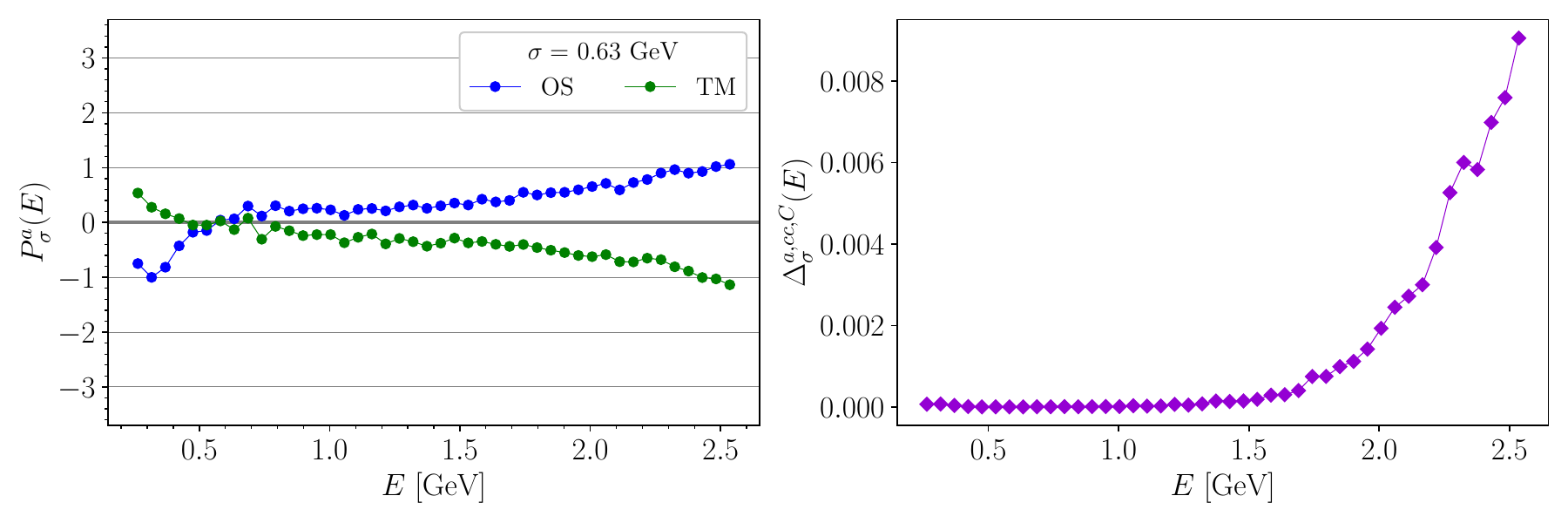}
	\caption{\label{fig:ccccontinuum} \emph{Top-panel}: Example of the continuum extrapolation of $R^{cc,C}_\sigma(E)$.\emph{Other panels}: see FIG.~\ref{fig:llccontinuum}.}
\end{figure}
The following discussion of the stability analysis and of the continuum extrapolations is analogous to the ones presented in the light-light and strange-strange cases. Similarly to the strange-strange contribution, the charm-charm correlator has been computed for different close-to-physical bare masses and the interpolation at the physical charm mass follows the same procedure used in the strange-strange case. Since no sizeable finite-volume effects are expected at high energy, and given the fact that we don't observe significant finite volume effects even in the light-light case, the charm-charm contribution has not been computed on the B96 ensemble. 

\emph{Stability analysis.} The top-panel of FIG.~\ref{fig:cccstability} shows an example of the stability analysis procedure for  $R^{cc,C,\mathrm{TM}}_\sigma(E)$ at $E=2.17$ GeV and $\sigma_1$ in the case of the C80 ensemble. The behaviour w.r.t. $d(\vec g^{\vec p})$ is the same observed for the other contributions, with the choice n$=2^-$ (blue points) providing the best stability among all the weighting functions considered in this work.
 
The other panels of FIG.~\ref{fig:cccstability} show the quantity $P_\sigma(E)$ for $\sigma_1$, $\sigma_2$ and $\sigma_3$ gathering together the three ensembles B64, C80 and D96. Unlike the light-light and strange-strange contributions, the region below 1 GeV in the B64  ensemble is systematics dominated. This is due to the facts that the charm-charm correlators are very precise and that $R_\sigma^{cc}(E)$ vanishes in this region in the $\sigma\mapsto 0$ limit. The first relevant charmonium state is the $J/\Psi$ resonance peaked around $3$~GeV and $R_\sigma^{cc,C}(E)$ is different from zero below this threshold only because of the sensitivity of the smeared kernels to the high energy region. In fact, for $E< 1.5$~GeV the contribution of $R_\sigma^{cc}(E)$ to $R_\sigma(E)$ is negligible w.r.t. the dominant light-light contribution (see FIG.~\ref{fig:ccfinal} below). Above $1$~GeV almost all the points are such that $\vert P_\sigma(E)\vert <1$. 

An example of the kernel reconstruction at  $E=2.17$ GeV and  $\sigma_1$ is given in  FIG.~\ref{fig:ccgaussians} for the different weighting functions. Although the reconstruction is expected to become more challenging at high energy and small $\sigma$, the difference between the target and  reconstructed kernels plotted in the bottom panel of the same figure shows that the reconstruction is excellent also in this case. Such a result is, again, due to the very good precision of the charm-charm correlator allowing to reach high energy at small $\sigma$ with an overall small error on the spectral reconstruction of $R_\sigma^{cc}(E)$.

\emph{Physical mass interpolation}
The charm-charm connected correlator has been computed for two values of bare masses on the D96 ensemble and three values of bare masses on the C80 and B64 ensembles. As in the strange-strange case, we followed both the strategy in which we interpolate the spectral reconstructed results and the one in which we interpolate the correlators. In the case of the ensemble D96 the ans\"atze used are the linear ones given in Eq.~(\ref{eq:massRlinear}) and Eq.~(\ref{eq:massVlinear}). In the other cases, where three bare masses are available, the ans\"atze are
\begin{equation}\label{eq:massRquadratic}
	R_\sigma(am_0) = A + B\, am_0 + C\, (am_0)^2
\end{equation}
and 
\begin{equation}\label{eq:massVquadratic}
	\log V(am_0) = A + B\, am_0 + C\, (am_0)^2.
\end{equation}
The top panel of FIG.~\ref{fig:cccmass} shows an example of the interpolation of $R_\sigma^{cc,C,\mathrm{TM}}(E;am_0)$ on the C80 ensemble at $E=2.32$ GeV and $\sigma_1$.  An example of interpolation at the physical mass of $V^{cc,C,\mathrm{TM}}(t;am_0)$ at $t=0.409$ fm is given in the second panel of FIG.~\ref{fig:cccmass} on the C80 ensemble. The comparison of the final continuum extrapolated results obtained by following the two procedures is in the other panels of FIG.~\ref{fig:cccmass} for $\sigma_1$, $\sigma_2$ and $\sigma_3$. The results are perfectly in agreement  within the errors at all energies and for all values of $\sigma$.

\emph{Continuum extrapolations.} 
All the results presented in this section correspond to the situation where the physical mass interpolation follows the application of the spectral reconstruction algorithm. The charm-charm connected contribution is the one expected to be affected from larger cutoff effects.  Nevertheless, the top panel of FIG.~\ref{fig:ccccontinuum} shows the continuum extrapolations, both constrained and unconstrained, for $\sigma_1$ and $E=2.48$GeV (a value very close to the largest energy we consider in this work) and, as it can be seen, these are perfectly under control.
The constrained and unconstrained fits are compatible in all cases, as summarized in the three bottom panels of FIG.~\ref{fig:ccccontinuum} where the quantity \(P_\sigma^a(E)\) defined in Eq.~(\ref{eq:apull}) is plotted together with our estimates of $\Delta^{a,cc,C}_\sigma(E)$. According to this analysis we therefore consider as central values and errors on $R_\sigma^{cc,C}(E)$ the ones obtained by the constrained linear fits and  the final results are shown in the top-panel of FIG.~\ref{fig:ccfinal} at all values of $E$ and $\sigma$. The precision of these results is remarkably good and, as expected, $R_\sigma^{cc,C}(E)$ represents a negligible contribution to the total \(R_\sigma(E)\) in the low energy regime (below $\sim 1.5$~GeV) at the values of \(\sigma\) considered here, see bottom-panel of FIG.~\ref{fig:ccfinal}.
\begin{figure}[h!]
	\includegraphics[width=\columnwidth]{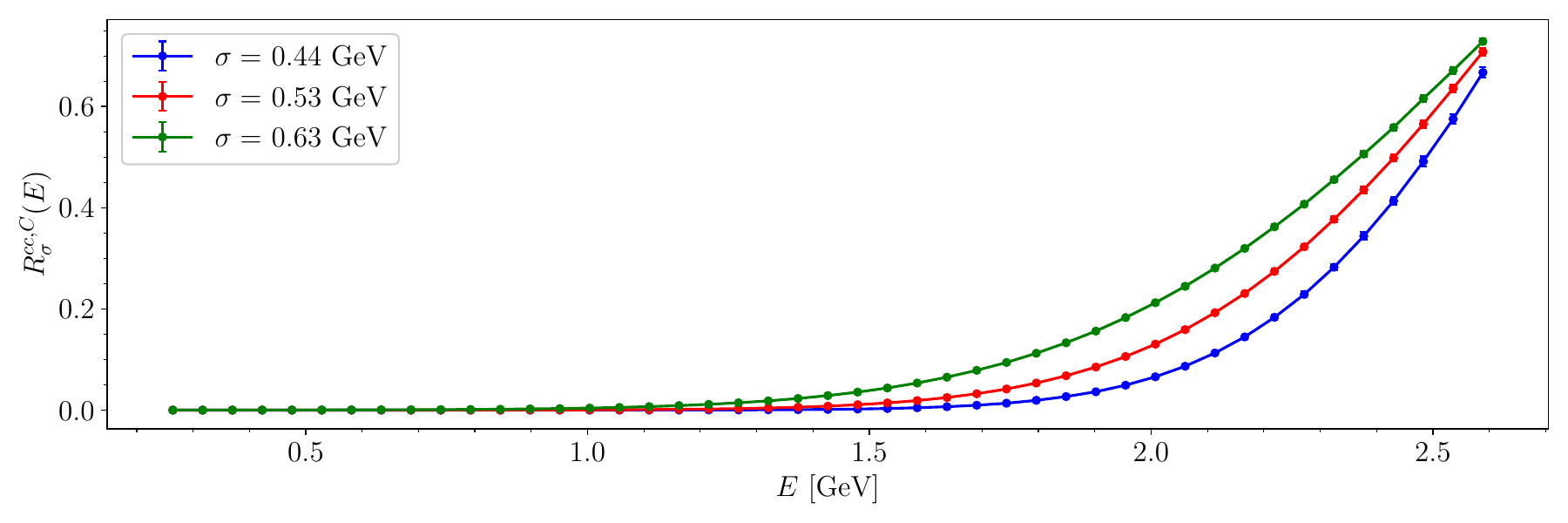}\\
	\includegraphics[width=\columnwidth]{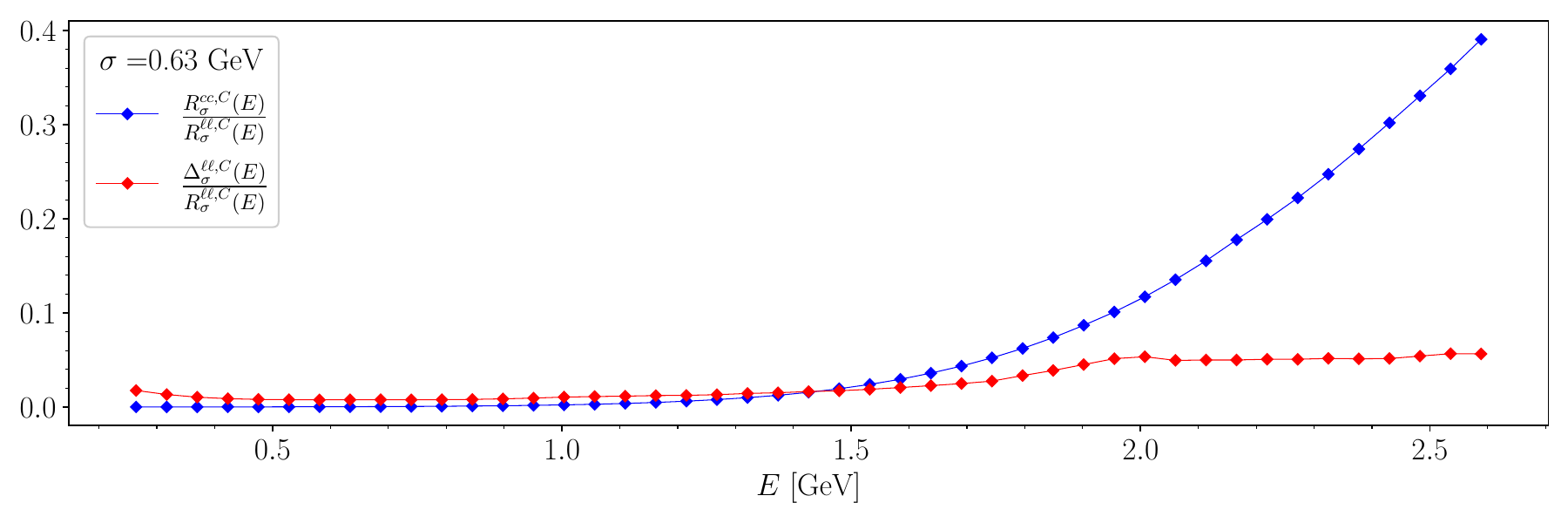}	
	\caption{\label{fig:ccfinal} \emph{Top-panel}: final results for $R^{cc,C}_\sigma(E)$. \emph{Bottom-panel}: comparison of $R_\sigma^{cc,C}(E)/R_\sigma^{\ell\ell,C}(E)$ (blue points) with the relative error of the light-light contribution, i.e. $\bar{\Delta}^{\ell\ell,C}_\sigma(E)/R_\sigma^{\ell\ell,C}(E)$, at $\sigma_3$ (red points). As it can be seen, $R_\sigma^{cc,C}(E)$ is negligible with respect to the dominant light-light contribution $R_\sigma^{\ell\ell,C}(E)$ for $E<1.5$~GeV. The same happens at $\sigma_1$ and $\sigma_2$.}
\end{figure}
%

\subsubsection{Disconnected contributions}
\begin{figure}[t!]
	\includegraphics[width=\columnwidth]{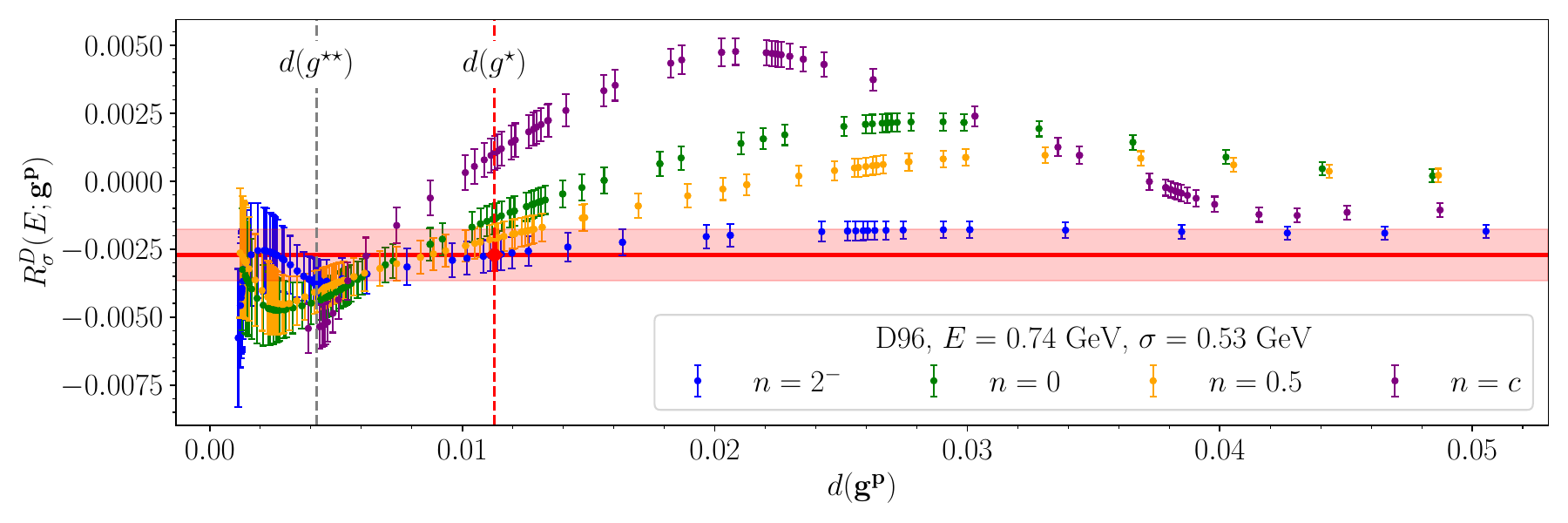}\\
	\includegraphics[width=\columnwidth]{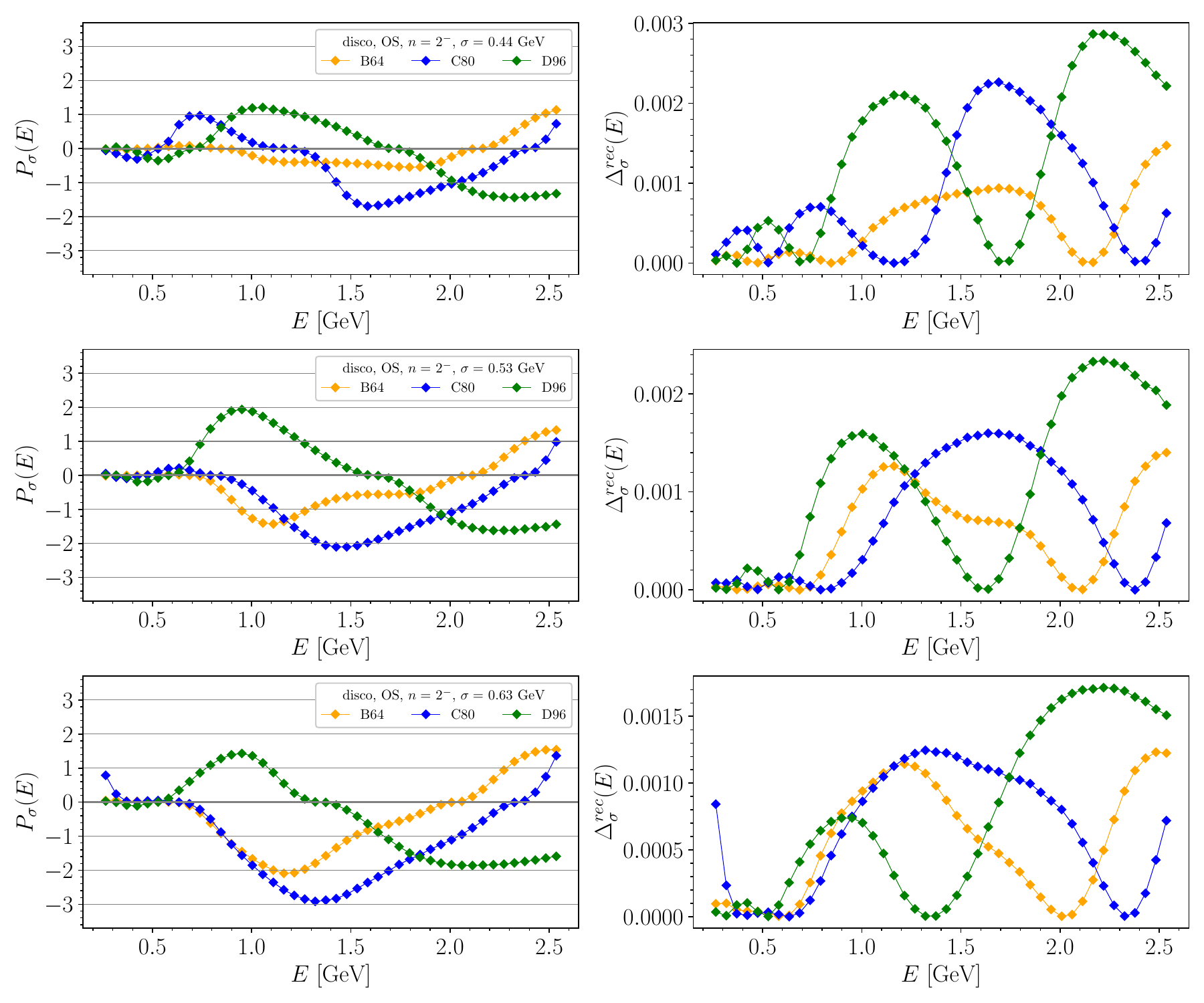}
	\caption{\label{fig:discocstability} \emph{Top-panel}: Example of the stability analysis procedure in the case of the disconnected contribution to $R_\sigma(E)$. \emph{Other panels}: See FIG.~\ref{fig:llcstability}. }
\end{figure}
\begin{figure}[t!]
	\includegraphics[width=\columnwidth]{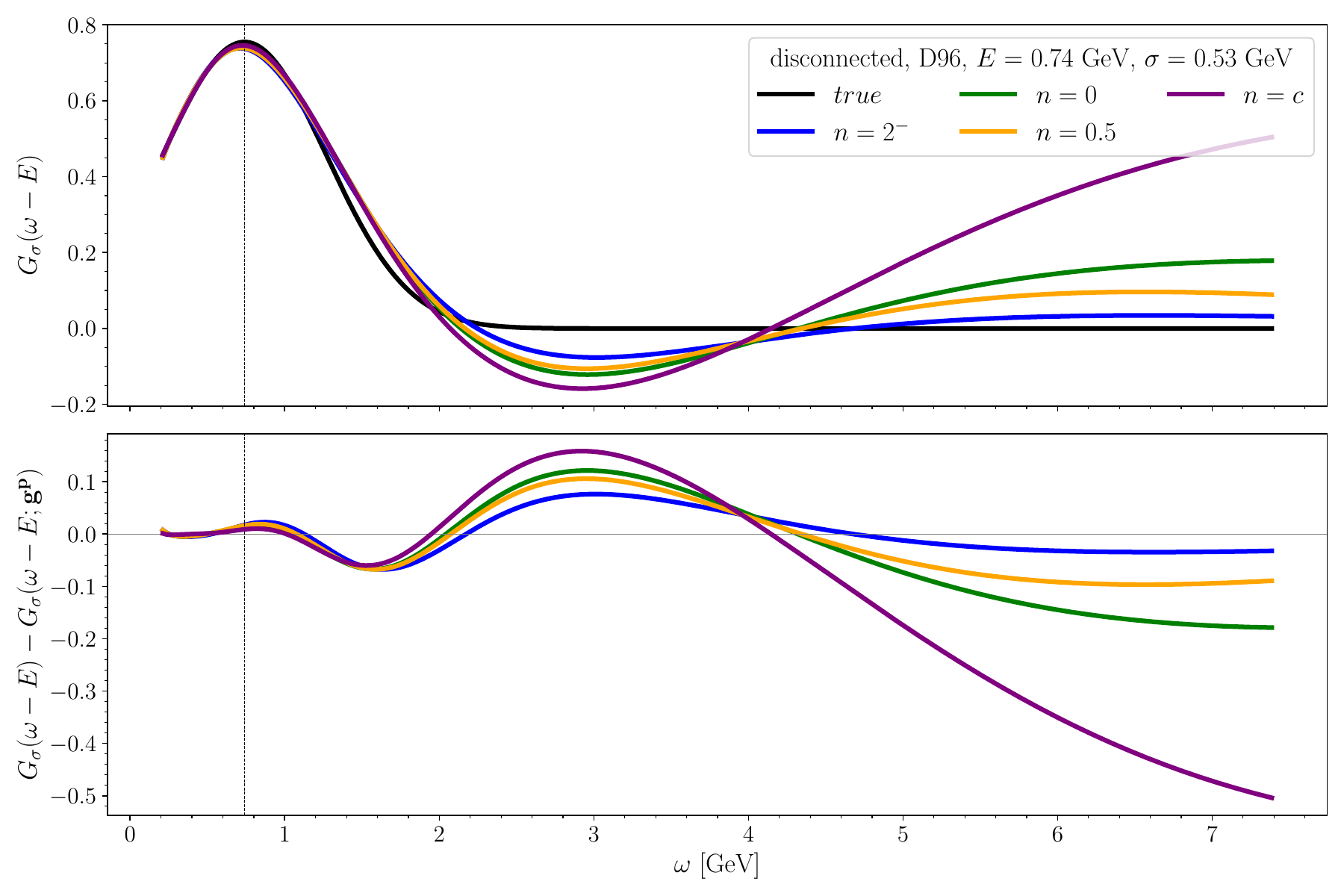}\\
	\caption{\label{fig:discogaussians} Reconstructed kernels at $d(\vec g^{\star})$ on the D96 ensemble at $\sigma_2$ and $E=0.74$~GeV. In both plots the results are shown for $\omega>E_0$ and the vertical lines mark the location of the peak of the target Gaussian.}
\end{figure}
\begin{figure}[t!]
	\includegraphics[width=\columnwidth]{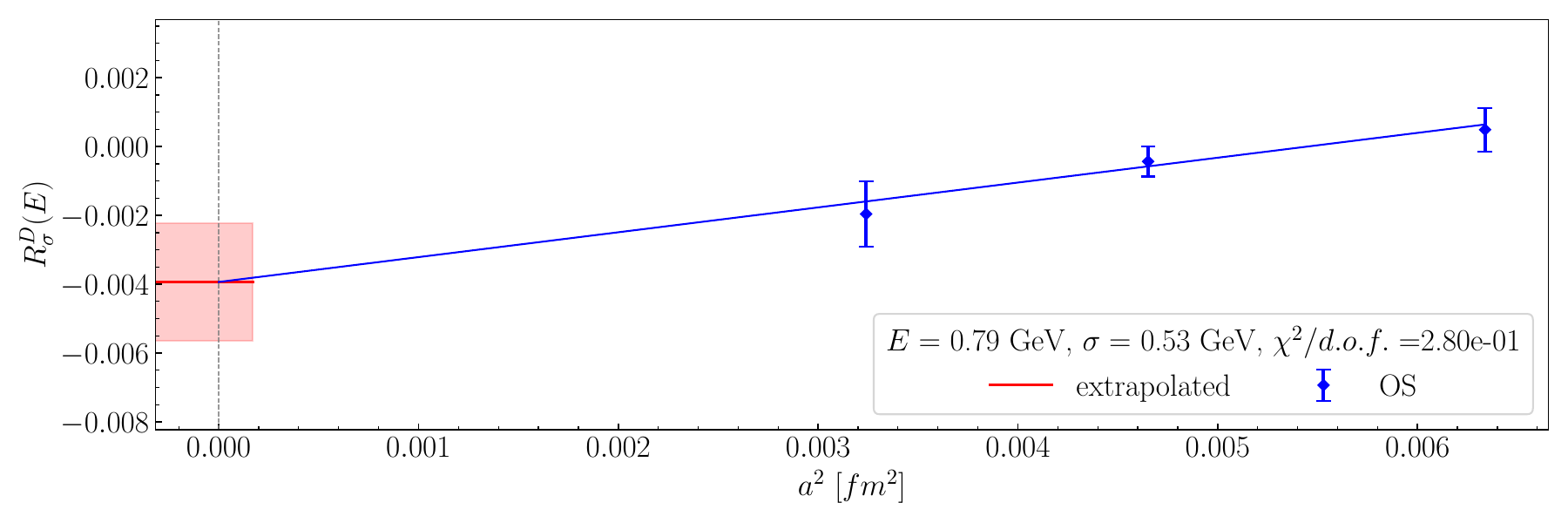}\\
	\includegraphics[width=\columnwidth]{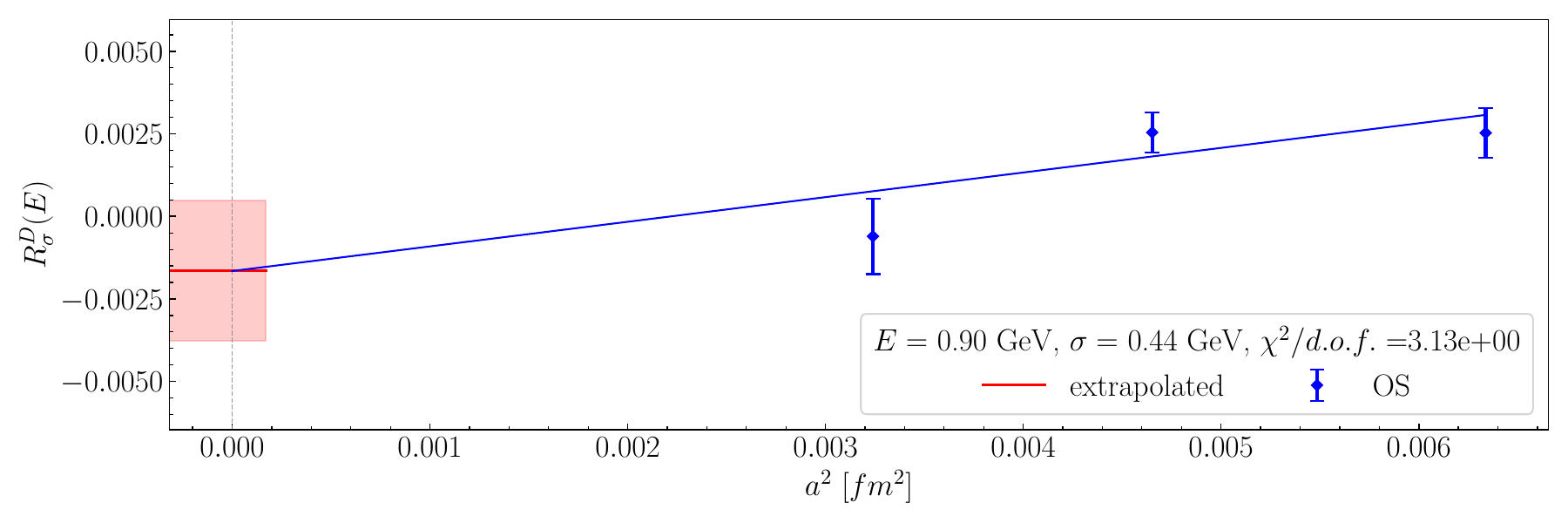}\\
	\includegraphics[width=\columnwidth]{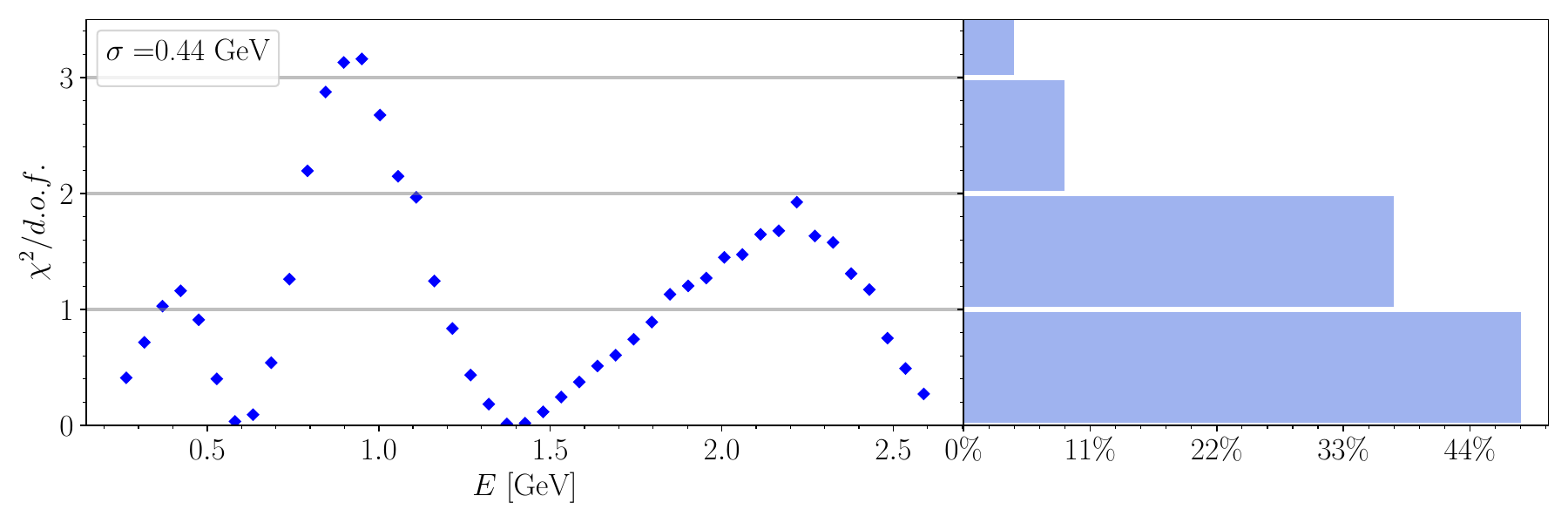}\\
	\includegraphics[width=\columnwidth]{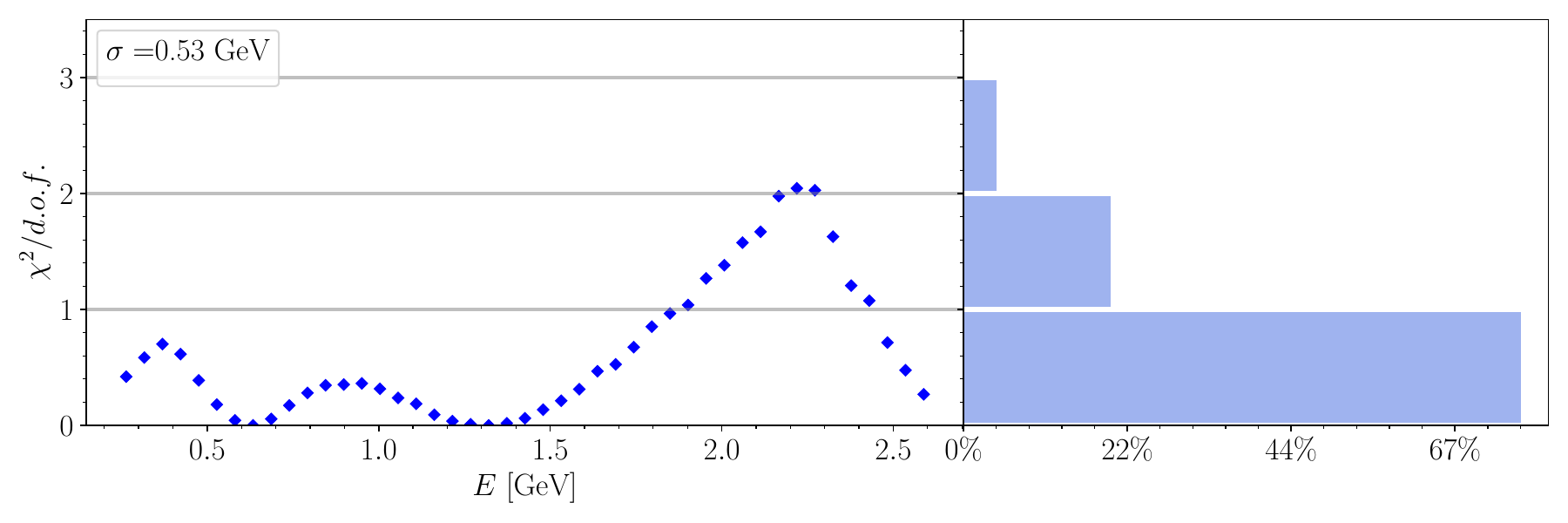}\\
	\includegraphics[width=\columnwidth]{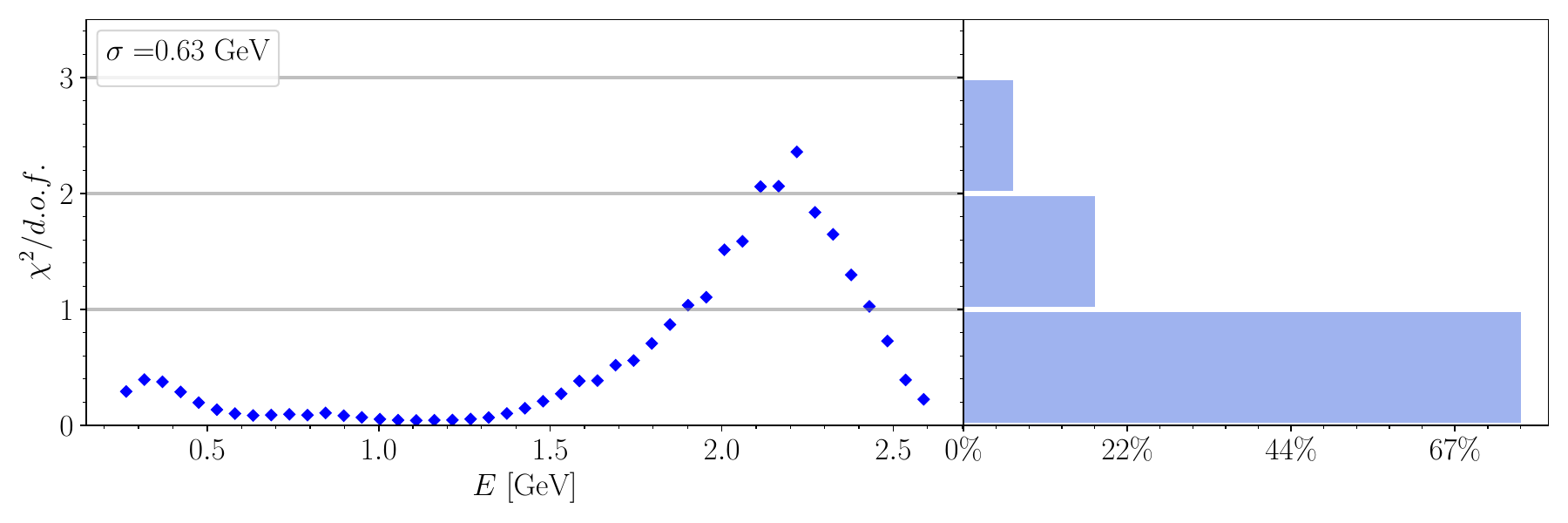}	
	\caption{\label{fig:discoccontinuum} \emph{First-panel}: Example of the continuum extrapolation of $R^{D}_\sigma(E)$ at $E=$0.79 GeV for $\sigma_2$.  \emph{Second-panel}: Example of the continuum extrapolation of $R^{D}_\sigma(E)$ at $E=$0.90 GeV for $\sigma_1$. 		\emph{Other panels}: reduced $\chi^2$ as a function of the energy for $\sigma_1$, $\sigma_2$ and $\sigma_3$. The histograms on the right give the percentage of points appearing in the corresponding left-plot with reduced $\chi^2$ in the intervals $[0,1)$, $[1,2)$, $[2,3)$ and $[3,\infty)$. }
\end{figure}
\begin{figure}[t!]
	\includegraphics[width=\columnwidth]{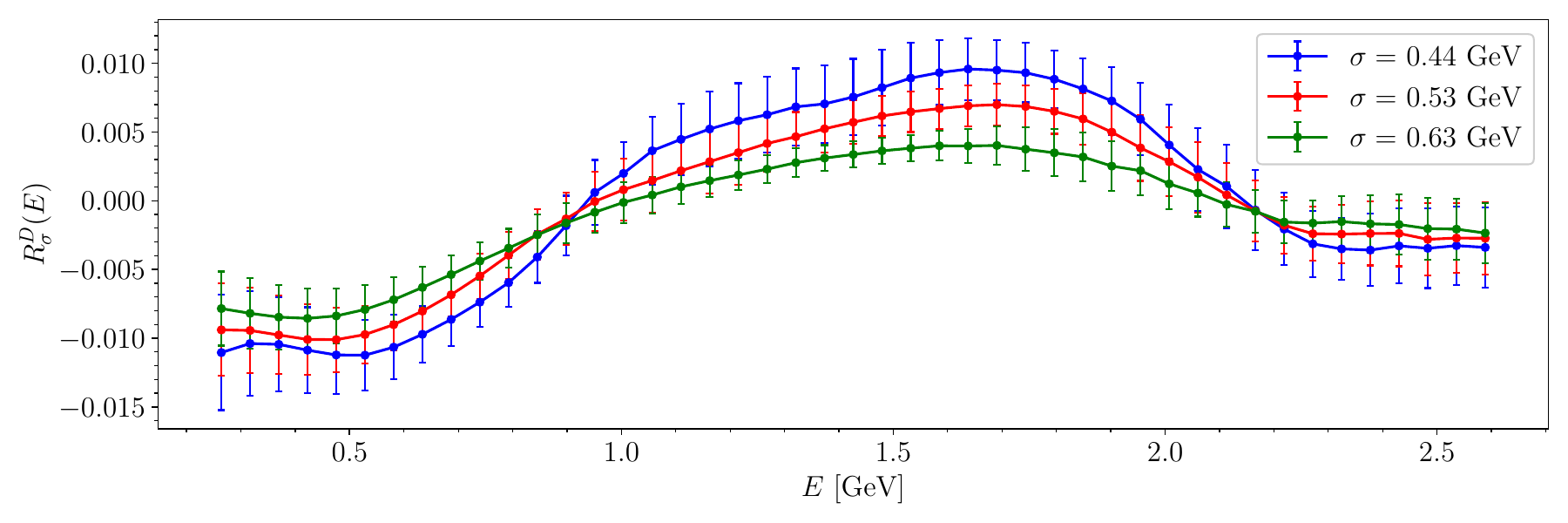}\\
	\includegraphics[width=\columnwidth]{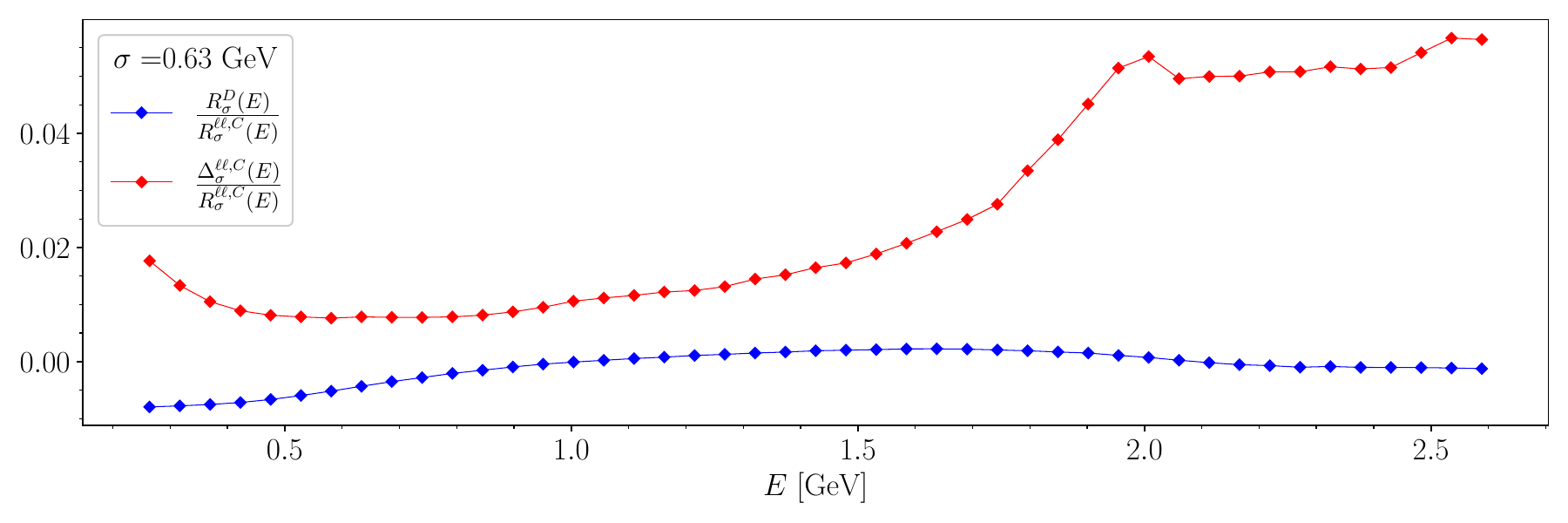}
	\caption{\label{fig:discofinal} \emph{Top-panel}: final results for $R^{D}_\sigma(E)$. \emph{Bottom-panel}: comparison of $R_\sigma^{D}(E)/R_\sigma^{\ell\ell,C}(E)$ (blue points) with the relative error of the light-light contribution, i.e. $\bar{\Delta}^{\ell\ell,C}_\sigma(E)/R_\sigma^{\ell\ell,C}(E)$, at $\sigma_3$ (red points). As it can be seen, $R_\sigma^{D}(E)$ is negligible with respect to the dominant light-light contribution $R_\sigma^{\ell\ell,C}(E)$ at all quoted energies. The same happens at $\sigma_1$ and $\sigma_2$.}
\end{figure}
 \begin{figure}[t!]
	\includegraphics[width=\columnwidth]{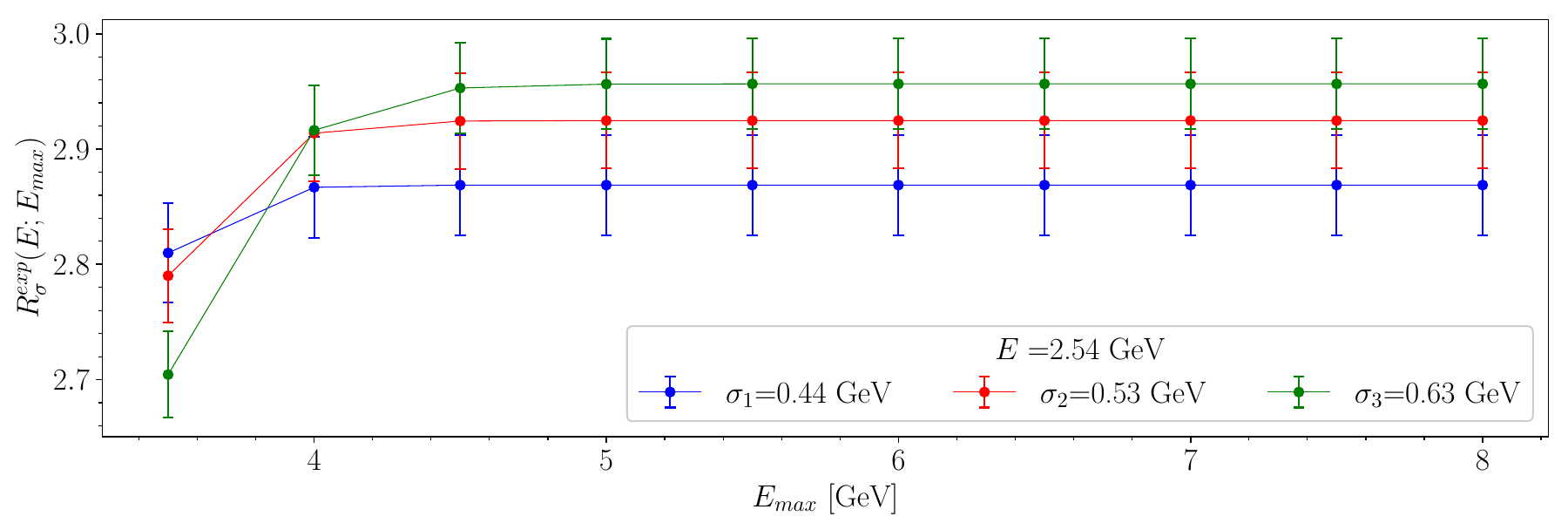}\\
	\includegraphics[width=\columnwidth]{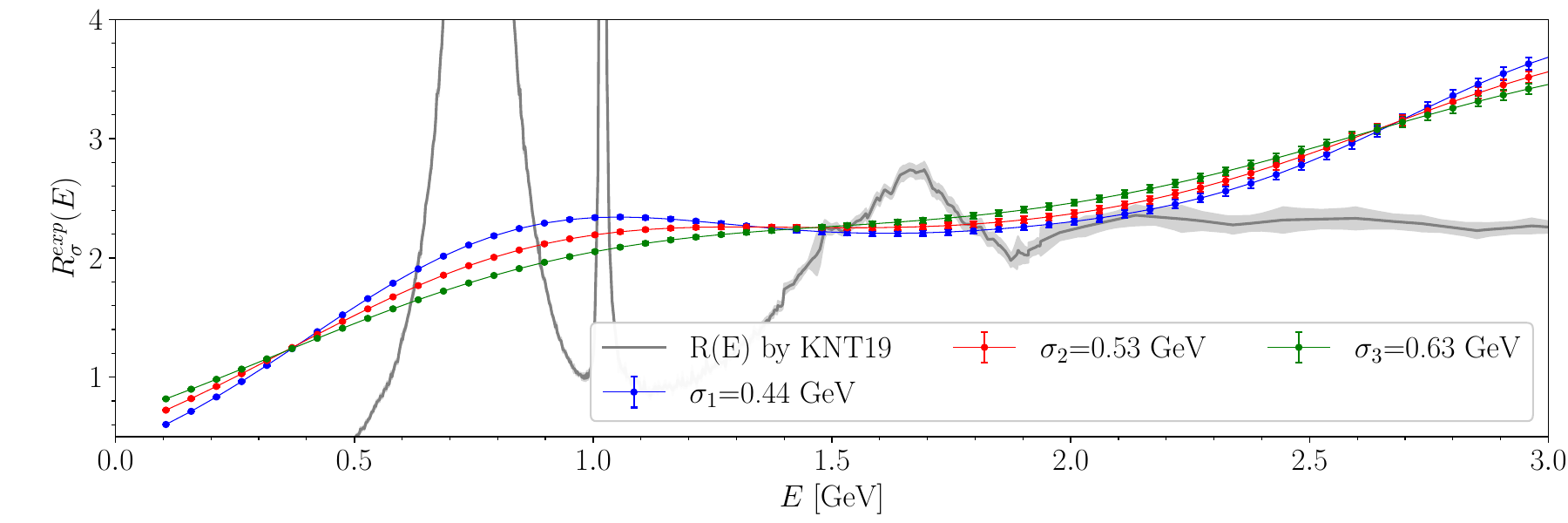}
	\caption{\label{fig:smeared} \emph{Top-panel}: Dependence on $E_\text{max}$ of $R_\sigma^\text{exp}(E;E_\text{max})$ at $E=2.54$ GeV for $\sigma_1$, $\sigma_2$ and $\sigma_3$. \emph{Bottom panels}: Final results with errors of the experimental $R$-ratio smeared with $\sigma_1$, $\sigma_2$ and $\sigma_3$.  }
\end{figure}

In this subsection we discuss the analysis of the disconnected contributions to $R_\sigma(E)$. These include the light-light, light-strange, light-charm, strange-strange, strange-charm and charm-charm fermionic disconnected Wick contractions that have been computed in the OS regularization (we therefore omit in the following the regularization tag) and linearly combined with the corresponding electric charge factors to build a single ``disconnected'' correlator. 

In Eq.~(\ref{eq:relative_normalization}) we defined the relative normalization between the norm and error functionals. We pointed out that this definition can always be reabsorbed in the unphysical parameter $\lambda$ and that, therefore, it can be arbitrarily changed without altering the physical meaning of the reconstructed spectral density. While the choice $\tau_\mathrm{norm}=1$ turned out to be particularly convenient for the connected contributions, it is not  suitable for the disconnected ones since $V^D(a) \ll V^C(a)$ by several orders of magnitude. We found that the effectiveness of the conditions of Eqs.~(\ref{eq:stars}) remains valid with the choice $\tau_\mathrm{norm}=0$ in the disconnected case. 

\emph{Stability analysis.}
The top panel of FIG.~\ref{fig:discocstability} shows an example of the stability analysis procedure for $R_\sigma^{D}(E)$ at $E=$0.74 GeV and $\sigma_2$ in the case of the D96 ensemble. Among the different weighting functions, the choice n$=2^{-}$ is still the most stable.  A summary of the quantity $P_\sigma(E)$ (see Eq.~(\ref{eq:syspull})) for $\sigma_1$, $\sigma_2$ and $\sigma_3$ is shown in the other panels of the same figure for the three ensembles. Almost all the points are such that $\vert P_\sigma(E)\vert <2$, that is, are in the statistically dominated regime. This is a solid evidence that the change of the relative normalization between the functionals is effective in the disconnected case. Few points at $\sigma_3$ on the C80 ensemble are in the systematically dominated regime, $\vert P_\sigma(E)\vert>2$. This can be explained by the fact that the C80 correlator is more precise and, at the same time, the spectral reconstruction gives smaller systematic errors at larger smearing radius. Nevertheless, even for the (few) points that are in the systematics dominated regime the procedure of Eq.~(\ref{eq:syserror}) gives a conservative estimate of the error. 

The top panel of FIG.~\ref{fig:discogaussians} shows an example of kernel reconstruction at $E=0.74$ GeV, $\sigma_2$ on the D96 ensemble. The kernel reconstruction is worse w.r.t. to the examples given in the previous subsections. Obtaining a very accurate result in the disconnected case is a particularly challenging task due to the fact that the signal-to-noise ratio of the disconnected correlator becomes very large after a handful of time-slices. On the one hand, in order to improve the situation, substantially more precise correlators are needed. On the other hand, the disconnected contribution is definitely negligible w.r.t. the dominant light-light contribution at the current level of precision. This can be seen in the bottom-panel of FIG.~\ref{fig:discofinal}. 

\emph{Continuum extrapolation}
Since the disconnected contributions are computed in the OS regularization only, we perform unconstrained linear continuum extrapolations. Two examples of continuum extrapolations are shown in the first two panels of FIG.~\ref{fig:discoccontinuum}. Given the fact that disconnected correlators are particularly noisy, these plots are particularly remarkable. Indeed, the three points at different lattice spacings come from different simulations and the spectral reconstruction algorithm provides results that are consistent with a linear behaviour in $a^2$. To quantify the quality of all the continuum  extrapolations, the other panels of the same figure show the reduced $\chi^2/d.o.f.$ as function of the energy for the three values of $\sigma$. Although in some of the cases $\chi^2/d.o.f.\sim 3$ (the worst case is shown in the second-panel plot) this has no significant effect on our final result for $R_\sigma(E)$. This can be understood by looking at the bottom-panel of FIG.~\ref{fig:discofinal} where we show, at $\sigma_3$, the comparison of the ratio $R_\sigma^D(E)/R_\sigma^{\ell\ell,C}(E)$ with the relative error of the dominant light-light contribution. As it can be seen, the disconnected contribution is negligible at all energies at the current level of precision. The same happens at the other values of $\sigma$ considered in this work. The final results for $R_\sigma^D(E)$ are shown in the top-panel of FIG.~\ref{fig:discofinal}.

\subsection{Experimental $R$-ratio}
In the main text we have compared our final results for $R_\sigma(E)$ with their experimental counterpart given by
\begin{flalign}
&R_\sigma^\text{exp}(E)=\lim_{E_\text{max}\mapsto \infty}R_\sigma^\text{exp}(E;E_\text{max})\;,
\nonumber \\
&R_\sigma^\text{exp}(E;E_\text{max}) = \int_{0}^{E_\text{max}} d\omega\, G_\sigma(E-\omega)R^\text{exp}(\omega),
\label{eq:exp_int}
\end{flalign}
where $G_\sigma(E-\omega)$ is the normalized Gaussian centred at $E$ and with width $\sigma$. 

For $R^\text{exp}(E)$ we used the KNT19 ~\citeSM{SMkeshavarzi2020g} compilation courteously provided by the authors. In this compilation the energy ranges from $0.216$~GeV to $11.2$~GeV and the central values of $R^\text{exp}(E)$ are provided together with the full covariance matrix that takes into account the correlations between different measurements.

In order to obtain $R_\sigma^\text{exp}(E)$ and estimate the associated errors, we generated 2000 bootstrap samples of $R^\text{exp}(E)$, each of which simulating an independent measurement, from a multivariate Gaussian distribution built by using the aforementioned central values and covariance matrix. Each sample has then been integrated with $G_\sigma(E-\omega)$ at fixed values of $E_\text{max}$ by using the trapezoid rule. The final results and the associated errors have finally been obtained by taking the boostrap average and standard deviation. The procedure has been repeated for $\sigma_1$, $\sigma_2$ and $\sigma_3$ and for all the central energy values $E$ at which we have computed $R_\sigma(E)$.

Since no experimental data are available above $11.2$~GeV, in order to take the $E_\text{max}\mapsto \infty$ limit we studied the dependence of the results thus obtained w.r.t. $E_\text{max}$. In fact, given our choice of smearing kernels, contributions to $R_\sigma^\text{exp}(E;\infty)$ coming from values of $\vert \omega-E\vert> 5\sigma$ are totally suppressed. The top-panel of FIG.~\ref{fig:smeared} shows how $R_\sigma^\text{exp}(E)$ at $E=2.54$ GeV, the largest value of center energy considered in this work, changes upon varying $E_\text{max}$. For all considered values of $\sigma$ a plateau is already reached at $E_\text{max}\sim5$ GeV. In view of this finding, we estimated $R_\sigma^\text{exp}(E;\infty)$ by the results obtained at $E_\text{max}=11.2$~GeV.
Our final estimates of $R_\sigma^\text{exp}(E)$ are shown in the bottom panel of FIG.~\ref{fig:smeared}.

\bibliographystyleSM{apsrev4-2}
\bibliographySM{SM}

\end{document}